\documentclass[aps,prd,reprint,superscriptaddress,nofootinbib]{revtex4-2}
\usepackage{blindtext}
\usepackage{titlesec}
\usepackage{amsfonts}
\usepackage{amsmath}
\usepackage{amssymb}
\usepackage{multirow}
\usepackage{array} \newsavebox\cellbox
\usepackage{calc}
\usepackage{float}
\usepackage{mathtools}
\usepackage[usenames, dvipsnames, table]{xcolor}
\usepackage{booktabs}
\usepackage{makecell, cellspace}
\usepackage{color, colortbl}
\usepackage[caption=false]{subfig}
\usepackage{hyperref}
\hypersetup{
    colorlinks,
    linkcolor={red!60!black},
    citecolor={green!50!black},
    urlcolor={blue!70!black}
}
\usepackage{lipsum}

\newcolumntype{q}[2]{%
>{\begin{lrbox}\cellbox}%
    l%
    <{\end{lrbox}%
\makebox[#2][#1]{\usebox\cellbox}}}

\newcommand{\deflen}[2]{%
    \expandafter\newlength\csname #1\endcsname
    \expandafter\setlength\csname #1\endcsname{#2}%
}
\deflen{tenth}{\textwidth/11}

\DeclareMathOperator*{\argmax}{arg\,max}

\newcommand{\sbt}{\,\begin{picture}(-1,1)(-1,-3)\circle*{3}\end{picture}\ }

\begin{document}

\title{What to expect from scalar-tensor space geodesy}
\author{Hugo L\'{e}vy}
\email{hugo.levy@onera.fr}
\affiliation{DPHY, ONERA, Universit\'{e} Paris Saclay
F-92322 Ch\^{a}tillon - France}
\affiliation{Sorbonne Universit\'{e}, CNRS, UMR 7095, Institut d’Astrophysique de Paris, 98 bis bd Arago, 75014 Paris, France}
\author{Jo\"{e}l Berg\'{e}}
\affiliation{DPHY, ONERA, Universit\'{e} Paris Saclay
F-92322 Ch\^{a}tillon - France}
\author{Jean-Philippe Uzan}
\affiliation{Sorbonne Universit\'{e}, CNRS, UMR 7095, Institut d’Astrophysique de Paris, 98 bis bd Arago, 75014 Paris, France}

\begin{abstract}

Scalar-tensor theories with screening mechanisms come with non-linearities that make it difficult to study setups of complex geometry without resorting to numerical simulations. In this article, we use the  \textit{femtoscope}  code that we introduced in a previous work in order to compute the fifth force arising in the chameleon model in the Earth orbit. We go beyond published works by introducing a departure from spherical symmetry \textemdash \ embodied by a mountain on an otherwise spherical Earth \textemdash \ as well as by implementing several atmospheric models, and quantify their combined effect on the chameleon field.
Building on the numerical results thus obtained, we address the question of the detectability of a putative chameleon fifth force by means of space geodesy techniques and, for the first time, quantitatively assess the back-reaction created by the screening of a satellite itself. We find that although the fifth force has a supposedly measurable effect on the dynamics of an orbiting spacecraft, the imprecise knowledge of the mass distribution inside the Earth greatly curtails the constraining power of such space missions. Finally, we show how this degeneracy can be lifted when several measurements are performed at different altitudes.

\end{abstract}

\maketitle

\section{Introduction}
\label{sec:intro}

Scalar fields appear in most of the extensions beyond the standard models. Theories involving extra dimensions, from Kaluza-Klein theories up to string theories in the low energy limit, predict the existence of a light spin-0 particle. Scalar fields are also key ingredients in cosmology phenomenology, in particular for the dark sector and  inflation. Coupling the scalar field to matter\footnote{From a quantum mechanical perspective, the introduction of a scalar field in the gravity sector \textit{always} generate interactions between this scalar and matter fields \cite{review-screened-modgrav}.} automatically gives rise to a so-called \textit{fifth force}, resulting in deviations from general relativity (GR) in gravitational phenomena. Evading the Solar system tests of GR and laboratory experiments \cite{Will2014} comes at the price of introducing non-linearities in the model which enable \textit{screening mechanisms} (e.g. Damour-Polyakov \cite{damour-polyakov}, chameleon \cite{Khoury&WeltmanPRD, Khoury&WeltmanPRL}, K-mouflage \cite{k-mouflage1, k-mouflage2}, or Vainshtein \cite{vainshtein1, vainshtein2}).

Although screening mechanisms are precisely designed to recover GR \textemdash \ and thus in the weak field regime, Newtonian gravity \textemdash \ at Solar system scales, they leave nonetheless a small imprint which we can attempt to measure. Tests can be performed in a very wide range of length scales, from laboratory experiments \cite{atom-interferometry, casimir, torsion-pendulum}, to spacecraft in orbit around the Earth \cite{microscope-final-results, LARES-2, lares-lageos-grace} or traveling through the Solar system \cite{cassini}, planetary motion \cite{LLR1, LLR2, moon-vainshtein}, and to astrophysical tests \cite{astro-test1, astro-test2, astro-test3} (see Refs.~\cite{review-screened-modgrav, compendium-chameleon-constraints} and references therein for a more comprehensive review). Here, we are interested in the category of space-based experiments, which have long been expected to provide new constraints in the case of the chameleon model \cite{Khoury&WeltmanPRL}. Several space missions were successfully launched in the past decades: MICROSCOPE \cite{microscope-final-results, Bergé_2023} for testing the weak equivalence principle (see Refs.~\cite{mpb1, mpb2} for how constraints on the chameleon model could be derived from those data), Gravity Probe A and B \cite{gravity-probe-b}, LAGEOS 1 and 2, LARES 1 and 2 \cite{LARES-2}.

Beside these space missions specifically tailored for fundamental physics, artificial satellites have also given rise to space geodesy. Initially, space geodesy primarily focused on measuring the Earth's shape and size, but technological advancements have propelled it into a realm of unprecedented accuracy and multifaceted applications. Cutting edge instruments onboard satellites allow for the implementation of complementary geodetic techniques such as laser and Doppler ranging, Global Navigation Satellite Systems, gravimetry (e.g. GOCE, CHAMP, GRACE-FO satellite missions), etc. The determination of the Earth's figure (mass distribution) constitutes an inverse problem: given the data $d_{\mathrm{obs}}$ collected by the various satellite missions and a model describing the laws of gravitation $\mathcal{M}$ with forward map $F_{\mathcal{M}}$, the goal is to determine the model parameters $p$ such that the residual $d_{\mathrm{obs}} - F_{\mathcal{M}}(p)$ is minimized (in some specific sense, e.g. least-squares or probabilistic approaches). In space geodesy, this inverse problem is solved with the central assumption that the governing equation is Newton's law of gravity (and $p$ would represent the distribution of mass) \cite{Moritz2015}.

The goal of the present article is to assess the pertinence of orbitography techniques to test screened scalar-tensor theories, illustrated with the chameleon model, and to characterize the \textit{best site} in the Solar system to perform such tests. This is a follow-up to our previous article Ref.~\cite{Levy_2022} where we laid the foundations in terms of numerical simulations. There, we saw that the unconstrained region of the chameleon parameter space (see Fig.~3 of Ref.~\cite{Yin2022-chameleon-constraints}) corresponds to a situation where the Earth is screened, i.e. where the chameleonic force is sourced only by its outer layers. This mere observation suggests that the local landform \textemdash \ specifically any local deviation from spherical symmetry \textemdash \ can leave a significant imprint on the chameleon profile. Consequently, if the chameleon's effects differ sufficiently from Newtonian gravity, it should leave a distinctive signature on the Earth's gravity.

Mountains and craters are typical examples of asphericities that can be sensed through space geodesy. Relative to the size of a planet, a mountain represents a spiky feature. Several works bring to light the parallel between chameleon (and symmetron) gravity in the screened regime and electrostatics: the behavior of the scalar field is roughly the same as the behavior of the electrostatic potential for a perfect conductor\footnote{Indeed, it can be shown that the equation of motion of the chameleon field in the quasi-static Newtonian limit with thin-shell can be well-approximated by the electrostatic potential equation. Then, same differential equations lead to same solutions.} \cite{electrostatic-analogy-1, electrostatic-analogy-2, electrostatic-analogy-3}. Taking the analogy a step further, Ref.~\cite{electrostatic-analogy-1} mentions the ``lightning rod effect" in electromagnetism, exhibited by needle-like conductors around which the electric field ($\propto$ gradient of the potential) is enhanced. In the case of the chameleon, the counterpart of the electric field would be the fifth force ($\propto$ gradient of the scalar field) \textemdash \ making the mountain an interesting case study. Nevertheless as Ref.~\cite{electrostatic-analogy-3} underlines, while this analogy provides valuable qualitative insights, numerical computations remain essential to establish a quantitative connection with real-world observations and experimental data. In that respect, we aim to address the long-standing question of how much an atmosphere \textit{smooths out} the mountain's contribution to the fifth force in space. More generally, existing work accounting for the atmosphere \cite{Khoury&WeltmanPRD, Khoury&WeltmanPRL, waterhouse2006, chameleon-scalar-waves, ss-constraints-chameleon, Mota2007} are, in our opinion, not extensive enough: the models are not realistic (one layer of constant density) and conclusions are drawn on qualitative arguments that can be misleading (see e.g. the introduction of Ref.~\cite{EP-cham}). We shall also pay attention to the influence of a spacecraft on the background field, and evaluate how this perturbation impacts the overall fifth force that it experiences.

The article is organized as follows. In Sec.~\ref{sec:methods}, we briefly recall the main equations describing both Newtonian and chameleon gravity, and give precise meaning to physical models outlined above, namely the modeling of the mountainous planet together with its atmosphere. In this setup, the total gravitational potential is computed numerically using \textit{femtoscope}, a code that was specifically designed to solve these equations with asymptotic boundary conditions \cite{Levy_2022}. It allows for the computation of both the Newtonian potential and the chameleon field in space. The numerical results are presented and discussed in Sec.~\ref{sec:profiles}. We explore a vast region of the chameleon parameter space and ascertain the influence of an atmosphere in several scenarios, making this a quite comprehensive study compared to what has been done in previous work. Finally, Sec.~\ref{sec:orbits} takes us back to space geodesy as we compare the dynamics of a spacecraft with and without a fifth force acting on it as it orbits the mountainous planet. We address the issue of being able to discriminate between the two in the presence of model uncertainties, and further suggest ways to break this source of degeneracy. These analyses pave the way to the design of orbitography experiments in the Solar system and their subtle interpretation. We conclude in Sec.~\ref{sec:discussion}.

\section{Model \& Numerical techniques}
\label{sec:methods}

\subsection{General equations}
\label{subsec:equations}

\subsubsection{Newtonian gravity}
\label{subsubsec:newton}

It is well known that, in the weak-field regime and when the sources are moving very slowly compared to the speed of light, GR  reduces to Newtonian gravity which is described by the \textit{Newtonian potential} $\Phi$ with dimension $[L^2 \cdot T^{-2}]$. For a static configuration, we define it as
\begin{equation}
    \Phi(\mathbf{x}) = - G \int_{\mathbb{R}^3} \frac{\rho(\mathbf{x'})}{\|\mathbf{x}-\mathbf{x'}\|} \, \mathrm{d}^3 x' \ ,
\label{eqn:newton-potential-integral}
\end{equation}
where $G$  is the Newtonian gravitational constant and $\rho = \rho(\mathbf{x})$ is the matter density function which depends on position $\mathbf{x}$. Assuming that the \textit{weak equivalence principle} holds perfectly (Ref.~\cite{microscope-final-results} shows that it holds at less than $10^{-15}$) and from a classical mechanics perspective, the gravitational acceleration undergone by a point-like particle is simply $\mathbf{a}_{\Phi} = - \boldsymbol{\nabla} \Phi$. Eq.~(\ref{eqn:newton-potential-integral}) provides a straightforward way of computing the Newtonian potential by evaluating some three-dimensional integral (see e.g. Ref.~\cite{Fukushima2016}). However, it may be more convenient from a numerical standpoint to solve the following Poisson's equation
\begin{equation}
    \Delta \Phi = 4 \pi G \rho \, ,
\label{eqn:poisson-potential}
\end{equation}
implied by the definition of $\Phi$. Indeed, on the one hand one has to evaluate the integral appearing in Eq.~(\ref{eqn:newton-potential-integral}) for each point $\mathbf{x}$ where the Newtonian potential is sought, whereas on the other hand solving the partial differential equation (PDE) (\ref{eqn:poisson-potential}) provides an approximation of $\Phi$ over the whole numerical domain.

Assuming that the mass density vanishes as one moves away from the source of gravity, the gravitational acceleration $\mathbf{a}_{\Phi} = - \boldsymbol{\nabla} \Phi$ is expected to decay to zero at infinity. The essential boundary condition is therefore defined at infinity and a very common choice for the constant of integration is
\begin{equation}
    \Phi(\mathbf{x}) \underset{\|\mathbf{x}\| \to +\infty}{\longrightarrow} 0 \, .
\label{eqn:newton-potential-BC}
\end{equation}

\subsubsection{Chameleon gravity}
\label{subsubsec:chameleon}

\begin{figure}
    \centering
    \includegraphics[width=0.9\linewidth]{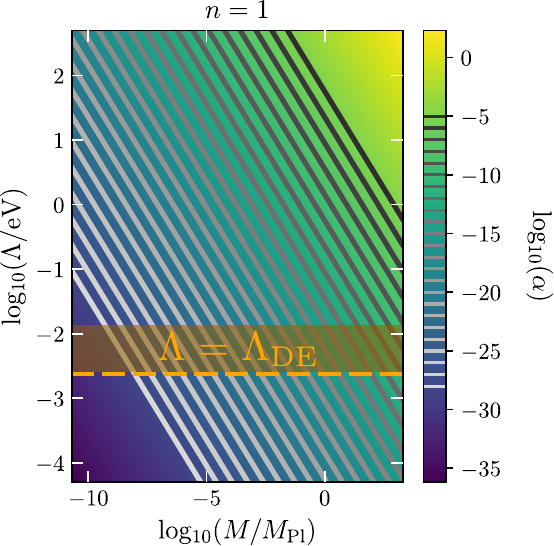}
    \caption{Mapping from the chameleon parameter space in the plane $n=1$ to the dimensionless parameter $\alpha$ appearing in Eq.~(\ref{eqn:KG-alpha}), where $M = M_{\mathrm{Pl}} / \beta$. The gray lines represent the iso-values of the $\alpha$ parameter covered in this study, ranging from $10^{-5}$ to $10^{-28}$. The orange horizontal dashed line corresponds to $\Lambda = \Lambda_{\mathrm{DE}} = 2.4 \times 10^{-3} \, \mathrm{eV}$, the dark energy scale.}
    \label{fig:param-space}
\end{figure}

In the Newtonian limit, the chameleon field $\phi$ is governed by a nonlinear Klein-Gordon equation which takes the form
\begin{equation}
  \Delta \phi = \frac{\mathrm{d}V_{\mathrm{eff}}}{\mathrm{d}\phi} = \frac{\beta}{\mathrm{M_{Pl}}} \rho - \frac{n \Lambda^{n+4}}{\phi^{n+1}} \, ,
\label{eqn:KG-dim}
\end{equation}
where $\mathrm{M_{Pl}} \equiv 1/\sqrt{8 \pi G}$ is the reduced Planck mass and $V_{\mathrm{eff}}$ is the so-called \textit{effective potential} of the scalar field. The model further has three parameters \textemdash \ $\beta$ a positive dimensionless constant which encodes the coupling of the scalar field to matter, $\Lambda$ a mass scale and $n$ a natural number. The 3-acceleration experienced by a point-like particle induced by its coupling to the chameleon field is proportional to the gradient of the scalar field and takes the form
\begin{equation}
    \mathbf{a}_{\phi} = - \frac{\beta}{\mathrm{M_{Pl}}} \boldsymbol{\nabla} \phi \, .
\label{eqn:a-phi}
\end{equation}
If we assume that the density uniformly decays to some vacuum density $\rho_{\mathrm{vac}}$ far away from the source, then the chameleonic acceleration is expected to decay to zero at infinity, just as in the Newtonian gravity case discussed above. Equating the r.h.s. of Eq.~(\ref{eqn:KG-dim}) to zero and solving for $\phi$ yields the following asymptotic boundary condition:
\begin{equation}
    \phi(\mathbf{x}) \underset{\|\mathbf{x}\| \to +\infty}{\longrightarrow} \left( \mathrm{M_{Pl}} \frac{n \Lambda^{n+4}}{\beta \rho_{\mathrm{vac}}} \right)^{\frac{1}{n+1}} \equiv \phi_{\mathrm{vac}} \, .
\label{eqn:chameleon-BC}
\end{equation}

In Ref.~\cite{Levy_2022}, we introduced \textit{femtoscope} \textemdash \ a \textsc{python} numerical tool based on the finite element method which enables us to solve Eq.~(\ref{eqn:KG-dim}) on spatially unbounded domains. We perform the same nondimensionalization as in Refs.~\cite{Levy_2022, Briddon_2021} by introducing (i) $\rho_0$ a characteristic density of the problem, (ii) $\phi_0 \equiv (n \mathrm{M_{Pl}} \Lambda^{n+4}/\beta \rho_0)^{1/(n+1)}$ the expectation value of the chameleon field in an ambient medium of density $\rho_0$ and (iii) $L_0$ a characteristic length scale of the system under study. Denoting the new dimensionless quantities with a tilde, trivial algebra leads to
\begin{equation}
\begin{aligned}
  & \alpha \Tilde{\Delta} \Tilde{\phi} = \Tilde{\rho} - \Tilde{\phi}^{-(n+1)} \, , \\
  \text{with} \ & \alpha \equiv \left( \frac{\mathrm{M_{Pl} \Lambda}}{\beta L_0^2 \rho_0} \right) \left( \frac{n \mathrm{M_{Pl}} \Lambda^3}{\beta \rho_0} \right)^{1/(n+1)} \, .
\end{aligned}
\label{eqn:KG-alpha}
\end{equation}
The mapping $(\beta, \Lambda) \mapsto \alpha$ for $n=1$ is illustrated in Fig.~\ref{fig:param-space}. Note that Eq.~(\ref{eqn:KG-alpha}) now only depends on two parameters, $\alpha$ and $n$, instead of the three initial ones, which allows for a more efficient numerical exploration of the chameleon parameter space\footnote{Naturally, the mapping $(\beta, \Lambda, n) \mapsto (\alpha, n)$ described above is not bijective.}. The chameleonic acceleration (\ref{eqn:a-phi}) then scales as
\begin{equation}
    \mathbf{a}_{\phi} \propto \Lambda^{\frac{n+4}{n+1}} \beta^{\frac{n}{n+1}} \Tilde{\boldsymbol{\nabla}} \Tilde{\phi} \, .
\end{equation}
We denote $a_0$ the multiplicative constant appearing in front of the dimensionless gradient, which reads
\begin{equation}
\begin{split}
  \displaystyle a_0 \, \mathrm{[m/s^2]} = ( \Lambda \, \mathrm{[eV]}  \times \mathfrak{e} \, & \mathrm{[J/eV]} )^{\frac{n+4}{n+1}} \\
  & \frac{\beta^{\frac{n}{n+1}}}{M_{\mathrm{Pl}} L_0} \left[ \frac{n M_{\mathrm{Pl}}}{\rho_0 (\hbar c)^3} \right]^{\frac{1}{n+1}} \, .
\end{split}
\label{eqn:a0}
\end{equation}
In Eq.~(\ref{eqn:a0}), physical quantities are expressed in SI units unless specified using square brackets and $\mathfrak{e} \sim 1.6022 \times 10^{-19} \, \mathrm{J/eV}$ is the conversion factor from electron-volts to joules. As a rule of thumb, the smaller $\alpha$, the more screened the setup. All physical results issued in this article are evaluated with $L_0 = R_{\oplus} = 6371 \, \mathrm{km}$ (the Earth radius) and $\rho_0 = 1 \, \mathrm{kg/m^3}$.

The Newtonian potential and the chameleon field do not have the same physical dimension. In order to be able to compare these two quantities, we define a new field
\begin{equation}
    \Psi = \frac{\beta}{M_{\mathrm{Pl}}} \phi
\end{equation}
which can be expressed in $\mathrm{m^2/s^2}$. We refer to $\Psi$ as the \textit{chameleon potential} since it plays the same role as $\Phi$. The total gravitational acceleration undergone by a point-like particle will simply be $-\boldsymbol{\nabla}(\Phi + \Psi)$. Furthermore, the term `fifth-force' will be used loosely throughout this article. Most occurrences of it should be taken as a synonym for `chameleon acceleration', i.e. a quantity homogeneous to an acceleration and not a force per say. Finally, we will often refer to the `screened regime' or to the `thin-shell of a body' in this article. These notions can be given precise meanings now that we have introduced the main notations. A macroscopic body is said to be \textit{screened} when the chameleon field reaches the value that minimizes its effective potential $V_{\mathrm{eff}}$ deep inside the body. In that case, the field remains essentially frozen in that body except in a (usually) thin surface layer, which is referred to as the thin-shell.

\subsection{Physical models}
\label{subsec:models}

\subsubsection{Mountains}
\label{subsubsec:mountains}

\begin{figure}
    \centering
    \includegraphics[width=\linewidth]{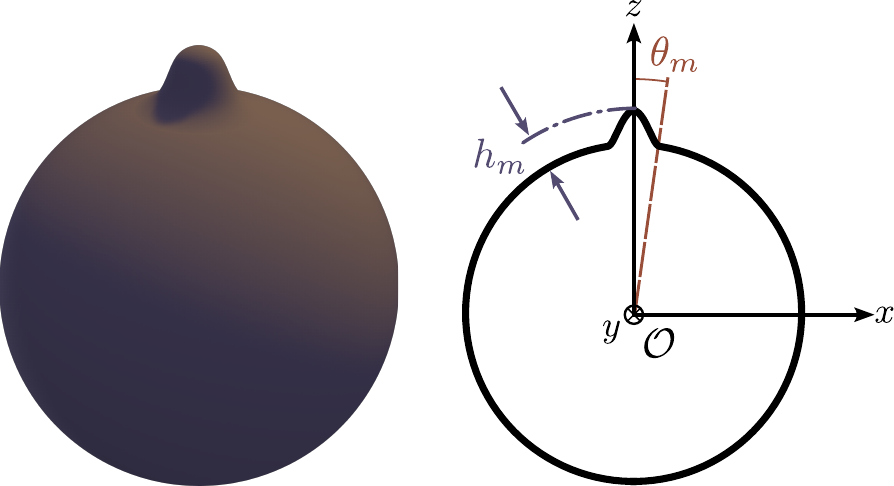}
    \caption{Mountain visualization and notations. The Cartesian frame $(\mathcal{O}, x, y, z)$ is centered at the geometric center of the sphere devoid of mountain. The actual mountain profile used in numerical computations is drawn using B-splines in polar coordinates so as to form a smooth manifold.}
\label{fig:mnt-views}
\end{figure}

At first order and seen from afar, planetary-mass objects have a rounded, ellipsoidal shape due to their self-gravity and rotation. It is only when we take a closer look at such bodies in the Solar System that smaller, more complex features become visible: mountains, ridges, craters, volcanoes, etc. This rich variety of topographies results in perturbations (with respect to the spherically symmetric case) in the gravitational field which, in the case of the Earth, can be measured by geodetic satellites. With a view to understand how $\mathrm{5^{th}}$-forces affect Newtonian gravity in the vicinity of these topographical features, it is desirable to first work with a simple toy-model. We thus consider a spherical body together with a single, axisymmetric mountain on top of it as depicted in Fig.~\ref{fig:mnt-views}. It is mainly described by two  dimensionless parameters:
\begin{itemize}
    \item[--] $h_m$, the height of the mountain divided by the radius $R_{\mathrm{body}}$ of the spherical body (which is unitary on Fig.~\ref{fig:mnt-views});
    \item[--] $\theta_m$, the mountain's half-angle.
\end{itemize}
Note that these two parameters are deliberately exaggerated on Fig.~\ref{fig:mnt-views} for better visualization, and are clearly not representative of any \textit{realistic} mountain in the Solar system \textemdash \ see Table~\ref{tab:SS-sites}. All numerical computations presented in this article were performed with $h_m = 10^{-2}$ and $\theta_m = 10^{-2}$ rad comparatively.
The resulting setup is itself axisymmetric which means FEM computations can be performed in two dimensions rather than three, greatly reducing computational complexity.

For the model to be complete, we further need to specify the density function $\rho(\mathbf{x})$ inside and outside the body. For the sake of simplicity, we assign a constant density to the body $\rho_{\mathrm{body}}$. The body may or may not be surrounded by an atmosphere. In either case, the density outside the body depends solely on the radial distance from the center $r$ and always goes down to a constant vacuum value $\rho_{\mathrm{vac}}$. For all FEM computations, we set
\begin{equation*}
    \Tilde{\rho}_{\mathrm{body}} = \frac{\rho_{\mathrm{body}}}{\rho_0} = 10^3 \ \text{and} \ \Tilde{\rho}_{\mathrm{vac}} = \frac{\rho_{\mathrm{vac}}}{\rho_0} = 10^{-15} \, .
\end{equation*}

Additionally, we will work most of the time with the dimensionless variable  $\Tilde{r} = r/L_0$, and set $L_0 = R_{\mathrm{body}}$. The various fields involved in this study (chameleon potential, Newtonian potential, together with their gradient) will be probed at fixed discrete values of $\Tilde{r}$ for the sake of consistency. We made the choice to show results for $\Tilde{r} \in \{1.059, 1.111, 1.314, 4.645, 6.617\}$, which for the case of the Earth corresponds roughly to peculiar orbits: the International Space Station, MICROSCOPE, a Medium Earth orbit, Galileo and geostationary satellites, respectively.

\begin{table*}
\caption{\label{tab:SS-sites}List of some peculiar mountains in the Solar system\footnote{Mainly based on \url{https://en.wikipedia.org/wiki/List\_of\_tallest\_mountains\_in\_the\_Solar\_System}, last visited: August $22^{\mathrm{th}}$, 2023}.}
\begin{tabular}{w{c}{2\tenth} w{c}{2\tenth} w{c}{1.25\tenth} w{c}{1.25
\tenth} w{c}{\tenth} w{c}{\tenth} w{c}{1.75\tenth}} \hline\hline
\multirow{2}{*}{Site} & \multirow{2}{*}{Body density [$\mathrm{kg/m^3}$]} & \multicolumn{2}{c}{Atmosphere} & \multicolumn{2}{c}{height (base to peak)} & \multirow{2}{*}{$\theta_m$ [$\mathrm{rad}$]}\\
& & \scriptsize density [$\mathrm{kg/m^3}$] & \scriptsize thickness [km] & \scriptsize [km] & \scriptsize $h_m$ & \\\midrule
\makecell{Earth\\ Mount Everest} & \makecell{$2.6 \times10^3$\\(Earth crust)} & 1.2 (sea level) & $\sim 100$ & 4.6 & $7.2 \times 10^{-4}$ & $\sim 10^{-3}$\\[10pt]
\makecell{Earth\\ Mauna Kea} & \makecell{$2.6 \times10^3$\\(Earth crust)} & 1.2 (sea level) & $\sim 100$ & 10.2 & $1.6 \times 10^{-3}$ & $\sim 10^{-2}$\\[10pt]
\makecell{Mars\\ Mons Olympus} & \makecell{2582\\ (Mars crust)} & $2 \times 10^{-2}$ (max.) & $\sim 10$ & 21.9 & $6.5 \times 10^{-3}$ & $\sim 9 \times 10^{-2}$\\[10pt]
\makecell{Moon\\ Mons Huygens} & \makecell{2550\\ (Moon crust)} & \multicolumn{2}{c}{no atmosphere} & 5.5 & $3.2 \times 10^{-3}$ & $\sim 6 \times 10^{-2}$\\[10pt]
\makecell{Io\\ Boösaule Montes} & \makecell{3500\\ (mean density)} & $< 10^{-6}$ & \textemdash & 18.2 & $10^{-2}$ & $\sim 1.5 \times 10^{-2}$\\[10pt]
\makecell{Vesta\\ Rheasilvia central peak} & \makecell{2800\\ (crust estimate)} & \multicolumn{2}{c}{no atmosphere} & 25 & $10^{-2}$ & $\sim 0.4$ \\[5pt]\hline\hline
\end{tabular}
\end{table*}

\subsubsection{Atmospheres}
\label{subsubsec:atm-models}

Some Solar system bodies are surrounded by an atmospheric layer \textemdash \ a gas envelop held in place by the gravity of the body. This slight over-density with respect to the case with no-atmosphere is expected to have an influence on the chameleon field profile and, therefore, on the $\mathrm{5^{th}}$-force in space \cite{Khoury&WeltmanPRD, Khoury&WeltmanPRL, Mota2007}. However, works that take account of the atmosphere often model it as an additional shell of matter with constant density satisfying $\rho_{\mathrm{body}} > \rho_{\mathrm{atm}} > \rho_{\mathrm{vac}}$, or at best as a constant piecewise function \cite{Khoury&WeltmanPRD, waterhouse2006, chameleon-scalar-waves, ss-constraints-chameleon, Mota2007}. It is actually difficult to be more precise than this using analytical techniques only. Here, we go a step further by taking advantage of \textit{femtoscope} to analyze the chameleon field profile in more realistic atmospheric setups. To avoid confusion, the requirement that the atmosphere must have a thin-shell stipulated in Ref.~\cite{Khoury&WeltmanPRD} only holds in the case of non-universal coupling, wherein unacceptably large violations of the weak equivalence principle would be observed in ground based experiments. Here, we work on the assumption of a universal coupling (characterized by a single dimensionless constant $\beta$) and so there is no particular reason for imposing this condition\footnote{The fact remains that, even in the case of a universal coupling, deviations from the inverse square law can be suppressed by the atmosphere.}.

\begin{figure}
    \centering
    \includegraphics[width=\linewidth]{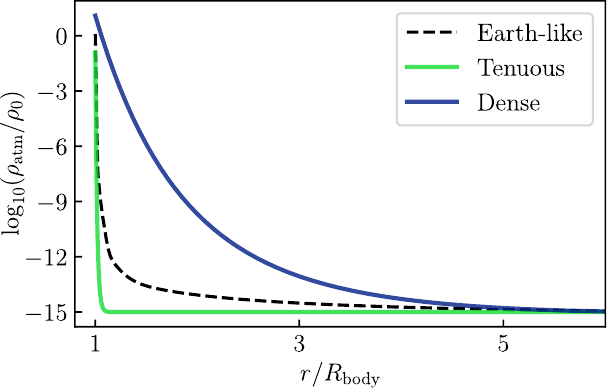}
    \caption{Atmospheric profiles investigated in this study.}
    \label{fig:density-profiles}
\end{figure}

Three atmospheric density profiles are considered in this study: \textit{Earth-like}, \textit{Tenuous} and \textit{Dense}. The Earth-like model is built from the 1976 version of the U.S. Standard Atmosphere model \cite{us76}, commonly known as the US76 model\footnote{Data downloaded from \url{http://www.braeunig.us/space/atmos.htm}, (especially for the density between $1000 \, \mathrm{km}$ - $36000 \, \mathrm{km}$ altitude). Last visited: June $\mathrm{1^{st}}$, 2022.}. It provides an estimate of the Earth atmospheric density $\rho_{\text{\tiny US}}$ as a continuous function of the altitude, up to $R_{\mathrm{atm}} \sim 36 \times 10^3 \, \mathrm{km}$. Because we want the minimum dimensionless density in the numerical domain to be exactly $\Tilde{\rho}_{\mathrm{vac}} = 10^{-15}$, we apply the following transformation on the original data:
\begin{equation} \notag
\begin{split}
    & \log \Tilde{\rho}_{\text{\tiny Earth-like}} = \log \Tilde{\rho}_{\text{\tiny US}} + k \Bigl[\log \Tilde{\rho}_{\text{\tiny US}} - \log (\min \Tilde{\rho}_{\text{\tiny US}})\Bigr] \\[5pt]
    & \text{with } k = \frac{\log(\Tilde{\rho}_{\text{\tiny US}} / \Tilde{\rho}_{\mathrm{vac}})}{\log(\max \Tilde{\rho}_{\text{\tiny US}} / \min \Tilde{\rho}_{\text{\tiny US}})}
\end{split}
\label{eqn:rho-earth-like}
\end{equation}
for $r < R_{\mathrm{atm}}$, which is nothing but an affine transformation on the logarithmic densities. Beyond $R_{\mathrm{atm}}$, we set $\Tilde{\rho}_{\text{\tiny Earth-like}} = \Tilde{\rho}_{\mathrm{vac}}$. The other two models \textemdash \ Tenuous and Dense \textemdash \ are purely empirical in the sense that they are not based on actual atmospheric data. Both are constructed via the expression
\begin{equation}
    \log \Tilde{\rho}(r) =
    \begin{cases}
        A \exp \left[ \dfrac{(r - R_{\mathrm{atm}})^2}{\sigma^2} \right] + B & \text{if } r<R_{\mathrm{atm}} \\
        \log \Tilde{\rho}_{\mathrm{vac}} & \text{otherwise}
    \end{cases} \, ,
\notag
\label{eqn:rho-tenuous-dense}
\end{equation}
where the parameters $(A, B, \sigma)$ are adjusted by hand to obtain either a very tenuous, thin atmosphere or a very dense, thick one instead. The resulting density profiles are depicted in Fig.~\ref{fig:density-profiles}.

\subsection{Decomposition of scalar fields into spherical harmonics}
\label{subsec:SH-conventions}
In geophysics and physical geodesy, the Earth gravitational potential is conveniently modeled as a spherical harmonics expansion \cite{EGM2008}. Any well-behaved function $f \colon \mathbb{R}^3 \to \mathbb{R}$ may be decomposed as
\begin{equation}
    f(r, \mathbf{n}) = \sum_{l=0}^{+\infty} \sum_{m=-l}^{+l} f_{lm}(r) Y_{lm}(\mathbf{n}) \, ,
\label{eqn:sph-decomposition-1}
\end{equation}
where $r$,  $\mathbf{n} = (\theta, \varphi)$ refer to spherical coordinates, $Y_{lm}$ is the \textit{real} spherical harmonic function of \textit{degree} $l$ and \textit{order} $m$ (see Ref.~\cite{SHTools} for its definition), and $f_{lm}$ are the spherical harmonic coefficients that only depend on the radial coordinate \textemdash \ they are referred to as the \textit{bare coefficients} in this article. There are several normalization conventions for an unequivocal definition of spherical harmonic functions. In this study, we stick to the \textit{orthonormalized} convention for which
\begin{equation}
    \int_{\mathcal{S}^2} \, Y_{lm}(\mathbf{n}) Y_{l'm'}(\mathbf{n}) \, \mathrm{d}^2\Omega = \delta_{ll'} \, \delta_{mm'} \, ,
\label{eqn:sph-orthonormalization}
\end{equation}
where $\mathcal{S}^2$ is the unit 2-sphere, $\mathrm{d}\Omega$ is the differential surface $\sin(\theta) \mathrm{d}\theta \mathrm{d} \varphi$ and $\delta_{ij}$ is the Kronecker delta function. The notations used to refer to the spherical harmonic coefficients of the Newtonian potential $\Phi$ and the chameleon potential $\Psi$ are gathered in Table~\ref{tab:sph-coefficients}.

\begin{table}
\def\arraystretch{1.4}
\caption{\label{tab:sph-coefficients} Notations for the spherical harmonic coefficients.}
\begin{ruledtabular}
\begin{tabular}{lcc}
   & Bare coefficients & Rescaled coefficients \\
 Newtonian potential & $\Phi_{lm}(r)$ & $y_{lm}^N$ \\
 Chameleon potential & $\Psi_{lm}(r)$ & $y_{lm}^C(r)$
\end{tabular}
\end{ruledtabular}
\end{table}

\subsubsection{Rescaled coefficients}
\label{subsubsec:rescaled-coeffs}
The bare spherical harmonic coefficients of the Newtonian potential $\Phi_{lm}$ further exhibit a scaling property. Let us denote by $\mu_{\mathrm{body}} \equiv G M_{\mathrm{body}}$ the standard gravitational parameter of the central body of mean radius $R_{\mathrm{body}}$ and mass $M_{\mathrm{body}}$. Then, the rescaled coefficients
\begin{equation}
    y_{lm}^N = \frac{r}{\mu_{\mathrm{body}}}  \left( \frac{r}{R_{\mathrm{body}}} \right)^l \Phi_{lm}(r)
\label{eqn:rescaled-newton-coeff}
\end{equation}
can be shown to be independent of $r$ \cite{Bergé_2018},\footnote{The numerical values of $\mu_{\mathrm{body}}$ and $R_{\mathrm{body}}$ could in theory be chosen arbitrarily. However the numerical values of the rescaled coefficients are tied to this choice.} owing to the specific form of the Newtonian potential (\ref{eqn:newton-potential-integral}). Such rescaled coefficients are thus \textit{universal} to the body under consideration. Similarly to Eq.~(\ref{eqn:rescaled-newton-coeff}), we denote by $y_{lm}^C(r)$ the rescaled coefficients of the chameleon potential which, for their part, have no particular reason to be independent of the radial distance. In that sense, Ref.~\cite{Bergé_2018} shows the explicit dependence of such coefficients with respect to $r$ in the case of a Yukawa interaction.

This relation can also serve as a means of checking the numerical results obtained for the Newtonian potential. This test is performed in Appendix~\ref{app:scaling-relation}.

\subsubsection{Recovery of the coefficients}
We use the software \texttt{SHTools} \cite{SHTools} to compute the spherical harmonic coefficients of the scalar fields of interest. The \textsc{Python} package \texttt{pyshtools} comes with the routine \texttt{SHGrid.expand} which calculates the coefficients by means of some numerical quadrature\footnote{In this study, we use a $\mathsf{N \times 2N}$ \textit{Driscoll and Healy} sampled grid.}. The only detail worth mentioning is the fact that this routine outputs separate variables for the \textit{cosine} $C_{lm}$ and \textit{sine} $S_{lm}$ coefficients (sometimes referred to as the \textit{Stokes coefficients}). The conversion from $(C_{lm}, S_{lm})$ to bare coefficients is outlined in Appendix~\ref{app:coefficient-conversion} \textemdash \ Eq.~(\ref{eqn:stokes2bare-summary}).

\subsection{Numerical techniques}
\label{subsec:num-tech}

\subsubsection{Using \it{femtoscope} to solve linear and nonlinear PDEs with asymptotic boundary conditions}
\label{subsubsec:femtoscope}

As mentioned earlier, \textit{femtoscope} is a ready-to-use \textsc{Python} program which plays a central role in this study as it enables us to compute both the Newtonian potential and the chameleon field by solving Eqs.~(\ref{eqn:poisson-potential}) and (\ref{eqn:KG-dim}) respectively. It is based on the finite element method \textemdash \ building on top of the open-source package \textit{Sfepy} \cite{sfepy} \textemdash \ and further implements techniques to deal with nonlinearities and asymptotic boundary conditions (\ref{eqn:newton-potential-BC}, \ref{eqn:chameleon-BC}).

The proper treatment of these asymptotic boundary conditions is of noticeable importance in this study. Indeed, it is tempting to simply truncate the numerical domain at a fixed radius and apply a homogeneous Dirichlet boundary condition on the artificial border resulting from that process. This procedure has several flaws:
\begin{enumerate}
    \item For the error that arise therefrom to be small, the domain must be sufficiently large, which translates to higher computational cost.
    \item Selecting the size of that domain is a \textit{blind experiment} in the sense that the dependence of the error on the truncation radius is not easily accessible without additional tricks.
    \item It wantonly imposes spherical symmetry on the solution as we approach the artificial boundary. This is particularly undesirable in this study where we are interested in the small deviations from spherical symmetry introduced by the presence of the mountain.
\end{enumerate}
This latter point is illustrated on Fig.~\ref{fig:newton-ortho-profiles-techniques-comparison} where it can be seen that, as we approach the artificial boundary, the truncation method (labeled `FEM bounded', dash-dotted pink line) exhibits a poor approximation.

Instead, we employ a technique based on the splitting of the numerical domain $\Omega$ into two subdomains $\Omega_{\mathrm{int}}$ and $\Omega_{\mathrm{ext}}$ such that $\Bar{\Omega} = \Bar{\Omega}_{\mathrm{int}} \cup \Bar{\Omega}_{\mathrm{ext}}$. $\Omega_{\mathrm{int}}$ is the bounded, interior domain, while $\Omega_{\mathrm{ext}}$ is the unbounded, exterior domain. An \textit{inversion} transform is then applied to $\Omega_{\mathrm{ext}}$, resulting in a bounded domain $\Tilde{\Omega}_{\mathrm{ext}}$ (called the \textit{inversed exterior domain}) which can be meshed on a computer. There are many possible numerical implementations based on this method, see e.g. Refs.~\cite{oh2003, boulmezaoud2005, kelvin-review}. In this study, we make use of the so-called \textit{virtual connection of DOFs} described in our previous work \cite{Levy_2022}.

\subsubsection{Numerical challenges and verification}
\label{subsubsec:challenges-verif}

\begin{figure}
    \centering
    \includegraphics[width=0.9\linewidth]{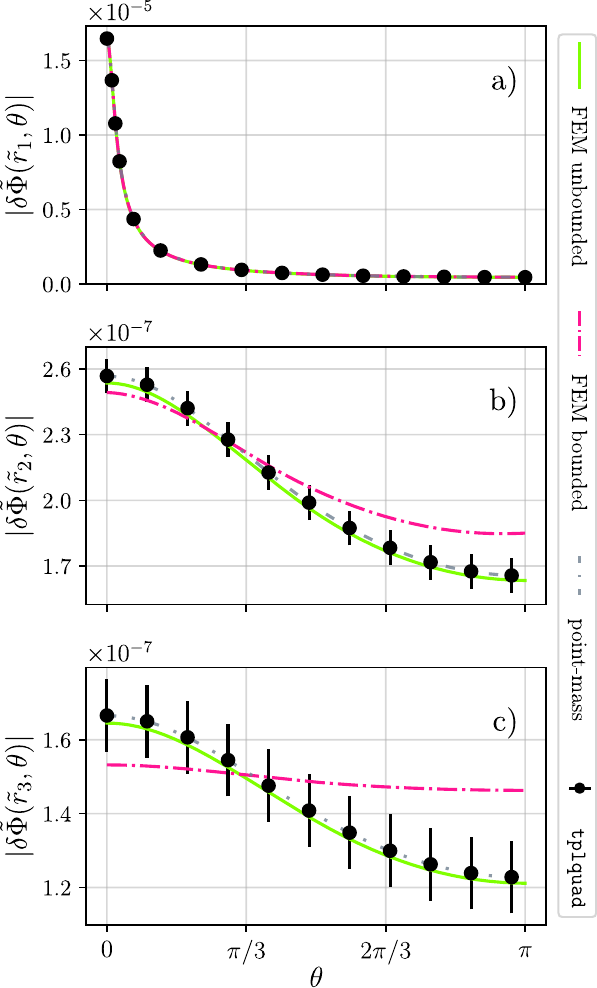}
    \caption{Orthoradial profiles of the dimensionless Newtonian potential $\delta \Tilde{\Phi}$ sourced by the mountain at three different altitudes, corresponding to $\Tilde{r}_1 = 1.059$, $\Tilde{r}_2 = 4.645$, and $\Tilde{r}_3 = 6.617$ (top, middle and bottom panels respectively). The black dots together with their error bar represent the benchmark solution, obtained through the computation of the integral Eq.~(\ref{eqn:newton-potential-integral}) with \texttt{scipy}'s \texttt{tplquad} routine. The pink dash-dotted line is obtained by solving Poisson's equation (\ref{eqn:poisson-potential}) with an homogeneous Dirichlet boundary condition applied at $\Tilde{r} = \Tilde{R}_{\mathrm{c}} = 7$ while the green solid line is the solution provided by \textit{femtoscope} with asymptotic boundary condition. Finally, the gray dash-dotted line is an analytical approximation where the mountain is replaced by a point-mass, whose location and mass were fitted to provide a good match with respect to the benchmark: $m_{\mathrm{mountain}}/M_{\mathrm{body}} = 2.23 \times 10^{-7}$ and $z/R_{\mathrm{body}} = 2.22 \times 10^{-3}$.}
    \label{fig:newton-ortho-profiles-techniques-comparison}
\end{figure}

There are several inconspicuous challenges associated with the numerical computation of the field profiles in the setup described in Sec.~\ref{subsec:models}. To start with, let us stress the fact that we are looking for small deviations from spherical symmetry, owing to the presence of a very localized over-density at the pole that we here call a mountain. Quantitatively speaking, a back-of-the-envelope calculation shows that \textemdash \ at a fixed altitude $h$ \textemdash \ the relative variation of the Newtonian potential  $\Phi(R_{\mathrm{body}}+h, \theta)$ along the latitudes with respect to its mean value at this altitude is no larger than a few $10^{-6}$. The higher we go, the smaller this ratio, which means our numerical approximations have to be correct up to at least seven significant digits to be deemed \textit{good}. This mere order-of-magnitude calculation raises an additional concern: how do we actually check that the numerical approximations we obtain are compliant with the required levels of precision?

The Poisson's equation (\ref{eqn:poisson-potential}) governing the Newtonian potential being linear, it is possible to apply the superposition principle, where the total field is simply the mountain's contribution on top of a spherically symmetric background: $\Phi_{\mathrm{tot}}(r, \theta) = \delta \Phi(r, \theta) + \Phi_0(r)$. Turning to the chameleon field, the nonlinearity in the r.h.s. of the Klein-Gordon equation (\ref{eqn:KG-dim}) prevents us from following the same path. Even if one were to decompose the chameleon field as $\phi_{\mathrm{tot}}(r, \theta) = \delta \phi(r, \theta) + \phi_0(r)$, the term $(\phi_0 + \delta \phi)^{-(n+1)}$ becomes linearizable only under the assumption that $\delta \phi \ll \phi_0$ \textit{everywhere}. Unfortunately, this assumption has no reason to hold in all scenarios, owing to the very nature of the screening mechanism. Indeed, it is far from being valid in the case where the mountain itself becomes screened, which turns out to be the most interesting case given the current constraints on the chameleon field \cite{Yin2022-chameleon-constraints}. For lack of a better workaround, we abandoned perturbation-based techniques and put our efforts into solving for the full field. It is therefore necessary to compare the FEM approximation obtained with \textit{femtoscope} against some benchmark. Failing to have an analytical solution for the Newtonian potential of a mountain, we can still resort to the numerical integration of Eq.~(\ref{eqn:newton-potential-integral}). In this respect, we use \texttt{scipy}'s \texttt{tplquad} routine \cite{scipy} to evaluate the integral with an estimated relative error of a few $10^{-9}$. This semi-analytical approach constitutes our benchmark and is depicted by the black dots together with their error bar in Fig.~\ref{fig:newton-ortho-profiles-techniques-comparison}. Note that while it takes only a few seconds to evaluate the potential at a single point with this method, it is not conceivable to construct a full map of the field in this way. Rather, this semi-analytical computation should be employed sparingly to assess the error of the FEM computations.

\begin{figure}
    \centering
    \includegraphics[width=0.95\linewidth]{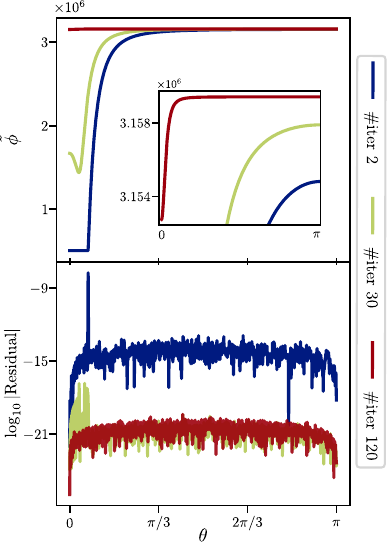}
    \caption{Evolution of the dimensionless chameleon profile $\Tilde{\phi}$ (top) and the associated residual (bottom) after different numbers of iterations of Newton's method (2, 30 and 120). These two quantities are displayed as a function of $\theta$ at fixed altitude $\Tilde{r} = 1.059$. The residual becomes stationary and thus no longer decreases after a sufficient number of iterations has been reached.}
    \label{fig:sol-res-iter}
\end{figure}

In contrast, the chameleon field does not enjoy a similar integral representation which in turns means that we cannot easily define a benchmark profile. Nonetheless, we came up with the following strategies:
\begin{itemize}
    \item[--] Select the set of FEM-related parameters (number and distribution of DOFs, order of the base functions, etc.) so that the FEM-approximation of the Newtonian potential matches the benchmark and use those parameters for the FEM computation of the chameleon field. The light green curves on Fig.~\ref{fig:newton-ortho-profiles-techniques-comparison} correspond to such FEM-approximations (using the aforementioned `virtual connection of DOFs' method) and show that it is indeed possible to reach a high level of accuracy as they stay within the error bars of the benchmarks.
    \item[--] It is also good practice to refer to pre-established \textit{FEM convergence curves}, which are simple charts relating the error to the number of DOFs \textemdash \ see e.g. Fig.~1 of Ref.~\cite{moriond2023}. We can then construct our meshes in an enlightened way, ensuring they are fine enough to meet the stated accuracy.
    \item[--] Evaluate the \textit{strong residual}, which can be done by inputting the FEM approximation obtained for the chameleon field into its equation of motion (\ref{eqn:KG-alpha}) \textemdash \ schematically: $\mathrm{Residual} = \alpha \Tilde{\Delta} \Tilde{\phi} - \Tilde{\rho} + \Tilde{\phi}^{-(n+1)}$. The closer the quantity is to zero, the better the numerical approximation. In order to make this criterion more quantitative, we can monitor (i) the strong residual's decrease throughout the Newton's iterations (see Fig.~\ref{fig:sol-res-iter}) and especially how small the final residual is compared to the initial one, and (ii) its size relative to the size of each term in it: the final residual should be at least a few orders of magnitude smaller than the dominant terms. This criterion is assessed on all 2D numerical computations of the chameleon field discussed in this article. As an example, Fig.~\ref{fig:residual-parts} in Appendix~\ref{app:checks} demonstrates that this criterion is indeed met on three distinct numerical solutions, at three altitudes.
\end{itemize}
Yet, formulating criteria based on the strong residual alone is not entirely satisfactory as it is an absolute quantity. Consequently, there is a priori no simple connection between the relative error committed on the approximation and the strong residual, since the latter quantity is dependent on the PDE's parameters (value of $\alpha$\footnote{In particular, we observed in Ref.~\cite{Levy_2022} that the 2-norm of the strong residual was increasing with $\alpha$, all other things being equal.}, density model, etc.). Computing a reduction factor, that is by how much the strong residual has decreased over the Newton's iterations, is not sufficient either as it depends on how well the initial guess has been chosen (see discussion in the next paragraph). One idea to break this deadlock is to compare our numerical approximations with the chameleon radial profile around a ball. Indeed, the spherically symmetric case is much more under control as we have analytical approximations at our disposal (see e.g. Refs.~\cite{Khoury&WeltmanPRD, mpb-thesis}) and the Klein-Gordon equation boils down to a one-dimensional ordinary differential equation (ODE) which can be solved numerically with a much higher density of DOFs and higher-order finite elements. In terms of residual, the numerical solutions are actually better than their analytical counterparts (see e.g. Tab.~II from Ref.~\cite{Levy_2022}), which is why we propose to use 1D numerical solutions as a benchmark for the spherically symmetric case. Because the addition of a mountain on top of the spherical planet is not expected to have a huge impact on the field's strength outside it, we can check that the evolution of the field along the outgoing radial direction follows that of the benchmark. We provide a quantitative way of assessing that statement in Appendix~\ref{app:checks}, which is applied for all the numerical solutions discussed in this article. Finally, the orthoradial variations of the field at fixed altitudes seems more difficult to verify. As a rough check, we can set $h_m = 0$ and verify that this leads to $\partial_{\theta} \phi \equiv 0$. In practice, we do not expect this equality to hold exactly so we rather make sure that the amplitude $\max_{\theta}\phi(r, \theta) - \min_{\theta}\phi(r, \theta)$ is much smaller in the case $h_m = 0$ compared to the case $h_m = 0.01$. Doing this sanity check on a handful of cases (doing it on all cases would have been too costly) consistently shows that the two quantities differ by at least two orders of magnitude, so that we can confidently state that the orthoradial profiles showed later on do not originate from numerical noise.

Ultimately, the most critical point in this FEM computation is the convergence of the Newton's iterations. Whether or not the method converges depends on a lot of factors. Unfortunately, there are no miracle techniques to address convergence issues but rather \textit{recipes} and good practices which we concisely report here. Perhaps the most important one is to start from a \textit{good} initial guess, i.e. an initial approximation that is as close as possible to the true solution. In most cases, we use a pre-computed 1D radial profile of the field to this end. Other common practices are to refine the meshes where the field is expected to vary quickly (large gradient) \textemdash \ that is near the transition between the inside and the outside of the body, and near the area representing spatial infinity in the inversed exterior domain $\Tilde{\Omega}_{\mathrm{ext}}$ \textemdash \ or to tweak the relaxation parameter \cite{HPL_fem}. Additionally in the particular case of the chameleon field entering the so-called screened regime, we can get rid of the region of the mesh $r<R_{\mathrm{screened}}$ where the field is screened (i.e. constant) and apply a Dirichlet boundary condition at $r=R_{\mathrm{screened}}$. When all the above failed, we resorted to so-called \textit{ramping} \cite{comsol1, comsol2} or \textit{numerical continuation} methods \cite{continuation-methods, HPL_fem}. For example in some cases, we would gradually vary the $\alpha$ parameter from Eq.~(\ref{eqn:KG-alpha}) from a value where the solution is known to the desired value which is problematic convergence-wise, using the solution at each intermediate step as an initial guess of the next one. In spite of all these additional tricks, some combinations of \{$\alpha$, atmosphere model\} resisted all our attempts and were thus discarded from this study. As a closing remark, let us emphasize the fact that we made use of many widely spread techniques in the literature for nonlinear FEM problems (see e.g. Ref.~\cite{HPL_fem} chapter 4), both for implementation and verification purposes. While we are unable to quantify the relative error made on each solution obtained in this study, we grant them a sufficiently high level of confidence that the orders of magnitude discussed hereafter are correct, leaving the physical conclusions unchanged.

In total, we ran FEM computations for four different density profiles outside the main body \textemdash \ the constant vacuum value $\Tilde{\rho}_{\mathrm{vac}} = 10^{-15}$ as well as the three atmospheric models depicted in Fig.~\ref{fig:density-profiles} \textemdash \ and for $\log_{10} \alpha \in \{-5, \dots, -28 \}$. This amounts to nearly a hundred problems to solve on meshes with roughly $10^6$ $\mathbb{P}_2$-triangles. The computations were performed on an ONERA's computing platform equipped with Broadwell and Cascade Lake nodes.

\section{Modified gravity around and above a mountain}
\label{sec:profiles}

In this section, we present and analyze simulations results. We start off with the atmosphere-free case before discussing the influence of each atmospheric models. Due to the parameters' degeneracy mentioned in Sec.~\ref{subsubsec:chameleon}, we decided to fix $\Lambda = \Lambda_{\mathrm{DE}}$ for all numerical results and figures presented in the following. One can refer to Fig.~\ref{fig:param-space} to get a better grasp of the $(\beta, \Lambda, n) \mapsto (\alpha, n)$ mapping.

\subsection{Simulation of an atmosphere-free planet}
\label{subsec:without-atmosphere}

\subsubsection{Gravitational potential profiles and spherical harmonics decomposition}
\label{subsubsec:profiles-sph}

The total gravitational potential is the sum of the Newtonian potential $\Phi$ and the chameleon potential $\Psi$ as defined in Sec.~\ref{subsec:equations}. These are the direct results of FEM computations, i.e. \textit{femtoscope}'s outputs. As this raw data can sometimes be noisy, we had recourse to smoothing splines notably for post-process operations requiring the evaluation of the fields outside mesh data points like the computation of spherical harmonic coefficients using \texttt{SHTools} \cite{SHTools}. Note that the azimuthal symmetry of our setup imposes that the only nonzero coefficients are the ones for which the order $m$ is equal to zero.

\begin{figure*}
    \centering
    \includegraphics[width=\textwidth]{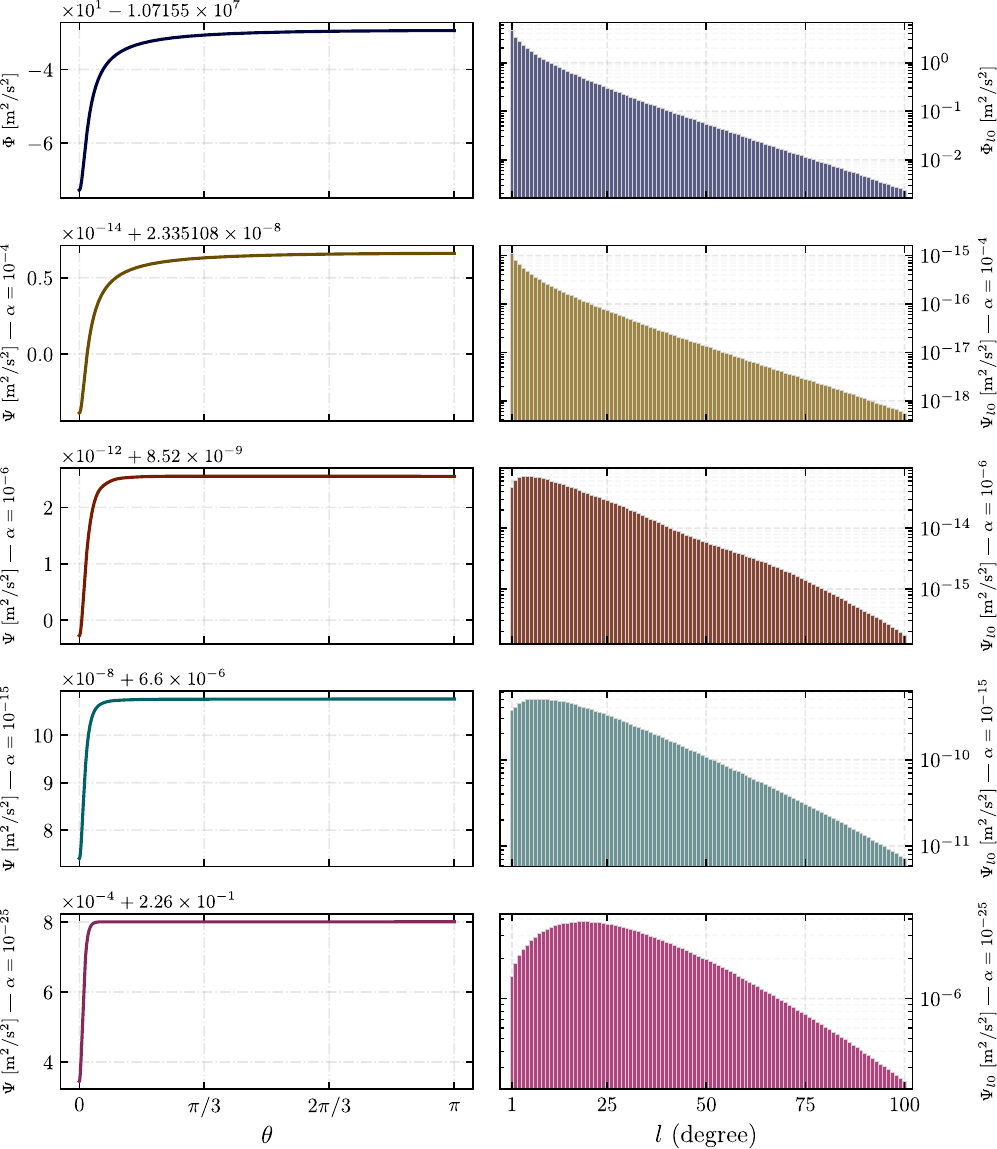}
    \caption{Newtonian and chameleon potential profiles (left column) together with their spherical harmonic coefficients up to degree 100 (right column) computed at $\Tilde{r} = 1.059$. The top row corresponds to the Newtonian case while the four remaining rows are associated with chameleon potentials with $\log_{10}\alpha \in \{-4, -6, -15, -25\}$ from top to bottom. The monopole ($\Phi_{00}$, $\Psi_{00}$) is deliberately excluded from the bar graphs because it is only dependent on the field's mean value. All quantities are expressed in $\mathrm{m^2/s^2}$.}
    \label{fig:profiles-sph}
\end{figure*}

In Fig.~\ref{fig:profiles-sph}, we represent the potential profiles as a function of the colatitude $\theta$ \textemdash \ the radial coordinate being fixed at $\Tilde{r} = 1.059$ \textemdash \ (left column) and their associated spherical harmonic coefficients for degrees $l \in \{ 1, \cdots , 100 \}$ (right column). The top row corresponds to the specific case of the Newtonian potential while the following four rows correspond to chameleon potentials for $\log_{10}\alpha \in \{-4, -6, -15, -25\}$ respectively. The Newtonian potential $\Phi$ is by far the largest contribution to the total potential: roughly $-10^7 \, \mathrm{m^2/s^2}$ compared to $0.2 \, \mathrm{m^2/s^2}$ for the chameleon potential in the $\alpha = 10^{-25}$ case (last row of the same figure). The variation of the potential with respect to $\theta$ around this mean value has the same kind of shape where both ends of the curves have a slope that goes down to zero for symmetry reasons. Note that the potential is always smaller at $\theta = 0$ than at $\theta = \pi$. This is because the mass excess that the mountain represents is located at $\theta = 0$, forming a deeper potential well.

While all four potential profiles share this apparently common trend, the spherical harmonic coefficients displayed on the right column reveal important differences and two types of spectrum emerge. On the one hand, the Newtonian potential and the chameleon potential for $\alpha = 10^{-4}$ have a similar, monotonically decreasing spectrum. This is due to the fact that here, the chameleon field is unscreened which means that all the mass of the main body contributes to the field just like in the Newtonian case. On the other hand, as soon as $\alpha < 10^{-5}$, the chameleon field enters the screened regime, changing the shape of the spectra. We recall that the smaller $\alpha$, the more screened the setup. These spectra all have a maximum for $l > 1$. This distinctive feature of screened chameleon potentials, which could not be seen by eye on the left-hand-side curves, is nevertheless small in front of the Newtonian potential's coefficients $\Phi_{l0}$.

As $\alpha$ decreases, the chameleon potential mean value increases. Indeed, we have $\Psi = K \Tilde{\phi}$ where $K \propto \alpha^{-(n+1)/(n+2)}$ at fixed $\Lambda$. As a result, the spherical harmonic coefficients also get amplified as $\alpha$ decreases, leading to a more disturbed gravitational potential.

Fig.~\ref{fig:sph-annexe} in Appendix~\ref{app:sph-altitudes} further shows how the spectra evolve as the altitude is increased for the Newtonian potential and the chameleon potential ($\alpha = 10^{-25}$).

\subsubsection{Newtonian gravity and fifth-forces}

\begin{figure}
    \centering
    \includegraphics[width=0.95\linewidth]{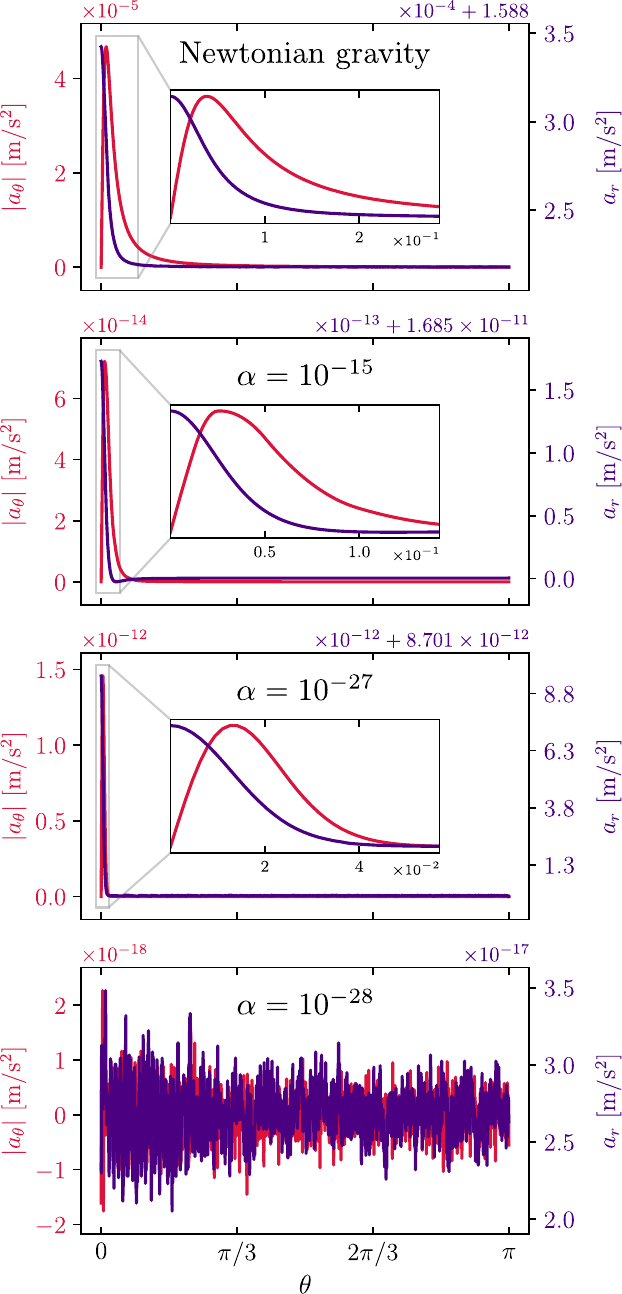}
    \caption{Gravitational field $\mathbf{a} = a_r \mathbf{e}_r + a_{\theta} \mathbf{e}_{\theta}$ in Newtonian gravity (top) and in the chameleon model for the set of parameters $\{ \alpha \in \{10^{-15}, 10^{-27}, 10^{-28}\}, n=1, \Lambda=\Lambda_{\mathrm{DE}}\}$ at $\Tilde{r} = 1.059$. The orthoradial acceleration $a_{\theta}$ is depicted by the red curve (left axis) while the radial acceleration $a_r$ is depicted by the purple curve (right axis).}
    \label{fig:acc-no-atm}
\end{figure}

Once the gravitational potential is known, the actual gravitational acceleration is easily derived by computing its gradient. It is convenient to decompose the acceleration vector $\mathbf{a}$ onto the unit vectors $(\mathbf{e}_r, \mathbf{e}_{\theta})$ (there is no component of the acceleration on $\mathbf{e}_{\varphi}$ due to rotational invariance) such that
\begin{equation*}
    \mathbf{a} = a_r \mathbf{e}_r + a_{\theta} \mathbf{e}_{\theta} \, .
\end{equation*}
In practice, the dimensionless gradient is computed numerically and then multiplied by the relevant coefficient $a_0$ with units $\mathrm{m/s^2}$ \textemdash \ see Eq.~(\ref{eqn:a0}). Fig.~\ref{fig:acc-no-atm} gives an overview of both Newtonian acceleration (top panel) and fifth-forces for $\log_{10} \alpha \in \{-15, -27, -28\}$. Specifically, we represent the component $a_r$ (purple curve) and $a_{\theta}$ (crimson curve) as a function of $\theta$ while the altitude is held fixed at $\Tilde{r} = 1.059$.

\begin{figure}
    \centering
    \includegraphics[width=\linewidth]{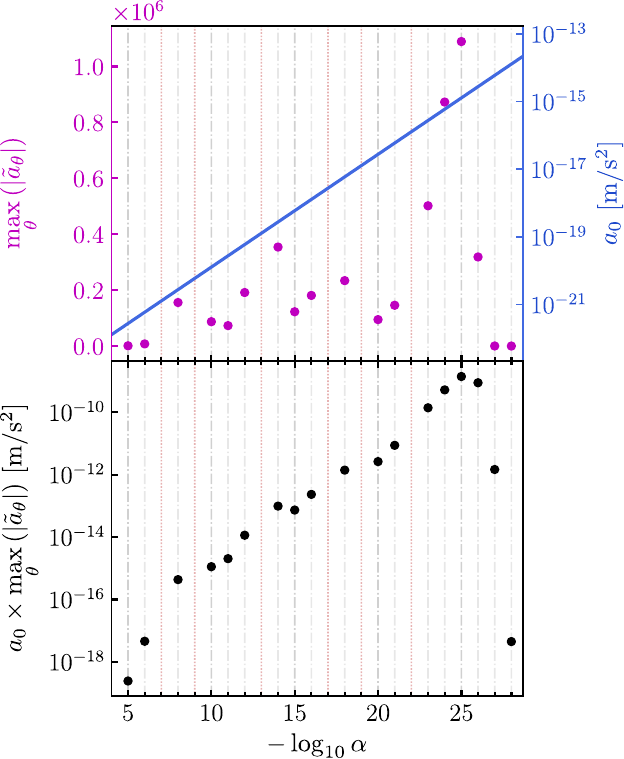}
    \caption{Study of the chameleon orthoradial acceleration $a_{\theta} = a_0 \times \Tilde{a}_{\theta}$ with respect to $\alpha$ at fixed $\Lambda$ and fixed altitude $\Tilde{r} = 1.059$. The top panel features each term separately, $\Tilde{a}_{\theta}$ in magenta dots (dimensionless) and $a_0$ as the blue curve (in $\mathrm{m/s^2}$). The bottom panel is simply the product of these two terms $a_{\theta}$ (in $\mathrm{m/s^2}$). Finally, the red vertical dotted lines correspond to values of $\alpha$ for which the FEM computation was deemed unsatisfactory (failure of Newton's method to converge or unacceptably large residuals).}
    \label{fig:atheta-alpha}
\end{figure}

An important point to discuss here is the limit $\alpha \to 0$. On the one hand, we have seen that for chameleon gravity, $a_0$ is proportional to $\alpha^{-(n+1)/(n+2)}$ at fixed $\Lambda$, and consequently
\begin{equation*}
    a_0 \underset{\ \alpha \to 0^+}{\longrightarrow} +\infty \text{ with constraint } \Lambda = \Lambda_{\mathrm{DE}} \, .
\end{equation*}
This gives the impression that one can make the fifth-force as large as desired simply by taking an ever-decreasing value of $\alpha$. On the other hand, we know that in the limit $\alpha = 0$, the chameleon field profile is trivially given by $\Tilde{\phi} = \Tilde{\rho}^{-1/(n+1)}$ (take $\alpha=0$ in Eq.~\ref{eqn:KG-alpha}). Yet for altitudes higher than the mountain's height, our models are such that $\partial_{\theta} \Tilde{\rho} \equiv 0$ so that $\Tilde{a}_{\theta}$ is expected to vanish for sufficiently small values of $\alpha$. In front of this \textit{apparent} paradox, we raise two points:
\begin{enumerate}
    \item Taking the limit $\alpha \to 0$ at fixed $\Lambda$ coerces $\beta \to +\infty$. Yet, a glimpse at the chameleon constraints plot (see e.g. Fig~3 from Ref.~\cite{compendium-chameleon-constraints}) reveals that chameleon models with $\beta > 10^{14}$ are ruled-out by precision atomic tests. In our case, this corresponds to $\alpha < 10^{-37}$, which is out of the range of $\alpha$ values covered in this study.
    \item Another argument that does not involve referring to current chameleon constraints can be made on the basis of Figs.~\ref{fig:acc-no-atm} and \ref{fig:atheta-alpha}. On Fig.~\ref{fig:atheta-alpha}, we decompose $a_{\theta}$ into the product $a_0(\alpha)$ times $\Tilde{a}_{\theta} = \partial_{\theta} \Tilde{\phi}/\Tilde{r}$. The two terms of this product both depend on $\alpha$: while $a_0$ is simply a power law of $\alpha$, $\Tilde{a}_{\theta}$ clearly exhibits the phenomenon aforementioned, namely that the dimensionless gradient \textemdash \ after reaching a peak for $\alpha = 10^{-25}$ \textemdash \ vanishes for $\alpha < 10^{-28}$. When multiplied together, these two quantities result in $a_{\theta}$ which is scattered in log-scale on the bottom panel of this figure. We recognize the power law behavior $a_{\theta} \propto \alpha^{-n/(n+2)}$ in the range $[10^{-10}, 10^{-21}]$ where $\Tilde{a}_{\theta}$ is roughly constant, followed by a sharp decline due to the vanishing of $\Tilde{a}_{\theta}$. This explains why on Fig.~\ref{fig:acc-no-atm}, the transition from $\alpha = 10^{-27}$ to $\alpha = 10^{-28}$ completely destroys $\boldsymbol{\nabla} \Psi$. There only remains numerical noise, whose amplitude has no genuine physical meaning.
\end{enumerate}

\begin{table}[]
    \centering
    \begin{ruledtabular}
    \begin{tabular}{l c c}
         & $\displaystyle \argmax_{\alpha}$ & value \\[8pt] \midrule
       $\Tilde{a}_r$ & $10^{-24}$ & $1.47 \times 10^{8}$ \\
       $a_r$ & $10^{-25}$ & $1.46 \times 10^{-7}$ $\mathrm{[m/s^2]}$ \\[5pt]
       $\Tilde{a}_{\theta}$ & $10^{-25}$ & $1.09 \times 10^{6}$ \\
       $a_{\theta}$ & $10^{-25}$ & $1.40 \times 10^{-9}$ $\mathrm{[m/s^2]}$ \\[5pt]
       $\|\Tilde{\mathbf{a}}\|$ & $10^{-24}$ & $4.58 \times 10^{9}$ \\
       $\|\mathbf{a}\|$ & $10^{-25}$ & $4.50 \times 10^{-6}$ $\mathrm{[m/s^2]}$ \\
    \end{tabular}
    \end{ruledtabular}
    \caption{Assessment of the maximum fifth-force at $\Tilde{r} = 1.059$}
    \label{tab:max-5th-force}
\end{table}

We also conducted the same analysis as performed in Fig.~\ref{fig:atheta-alpha} for the radial component of the acceleration vector as well as for its norm (still for $\Tilde{r} = 1.059$). The results are reported in Table~\ref{tab:max-5th-force}. In terms of dimensionless quantities, the orthoradial component $\Tilde{a}_{\theta}$ is maximum for $\alpha = 10^{-25}$ while the radial component and the norm of the gradient are both maximized for $\alpha = 10^{-25}$. When these quantities are expressed in units homogeneous to an acceleration $(\mathrm{m/s^2})$, $\alpha = 10^{-25}$ is again the argument that maximizes them all. This corresponds to the traditional parameters $(\beta, \Lambda, n) \sim (10^6, \Lambda_{\mathrm{DE}}, 1)$

Besides, we can discuss the direction of the acceleration vector by computing the ratio $a_{\theta} / a_r$ for both Newtonian and chameleon gravity. We conclude that the Newtonian part of the total acceleration vector is very radial, with $\max_{\theta} \left( a_{\theta} / a_r \right) \leq 3 \times 10^{-5}$, whereas the chameleon acceleration has a more significant orthoradial component since $\max_{\theta} \left( a_{\theta} / a_r \right) \leq 10^{-2}$ for $\log_{10} \alpha = -25$. The physical interpretation for this discrepancy is that in the screened regime, the chameleon acceleration is sourced only by a thin outer layer of the planet which is commonly referred to as the \textit{thin-shell} \cite{Khoury&WeltmanPRD}.

Finally, one may be surprised by the fact that maximum fifth forces are obtained for values of the parameters $(\beta, \Lambda, n)$ which belong to the thin-shell regime; as this appears to contradict the usual rule of thumb that fifth forces should be suppressed in this regime. There is actually no contradiction, provided we clearly define the context. Indeed, the total fifth force acting on a given \textit{macroscopic} body can be computed via the integration of the gradient of the field on its whole volume \textemdash \ as done later in Sec.~\ref{subsubsec:full-field} for instance. It is true that, if the macroscopic object at stake has a thin-shell, the integral of the gradient of the field vanishes everywhere but in that thin-shell, greatly reducing the overall fifth force experienced by that body. Here however, the situation is radically different: we are interested in the fifth force \textit{experienced by} a point-mass (which by essence, cannot possess a thin-shell) \textit{sourced by} a mountainous planet. In this framework, we merely observe that increasing the value of the coupling constant $\beta$ in a given range, while keeping $n$ and $\Lambda$ fixed, results in greater fifth forces. Incidentally, increasing $\beta$ while keeping $n$ and $\Lambda$ fixed means decreasing $\alpha$ (see Fig.~\ref{fig:param-space}) and results in a more screened body. This phenomenon was already observed in Figs.~14 and 15 of our previous work \cite{Levy_2022} and in Fig.~7 of Ref.~\cite{Burrage2015} for instance. This can also be understood in the framework of the analytical approximation of the chameleon fifth force for spherical objects. Taking Eq.~(2.64) from Ref.~\cite{mpb-thesis} reads
    \begin{equation*}
        a_{\phi} = 3 \beta^2 \frac{\Delta  R}{R} \frac{G M_{\mathrm{ball}}}{r^2} (1 + m_{\phi}) \mathrm{e}^{- m_{\phi} (r - R_{\mathrm{ball}})}  \, .
    \end{equation*}
In this expression, $\Delta R/R \propto \beta^{-1}$ and $m_{\phi} \propto \beta^{\frac{n+2}{2(n+1)}}$ so that, at fixed $r > R_{\mathrm{ball}}$, the function $\colon  \beta \mapsto a_{\phi}$ is increasing on the interval $]0, \beta^*[$ and decreasing on $]\beta^*, +\infty[$, for a certain parameter $\beta^* > 0$. We also see that $a_{\phi} \to 0$ when $\beta \to \infty$, because of the exponential term. This is exactly the phenomenology that we observe on numerical simulations: beyond a certain value of $\beta \sim 10^8$ (i.e. below a certain value of $\alpha$, which is around $10^{-28}$), the fifth force vanishes \textemdash  \ see Figs.~\ref{fig:acc-no-atm} and \ref{fig:atheta-alpha}.

\subsection{Adding an atmosphere}
\label{subsec:with-atmosphere}

\begin{figure*}
    \centering
    \includegraphics[width=\textwidth]{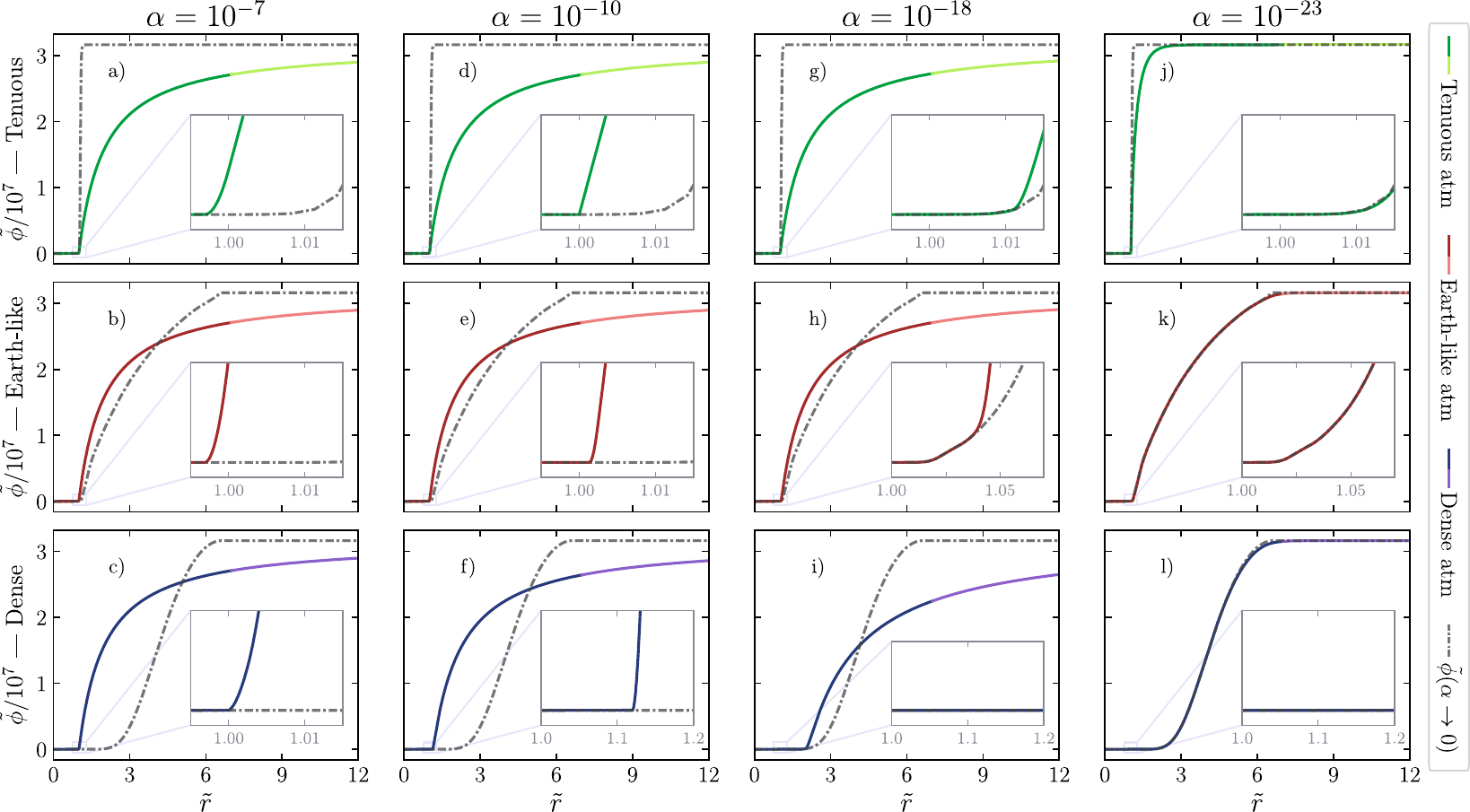}
    \caption{Radial profiles of the chameleon field for $\log_{10} \alpha \in \{-7, -10, -18, -23\}$ (columns) and for all three atmospheric models defined in Sec.~\ref{subsubsec:atm-models}, namely \textit{Tenuous}, \textit{Earth-like} and \textit{Dense} (rows). On each sub-panel, the grey dash-dotted line corresponds to \textit{fully screened} profile, that is obtained in the limit $\alpha = 0$ and is given by $\Tilde{\phi}(\alpha \to 0) = \Tilde{\rho}^{-1/(n+1)}$. The radial chameleon profile is depicted by the bi-color solid line, where the transition from the darker color to the lighter one occurs at the chosen interface ($\Tilde{r} = 7$) between the interior domain and the kelvin-inversed exterior domain (see Ref.~\cite{Levy_2022} for more details).}
    \label{fig:1D-phi-atm}
\end{figure*}

Here we study the influence of adding an atmosphere to the density model. To the best of our knowledge, only a handful of studies deal with the influence of the atmosphere \cite{Khoury&WeltmanPRD, Khoury&WeltmanPRL, waterhouse2006, chameleon-scalar-waves, ss-constraints-chameleon}. In this section, we address simple questions: how is the fifth-force mitigated by the presence of an atmosphere? Can the mountain still be somehow \textit{seen} in the field profile? How does all of this depend on the atmospheric model?

Part of the answer can be unveiled by first studying a simplified version of the setup. More precisely, we got rid of any orthoradial dependence in the density distribution function \textemdash\ which amounts to taking the mountain out of our model to end up with a purely radial setup. This simplification allows us to perform computationally inexpensive 1D FEM simulations with \textit{femtoscope} and still get valuable insight into how the chameleon field behaves in the presence of an atmosphere. We ran computations for all atmospheric models outlined in Sec.~\ref{subsubsec:atm-models} and for all values of $\alpha \in \{10^{-5}, \cdots , 10^{-29}\}$. Part of this simulation campaign has been compiled into Fig.~\ref{fig:1D-phi-atm}, where the vast range of $\alpha$ values explored has been boiled down to only four distinct values for the sake of clarity and conciseness. On each sub-panel, the grey dash-dotted line is associated with the \textit{fully screened} profile obtained in the limit $\alpha = 0$ which is given by $\Tilde{\phi}(\Tilde{r}) = \Tilde{\rho}(\Tilde{r})^{-1/(n+1)}$. Contrary to the previous atmosphere-free case, where the radial density would have been a mere Heaviside step function, the atmospheric density function smoothly interpolates between $\Tilde{\rho}_{\mathrm{atm}}(\Tilde{r}=1)$ to $10^{-15}$  such that the asymptotic profile's gradient $\partial_r \Tilde{\phi}(\alpha \to 0)$ is not identically zero.

\begin{table*}
\renewcommand{\arraystretch}{1.3}
\newcolumntype{C}{@{\extracolsep{5pt}}c@{\extracolsep{0pt}}}%
\begin{tabular}{l C c c c c C c c c c}  
\toprule
& \multicolumn{4}{c}{$\Tilde{r}=1.059$} & \multicolumn{4}{c}{$\Tilde{r}=1.314$} \\
\cmidrule(lr){2-5} \cmidrule(lr){6-9}
& $\alpha = 10^{-6}$ & $\alpha = 10^{-10}$ & $\alpha = 10^{-15}$ & $\alpha = 10^{-20}$ & $\alpha = 10^{-6}$ & $\alpha = 10^{-10}$ & $\alpha = 10^{-15}$ & $\alpha = 10^{-20}$ \\
Tenuous & \scriptsize(1.00 \textendash \ 1.00) & \scriptsize(1.00 \textendash \ 0.89) & \scriptsize(1.00 \textendash \ 0.15) & \scriptsize(1.03 \textendash \ N/A) & \scriptsize(1.00 \textendash \ 0.99) & \scriptsize(1.00 \textendash \ 0.72) & \scriptsize(1.01 \textendash \ 0.11) & \scriptsize(1.02 \textendash \ N/A) \\
Earth-like & \scriptsize(1.00 \textendash \ 1.00) & \scriptsize(1.00 \textendash \ 0.76) & \scriptsize(1.01 \textendash \ N/A) & \scriptsize(0.07 \textendash \ N/A) & \scriptsize(1.00 \textendash \ 0.99) & \scriptsize(1.00 \textendash \ 0.61) & \scriptsize(1.02 \textendash \ N/A) & \scriptsize(1.03 \textendash \ N/A) \\
Dense & \scriptsize(1.00 \textendash \ 0.99) & \medmuskip=0mu \scriptsize($7 \times 10^{-7}$ \textendash \ N/A) & \medmuskip=0mu \scriptsize($7 \times 10^{-7}$ \textendash \ N/A) & \medmuskip=0mu \scriptsize($6 \times 10^{-7}$ \textendash \ N/A) & \scriptsize(1.00 \textendash \ 0.98) & \scriptsize(1.12 \textendash \ N/A) & \medmuskip=0mu \scriptsize($6 \times 10^{-5}$ \textendash \ N/A) & \medmuskip=0mu \scriptsize($6 \times 10^{-5}$ \textendash \ N/A) \\
\bottomrule
\end{tabular}
\caption{\label{tab:att-coeffs} Attenuation coefficients of the radial and orthoradial component of the chameleon acceleration with an atmosphere compared to the atmosphere-free case. The first number of each pair corresponds to the radial part and is computed as $\max_{\theta} a_r^{\text{with-atm}} (\Tilde{r}, \theta) / \max_{\theta} a_r^{\text{no-atm}} (\Tilde{r}, \theta)$. Similarly, the second figure of each pair is the orthoradial attenuation factor and is defined by $\max_{\theta} a_{\theta}^{\text{with-atm}} (\Tilde{r}, \theta) / \max_{\theta} a_{\theta}^{\text{no-atm}} (\Tilde{r}, \theta)$. These attenuation factors are computed for $\Tilde{r} \in \{1.059, 1.314\}$.}
\end{table*}
It is only when we put these 1D chameleon profiles into perspective with the full 2D simulation's results that a clear understanding of the influence of the atmosphere emerges. Starting from $\alpha = 10^{-5}$ and gradually decreasing the value of this parameter, we witness the succession of several regimes:
\begin{enumerate}
    \item For the larger values of $\alpha$, the planet is not fully screened, i.e. there is still a thin-shell. This can be seen on the first column of Fig.~\ref{fig:1D-phi-atm} ($\alpha = 10^{-7}$) where the chameleon \textit{kicks in} (i.e. departs from limit profile $\Tilde{\phi}(\alpha \to 0)$) before $\Tilde{r} = 1$ (which corresponds to the transition between the planet and the atmosphere). This regime is particularly visible on sub-panels a) to d). The impact of the atmosphere on the fifth-force at higher altitudes is then minor \textemdash \ see Tab.~\ref{tab:att-coeffs} thereafter where we compare the amplitude of the fifth-force with and without  atmosphere at $\Tilde{r} \in \{1.059, 1.314\}$.
    \item At some point when decreasing $\alpha$, the lowest part of the atmosphere  becomes screened itself. This is especially visible on sub-panels e) to h). We provide a zoomed-in view of this very region in order to be able to compare the fraction of the atmosphere that is screened against the relative size of the mountain $\Tilde{h}_m = 0.01$. As soon as the screened area overflows the mountain, i.e. everything below $\Tilde{r} = 1.01$ is screened, the imprint of the mountain of the chameleon field is definitely lost at higher altitudes. In other words, the orthoradial acceleration vanishes, giving way to numerical noise. This is why some entries of Table~\ref{tab:att-coeffs} are set to N/A. When it comes to radial component of the fifth-force, it is hardly modified compared to the scenario without atmosphere.
    \item For even smaller values of $\alpha$, the screening eventually reaches the probed region at high altitude. This is particularly clear in sub-panels i) to l), where the chameleon field profile is getting closer to the limit profile (grey dash-dotted line). Here, the orthoradial acceleration remains drowned in the numerical noise while the radial acceleration is fully dictated by the density profile. This is why in some cases, $a_r$ can even become larger with an atmosphere than without (see entries of Table~\ref{tab:att-coeffs} greater than unity).
\end{enumerate}
Once we know that these three regimes exist regardless of the specific form of the atmospheric profile (as long as density decreases with altitude), we can start to be more quantitative by
\begin{itemize}
    \item[--] specifying where the transition between each regime occurs for the atmospheric models at stake;
    \item[--] computing the attenuation factor on the fifth-force for the different density models.
\end{itemize}
These quantitative results are reported in Table~\ref{tab:att-coeffs} where each entry is a pair $(\mu_r, \mu_{\theta})$ defined as
\begin{equation}
\begin{split}
    \mu_r &= \max_{\theta} a_{r}^{\text{with-atm}} (\Tilde{r}, \theta) / \max_{\theta} a_{r}^{\text{no-atm}} (\Tilde{r}, \theta) \\
    \mu_{\theta} &= \max_{\theta} a_{\theta}^{\text{with-atm}} (\Tilde{r}, \theta) / \max_{\theta} a_{\theta}^{\text{no-atm}} (\Tilde{r}, \theta)
\end{split}
\label{eqn:att-factor}
\end{equation}
at a specific radial coordinate $\Tilde{r}$. We refer to these coefficients as the attenuation factors, which are of course dependent on the atmospheric model as well the altitude at which they are computed.

The take home message from this study of atmospheric models is that the presence of an atmosphere, as tenuous as it may be, prevents access to the biggest fifth-force attainable without atmosphere. Indeed, we saw earlier that, around $\Tilde{r} = 1.059$, the fifth-force was reaching its maximum value for $\alpha$ in the order of $10^{-25}$. Yet in all three atmospheric models under study, the screening of the atmosphere at low altitude occurs for much bigger values of $\alpha$, putting a lower threshold on the maximum accessible fifth-force. When put into perspective with current bounds on $n=1$ chameleon theory, these results show that the largest part of the unconstrained region maps to a screened atmosphere in the LEO altitude range. All other things remaining equal, the radial component of the fifth force can be recovered by going higher up in altitude, where the atmospheric density is lower.

\section{Influence on spacecraft trajectory}
\label{sec:orbits}

In this section, we shift our focus to how geodesics get modified in the presence of a putative chameleon fifth-force with respect to the purely Newtonian case. We want to ascertain the effects of the fifth force in a rather quantitative way: is the deviation from Newtonian dynamics large enough to be detected by current satellite technology? Is it possible to discriminate the presence of a fifth force from the imperfect knowledge of the model at stake or small perturbations of the initial conditions? When does a satellite in Low Earth Orbit (LEO) become screened? Besides, we will refrain from commenting too much on secular drifts that can arise between modified gravity and Newtonian gravity. That is because in any realistic scenario \textemdash \ where many additional forces of different nature come into play \textemdash, it would be merely impossible to discriminate the fifth force from such forces. We thus keep our analysis \textit{local}, by focusing our attention on the dynamics at the passage over the mountain.

\subsection{Screening of the fifth force by the spacecraft}
\label{subsec:satellite-screening}

\subsubsection{Existing criteria}
\label{subsubsec:screening-criteria}

We stress that modeling a spacecraft by a material point (in the framework of chameleon gravity) roughly amounts to making the hypothesis that it does not possess a thin-shell. Ref.~\cite{Khoury&WeltmanPRD} derives an analytical criterion for a typical satellite in low Earth orbit not to have a thin-shell (see Eq.~80 of this reference). Applying this criterion with the density values employed in our study (except for that of the satellite itself which is set at $8 \times 10^3 \, \mathrm{kg/m^3}$) leads straightforwardly to the requirement that $\beta \lesssim 2 \times 10^2 \iff \alpha \gtrsim 3 \times 10^{-20}$ (for $\Lambda = \Lambda_{\mathrm{DE}}$ and $n=1$). The following orbit propagation results being performed with $(n=1, \Lambda = \Lambda_{\mathrm{DE}}, \alpha = 10^{-25})$, the satellite would be partially screened according to this criterion and the chameleon effects would thus be smaller than presented. 

Ref.~\cite{mpb1} derives another criterion based on numerical simulations claiming that the satellite will be fully screened when the thickness of its walls is larger than $100 \lambda_{\mathrm{c, wall}}$, where $\lambda_{\mathrm{c, wall}}$ refers to the Compton wavelength in the wall. However, this criterion must be taken with a grain of salt as it was derived for a density contrast $\rho_{\mathrm{vac}} / \rho_{\mathrm{wall}} = 10^{-3}$, far from a realistic setup. Still, applying this second criterion for a wall of thickness $10 \, \mathrm{cm}$ leads to the fact that the satellite will not have a thin-shell if $\beta \lesssim 2 \times 10^{-2} \iff \alpha \gtrsim 4 \times 10^{-14}$.

\begin{figure*}
    \centering
    \includegraphics[width=\linewidth]{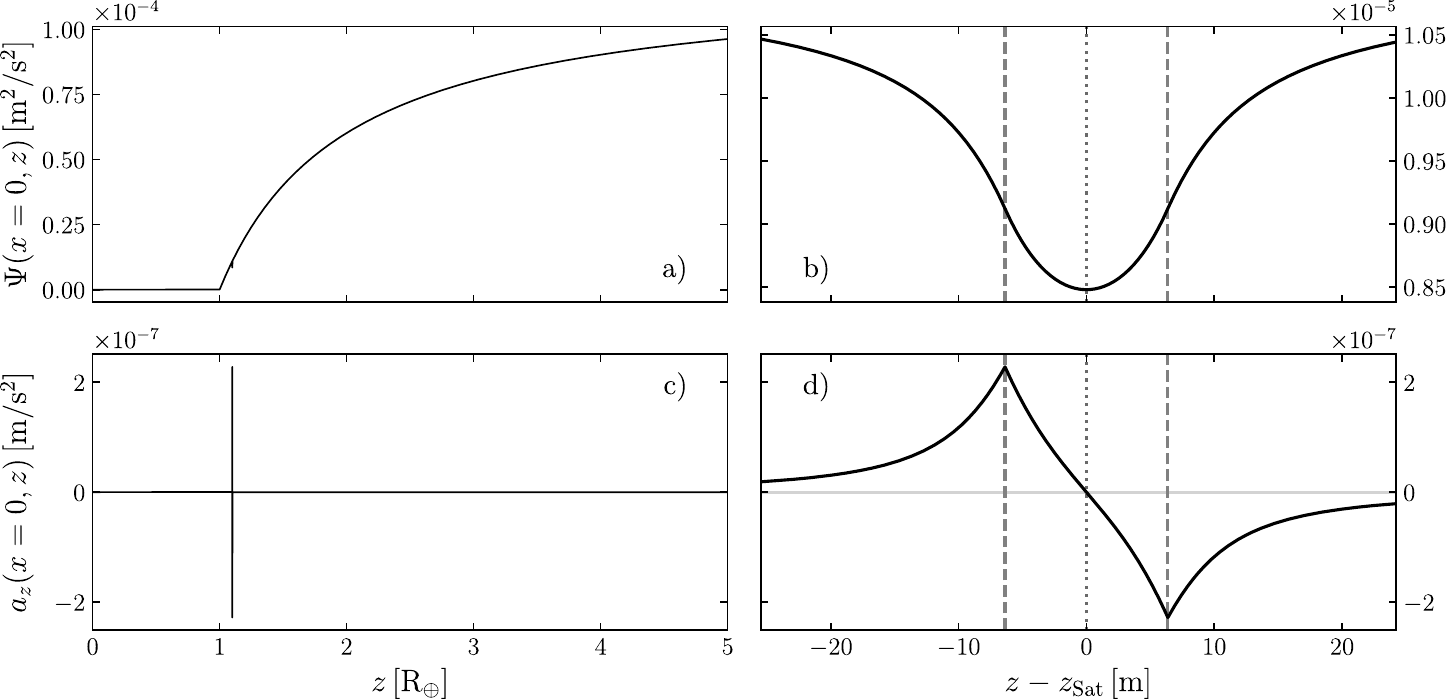}
    \caption{Chameleon potential $\Psi$ (top row) and acceleration $a_z$ (bottom row) along the $z$-axis. The panels b) and d) are a zoomed version around $z_{\mathrm{Sat}} = 1.1 \, R_{\oplus}$ of panels a) and c) respectively. The parameters used to produce this figure are: $\rho_{\oplus} = 10^3 \, \mathrm{kg/m^3}$, $\rho_{\mathrm{vac}} = 10^{-15} \, \mathrm{kg/m^3}$, $\rho_{\mathrm{Sat}} = 10^3 \, \mathrm{kg/m^3}$, $L_{\mathrm{Sat}} = 2 \times 10^{-6} \, R_{\oplus} \sim 12.7 \, \mathrm{m}$, $z_{\mathrm{Sat}} = 1.1 \, R_{\oplus} \sim 7 \times 10^3 \, \mathrm{km}$, $\alpha = 10^{-15}$, $n=1$, $\beta  = 0.24$, $\Lambda = \Lambda_{\mathrm{DE}}$. On panels b) and d), the dotted line is centered at $z = z_{\mathrm{Sat}}$ while the dashed lines represent the extent of the satellite.}
    \label{fig:screening-sat}
\end{figure*}

\subsubsection{Computation of the field with femtoscope}
\label{subsubsec:full-field}

Although quite qualitative, these two criteria provide us with a comprehension of how the parameters in our model affect the screening of the satellite. For instance, increasing the overall density of the satellite (other things being equal) results in more screening. Ideally, one would compute the scalar field profile sourced by the Earth and the spacecraft all at once \textemdash \ which would avoid having to rely on such criteria and provide a definitive answer. The problem then becomes a numerical one, because the simulation should accommodate a thousand-kilometer-size object (a planet) together with a meter-size object (a satellite). We create a mesh using the Gmsh software \cite{gmsh} that captures both scales (whose ratio is equal or less than $10^{-6}$) thanks to \textit{h-adaptivity} \textemdash \ a technique that adjusts the mesh resolution by refining or coarsening elements to focus computational resources where they are most needed. The setup is as follows: we place a cylindrical object centered at coordinates $(\Tilde{x}_{\mathrm{Sat}}, \, \Tilde{z}_{\mathrm{Sat}}) = (0, \, 1.1)$ whose axis is aligned with the $z$-axis (Fig.~\ref{fig:mnt-views}). The diameter and height of the cylinder are set equal to $L_{\mathrm{Sat}}$ and we denote by $\rho_{\mathrm{Sat}}$ its density. In order to get an order of magnitude of a satellite mean density, we take the example of a CubeSat\footnote{CubeSats have a form factor of 10 cm cubes and have a mass of no more than $2 \, \mathrm{kg}$.} whose density is around $\sim 10^3 \, \mathrm{kg/m^3}$. We then compute the chameleon field map in the $(x, \, z)$-plane for various combinations of $\rho_{\mathrm{Sat}}$, $L_{\mathrm{Sat}}$ and $\alpha$. The global acceleration undergone by the cylindrical satellite $\mathbf{a}_{\mathrm{cham}}^{\mathrm{tot}}$ is obtained by integrating the gradient of the scalar field over its whole volume. Under the assumption that the satellite is made of a material of constant density, one gets
\begin{equation}
    \mathbf{a}_{\mathrm{cham}}^{\mathrm{tot}} = - \frac{1}{V} \int_{V} \boldsymbol{\nabla} \Psi \, \mathrm{d}V = - \frac{\beta}{V M_{\mathrm{Pl}}} \int_V \boldsymbol{\nabla} \phi \, \mathrm{d}V \, ,
\label{eqn:int-acc-cham-1}
\end{equation}
where $V = \pi L_{\mathrm{Sat}}^3 / 4$ is the volume of the cylinder. Now because the setup admits $\mathcal{O}_{xz}$ and $\mathcal{O}_{yz}$ as planes of symmetry, $a_z^{\mathrm{tot}} = \mathbf{a}_{\mathrm{cham}} \cdot \mathbf{e}_z$ is the only non-zero component of the acceleration vector. Setting $x_{\mathrm{max}} = L_{\mathrm{Sat}}/2$, $z_{\pm} = z_{\mathrm{Sat}} \pm L_{\mathrm{Sat}}/2$, the calculation thus simplifies to
\begin{equation}
  \begin{split}
      a_z^{\mathrm{tot}} &= - \frac{1}{V} \int_{0}^{2 \pi}  \int_{0}^{x_{\mathrm{max}}} \int_{z_{-}}^{z_{+}} x \partial_z \Psi \, \mathrm{d}z \, \mathrm{d}x \, \mathrm{d} \theta \\[5pt]
          &= \left(\frac{2}{L_{\mathrm{Sat}}}\right)^3 \int_{0}^{x_{\mathrm{max}}} x \left[ \Psi(x, z_{-}) - \Psi(x, z_{+}) \right] \, \mathrm{d}x \, .
  \end{split}
\label{eqn:int-acc-cham-2}
\end{equation}
The resulting 1D integral can easily be computed using any numerical integration routine.

Fig.~\ref{fig:screening-sat} shows the chameleon potential $\Psi$ (top row) together with the elementary acceleration $a_z = - \partial_z \Psi$  (bottom row) along the axis of the cylinder that passes through the Earth. On panel a), we recognize the customary chameleon field profile of a screened ball, perturbed nearby $z = 1.1 \, R_{\oplus}$ by the presence of the satellite. When we zoom-in, we see the potential well imputed to the satellite in panel b). This localized variation of the chameleon field results in a large gradient in absolute value (bigger than anywhere else in the numerical domain). However big the field's gradient may be, looking at panel d) with naked eyes could lead us to believe that it is an odd function with respect to $z=z_{\mathrm{Sat}}$. If that turned out to be the case, then  performing the integration (\ref{eqn:int-acc-cham-1}) would result in a net zero acceleration and the satellite's trajectory would coincide with GR geodesic (in the absence of any non-gravitational perturbation).

\begin{table*}
\renewcommand{\arraystretch}{1.3}
\newcolumntype{C}{@{\extracolsep{10pt}}c@{\extracolsep{0pt}}}%
\begin{tabular}{p{1.0cm}  p{1.6cm}  p{1.6cm}  p{1.6cm}  p{1.6cm}  p{1.6cm}  p{1.6cm}  p{1.6cm}  p{1.6cm}  p{1.6cm}}  
\toprule
& \multicolumn{3}{c}{\makecell{\textit{Case 1 \textemdash \ benchmark} \\[2pt] \footnotesize $\left( \rho_{\mathrm{Sat}} = 10^3 \, \mathrm{kg/m^3}, \ L_{\mathrm{Sat}} = 2 \times 10^{-6} \, R_{\oplus} \right)$}} & \multicolumn{3}{c}{\makecell{\textit{Case 2 \textemdash \ denser} \\[2pt] \footnotesize $\left( \rho_{\mathrm{Sat}} = 10^4 \, \mathrm{kg/m^3}, \ L_{\mathrm{Sat}} = 2 \times 10^{-6} \, R_{\oplus} \right)$}} & \multicolumn{3}{c}{\makecell{\textit{Case 3 \textemdash \ smaller} \\[2pt] \footnotesize $\left( \rho_{\mathrm{Sat}} = 10^3 \, \mathrm{kg/m^3}, \ L_{\mathrm{Sat}} = 5 \times 10^{-7} \, R_{\oplus} \right)$}} \\
\cmidrule(lr){2-4} \cmidrule(lr){5-7} \cmidrule(lr){8-10}
$\alpha$ & \centering $10^{-14}$ & \centering $10^{-15}$ & \centering $10^{-16}$ & \centering $10^{-14}$ & \centering $10^{-15}$ & \centering $10^{-16}$ & \centering $10^{-14}$ & \centering $10^{-15}$ & \multicolumn{1}{c}{$10^{-16}$}\\[3pt]
$\left|a_{z}^{\mathrm{tot}}\right|$ & \centering \scriptsize $7.25 \times 10^{-12}$ & \centering \scriptsize $1.56 \times 10^{-11}$ & \centering \scriptsize $\sim 0$ & \centering \scriptsize $7.25 \times 10^{-12}$ & \centering \scriptsize $\sim 0$ & \centering \scriptsize $\sim 0$ & \centering \scriptsize $7.25 \times 10^{-12}$ & \centering \scriptsize $1.56 \times 10^{-11}$ & \multicolumn{1}{c}{\scriptsize $3.36 \times 10^{-11}$} \\
$\left|a_{z}\right|$ & \centering \scriptsize $7.25 \times 10^{-12}$ & \centering \scriptsize $1.56 \times 10^{-11}$ & \centering \scriptsize $3.37 \times 10^{-11}$ & \centering \scriptsize $7.25 \times 10^{-12}$ & \centering \scriptsize $1.56 \times 10^{-11}$ & \centering \scriptsize  $3.37 \times 10^{-11}$ & \centering \scriptsize $7.25 \times 10^{-12}$ & \centering \scriptsize $1.56 \times 10^{-11}$ & \multicolumn{1}{c}{\scriptsize $3.37 \times 10^{-11}$}\\
\bottomrule
\end{tabular}
\caption{\label{tab:sat-acc-cham} Total chameleon acceleration undergone by a satellite (extended object) $\left|a_{z}^{\mathrm{tot}}\right|$ compared to that of a point-like particle $\left|a_{z}\right|$. The accelerations are expressed in $\mathrm{m/s^2}$. Each of the three cases corresponds to three different satellites: \textit{Case 1} is a benchmark, \textit{Case 2} represents a 10 times denser satellite, \textit{Case 3} represents a 4 times smaller satellite. As long as the satellite is not screened, $\left|a_{z}^{\mathrm{tot}}\right| \simeq \left|a_{z}\right|$. When the satellite is screened (which occurs at a different $\alpha$ depending on the satellite's characteristics), $\left|a_{z}^{\mathrm{tot}}\right|$ drops down to nearly zero. Note that Fig.~\ref{fig:screening-sat} corresponds to \textit{Case 1} with $\alpha = 10^{-15}$.}
\end{table*}

We tackle this issue by computing $a_z^{\mathrm{tot}}$ using Eq.~(\ref{eqn:int-acc-cham-2}) for several physical parameters ($\rho_{\mathrm{Sat}}$, $L_{\mathrm{Sat}}$) and several chameleon parameters $\alpha$. From there, the whole question is to determine how the total chameleon acceleration undergone by the satellite compares against that of a point-like particle not affecting the background field. The results set out in Table~\ref{tab:sat-acc-cham} provide some answers. We consider three cases which correspond to three satellites with distinct characteristics, namely different length scale and density. For each case, we vary $\alpha \in \{10^{-14}, 10^{-15}, 10^{-16}\}$ and compute the total chameleon acceleration undergone by the satellite (extended object) $\left|a_{z}^{\mathrm{tot}}\right|$ and that of a point-like particle $\left|a_{z}\right|$. Surprisingly, the outcome of this experiment is binary:
\begin{itemize}
    \item[--] When the satellite is unscreened \textemdash \ that is when the scalar field does not reach the value that minimizes the effective potential inside the cylinder $\phi_{\mathrm{Sat}}$ \textemdash \ we find that the total chameleon acceleration it undergoes is equal to that of a test particle placed at $z_{\mathrm{Sat}}$. This is a remarkable fact, which we did not anticipate by simply looking at Eq.~(\ref{eqn:int-acc-cham-2}) and we thus provide an attempt to explain this phenomenon in Appendix~\ref{app:remarkable}. In other words, the satellite \textit{feels} the fifth force sourced by the Earth as if it did not perturb the field at all. Consequently, it behaves as a point-like particle and will follow the geodesics of the Jordan frame metric $\Tilde{g}_{\mu \nu} = \exp \left( 2 \beta \phi / M_{\mathrm{Pl}} \right) g_{\mu \nu}$, where $g_{\mu \nu}$ refers to the Einstein frame metric.
    \item[--] When the satellite is screened, the integral of the field's gradient over the volume occupied by the satellite vanishes almost completely \textemdash \ essentially because there, the gradient is null. The satellite only \textit{feels} the Newtonian part of the gravitational force and thus follows the geodesics of the Einstein frame metric $g_{\mu \nu}$. 
\end{itemize}
Of course, there actually exists an intermediate case where the satellite would only be \textit{partially} screened, i.e. where the field would indeed reach $\phi_{\mathrm{Sat}}$ deep inside the cylinder while still having some space to vary in its outermost regions. In this specific case, the ratio $\left|a_{z}^{\mathrm{tot}}\right| / \left|a_{z}\right|$ lies somewhere between 0 and 1. However, the results reported in Table~\ref{tab:sat-acc-cham} suggest that the transition from the unscreened case and the fully screened case does not cover a wide region of the chameleon parameter space. Indeed, taking the \textit{Case 1} as an example, the transition occurs between $\alpha = 10^{-15}$ and $\alpha = 10^{-16}$ \textemdash \ refer to Fig.~\ref{fig:param-space} to get a better idea of the narrowness of this region in the chameleon parameter space.

\subsubsection{Discussion}
\label{subsubsec:discussion}

We can check that the reported results are in accordance with the qualitative predictions made by the first two criteria discussed earlier. They both predict that increasing the density and/or the length of the satellite should make it more likely to be screened. This is in agreement with our findings: (i) going from \textit{Case 1} to \textit{Case 2} shows the effect of an increase by one order of magnitude of the satellite's density, (ii) going from \textit{Case 3} to \textit{Case 1} illustrates the effect of increasing the satellite's overall size. A follow-up question is whether it is possible to find a distribution of mass inside the satellite $\rho_{\mathrm{Sat}}(x, z)$ such that $\left|a_{z}^{\mathrm{tot}}\right| > \left| a_z \right|$. The simple tests we performed so far \textemdash \ for instance, setting different densities for the upper and lower halves of the satellite \textemdash \ all resulted in $\left|a_{z}^{\mathrm{tot}}\right| \leq \left| a_z \right|$. The question remains open. Additionally, dealing with an extended object means that new rotational degrees of liberty can enter the scene and it would be interesting to look at similar optimization process in order to find the maximum torque (note that Refs.~\cite{electrostatic-analogy-1, electrostatic-analogy-3} mention this effect and highlight the fact that it can stand out from Newtonian gravity).

Although the satellite model implemented in this section is very simple, this study shows that it is possible for a realistic satellite not to be screened in parts of the chameleon parameter space. This has implications for space-based tests of gravity. For instance, in chameleon models where the scalar field does not couple universally to all matter fields, violations of the weak equivalence principle are not necessarily suppressed by the satellite walls or the experimental setup (as opposed to what was claimed in Ref.~\cite{mpb2}). Another example (which holds for a universal coupling constant $\beta$) is that of an accelerometer with a screened test mass onboard an unscreened satellite: the accelerometer would measure a force akin to a bias.

\subsection{Orbital dynamics of an artificial satellite}
\label{subsec:orbital-mechanics}

Sec.~\ref{subsec:satellite-screening} made it clear that in practice, a satellite orbiting some planetary body in the framework of chameleon gravity cannot be treated as a point-like particle in the entire parameter space. We have highlighted that there is a narrow transition zone beyond which the satellite becomes fully screened and the net fifth force acting on it vanishes almost completely. In what follows however, we make the assumption that we can treat the satellite as a point-like particle. This is justified by at least two reasons:
\begin{enumerate}
    \item This is a valid approximation in parts of the parameter space (see Table~\ref{tab:sat-acc-cham}).
    \item One can always, at least at the thought experiment stage, make the satellite smaller or less dense so that is it not subject to screening.
\end{enumerate}
That being said, in all the orbit propagation results presented in the following, we choose the chameleon parameters that produce the strongest fifth force at $\Tilde{r} = 1.059$ (which represents an altitude of approximately $376 \, \mathrm{km}$) in the absence of atmosphere: $(n=1, \, \Lambda=\Lambda_{\mathrm{DE}}, \, \beta=1.1 \times 10^6)$. Notice that this point of the parameter space is already constrained by atom interferometry, see e.g. Ref.~\cite{Burrage2018}.

Suppose a point-like particle is placed in a gravitational potential $U$. The equations of motion in spherical coordinates are
\begin{equation}
\begin{split}
    \Ddot{r} - r\left( \Dot{\theta}^2 + \Dot{\varphi}^2 \sin^2\theta \right) &= - \partial_r U \\
    r\left( \Ddot{\theta} + 2 \frac{\Dot{r}}{r} \Dot{\theta} - \Dot{\varphi} \sin \theta \cos \theta \right) &= -\frac{1}{r} \partial_{\theta} U \\
    r \sin \theta \left( \Ddot{\varphi} + 2 \frac{\cos \theta}{\sin \theta} \Dot{\theta} \Dot{\varphi} + 2 \frac{\Dot{r}}{r} \Dot{\varphi} \right) &= - \frac{1}{r \sin \theta} \partial_{\varphi} U
\end{split}
\, ,
\label{eqn:orbital-dynamics-1}
\end{equation}
where dots refer to time derivatives. The massic energy is given by
\begin{equation}
    \mathcal{E} = \frac{1}{2} \left( \Dot{r}^2 + r^2 \Dot{\theta}^2 + r^2 \sin^2 \theta \Dot{\varphi}^2 \right) + U
\label{eqn:massic-energy}
\end{equation}
and it is conserved along the trajectory, i.e. $\Dot{\mathcal{E}} \equiv 0$. Our setup being axisymmetric, we can get rid of the $\varphi$-dependence. Then, note that Eq.~(\ref{eqn:orbital-dynamics-1}) implies that the angular momentum $L \equiv r^2 \Dot{\theta}$ satisfies
\begin{equation}
  \Dot{L} = - \partial_{\theta} U \, .
\label{eqn:angular-momentum}
\end{equation}

The problem at stake is a perturbed Kepler problem (the mountain and fifth force contributions are small compared to the central force), whose total gravitational potential $U$ can therefore be decomposed into 
$$U = -\mu / r + \delta U \, ,$$
where $\mu \equiv G M_{\mathrm{body}}$ is the standard gravitational parameter of the main body (note that $\mu$ does not encompass the mass contained in the mountain itself). The perturbation $\delta U$ is the sum of the Newtonian potential of the mountain $\delta \Phi$ and the chameleon potential $\Psi$ of the whole system. With this in mind, it is also more appropriate to decompose the motion into a Keplerian part \textemdash \ that we assume to be circular \textemdash \ and a perturbed part, reading
\begin{equation}
    \begin{aligned}
  r &= a + \delta r    &  \hphantom{hphantom}  \theta &= \theta_0 + \omega t +\delta \theta \\
  \Dot{r} &= \Dot{\delta r}   &   \Dot{\theta} &= \omega + \Dot{\delta \theta} \\
  \Ddot{r} &= \Ddot{\delta r} &   \Ddot{\theta} &= \Ddot{\delta \theta} \\
  L &= L_0 + \delta L         &   \Dot{L} &= \Dot{\delta L}
\end{aligned} \, .
\label{eqn:keplerian-decomposition}
\end{equation}
In the above, $a$ is the radius of the circular orbit and $\omega$ is the Keplerian pulsation, satisfying $\omega^2 = \mu / a^3$. $L_0$ is the initial angular momentum with $L_0^2 = \mu a$ and $\theta_0$ is the initial co-latitude for a Keplerian motion. This lets us rewrite the equations of motion (\ref{eqn:orbital-dynamics-1}) as
\begin{equation}
    \Ddot{\delta r} = L^2 / r^3 - \partial_r U \, , \hspace{0.8em} \Dot{\delta \theta} = L / r^2 - \omega \, , \hspace{0.8em} \Dot{\delta L} = - \partial_{\theta} U \, ;
\label{eqn:orbital_dynamics2}
\end{equation}
while the energy conservation reads
\begin{equation}
     (\Dot{\delta r})^2 + (L/r)^2 + 2 U - 2 \mathcal{E} = 0 \, .
\label{eqn:energy-conservation}
\end{equation}
Note that at the $\mathrm{0^{th}}$-order, Eq.~(\ref{eqn:energy-conservation}) boils down to the usual energy conservation in a circular Keplerian orbit
\begin{equation*}
    ( a \Dot{\theta} )^2 - \frac{2 \mu}{a} - 2 \mathcal{E} = 0 \, .
\end{equation*}

\subsection{Numerical integration with energy conservation}
\label{subsec:eom-integration}

The state vector that we wish to propagate over time is $\mathbf{X} = (\delta r, \Dot{\delta r}, \delta \theta , \delta L) \in \mathbb{R}^4$. It is governed by the ordinary differential equation (ODE) $\Dot{\mathbf{X}} = F(t, \mathbf{X})$, where $F \colon \mathbb{R} \times \mathbb{R}^4 \to \mathbb{R}^4$ is given by Eq.~(\ref{eqn:orbital_dynamics2}). Note that while energy conservation (\ref{eqn:energy-conservation}) is derivable from the ODE itself, there is no \textit{a priori} reason for it to hold on the numerical approximation. For one thing, the energy might fluctuate on short time scales depending on the numerical integrator employed, leading to an increase or decrease over longer time scales. Additionally, the r.h.s. of ODE (\ref{eqn:orbital_dynamics2}) is obtained through FEM computation and is hence \textit{noisy}, meaning that even so-called \textit{energy-preserving integrators} would exhibit the energy-drift phenomenon. 

Appending the energy conservation (\ref{eqn:energy-conservation}) to the ODE defines an over-determined differential-algebraic system of equations (DAE). One convenient way to preserve first integrals such as energy conservation when numerically integrating dynamical systems is to resort to \textit{projection techniques}. The idea behind this class of techniques is to slightly perturb the state after each solver's step so that the energy remains constant. This technique is described in Ref.~\cite{Hairer2006}. As for the implementation, one can easily modify any existing general purpose ODE solver to perform this projection. We provide a minimally modified version of \texttt{scipy}'s Runge-Kutta solvers that was used for the numerical integration as supplementary material.

The simulations presented below are performed at $\Tilde{r} = 1.059$, which corresponds to an altitude of roughly $376 \, \mathrm{km}$. The Newtonian potential and its gradient are evaluated using the point-mass approximation introduced in Sec.~\ref{subsubsec:challenges-verif} as it is hardly distinguishable from the semi-analytical solution. As for the chameleon field, we have the freedom to select an operating point in the parameter space. We choose $(n=1, \Lambda = \Lambda_{\mathrm{DE}}, \alpha = 10^{-25}, \beta = 1.1 \times 10^{6})$ which we have identified as the point that concurrently results in the strongest fifth force and the greatest field's strength (in the atmosphere-free scenario, see Sec.~\ref{subsec:without-atmosphere}). The experiment performed in Sec.~\ref{subsec:satellite-screening} indicates that any medium to large size satellite would presumably be screened in this case. Consequently, the point-like approximation we adopt can be understood as a \textit{best case scenario}, i.e. an upper bound on the maximum fifth force. Indeed, escaping the screened regime comes at the prize of restricting the allowed range of parameter $\alpha$ to $\alpha > \alpha_{\mathrm{screened}}$, which limits the maximum fifth force \textemdash \ see Sec.~\ref{subsec:without-atmosphere}. In terms of initial conditions, the point-like particle is set in a Keplerian motion so that the initial state vector reads $\mathbf{X}(t=0) = \mathbf{0} \in \mathbb{R}^4$, with $\theta_0 = \pi$. 

\subsection{Results and discussion}
\label{subsec:GFO}

Here we present and discuss the orbit propagation results. For the sake of clarity and concision, we denote by $\mathbf{X}^{\mathrm{New}}$ and $\mathbf{X}^{\mathrm{Cham}}$ the state vectors in the purely Newtonian case and in the modified gravity (i.e. the sum of chameleon and Newtonian gravity) respectively.

\subsubsection{Results of the simulations}
\label{subsubsec:geodesic-deviation}

\begin{figure*}
    \centering
    \includegraphics[width=\textwidth]{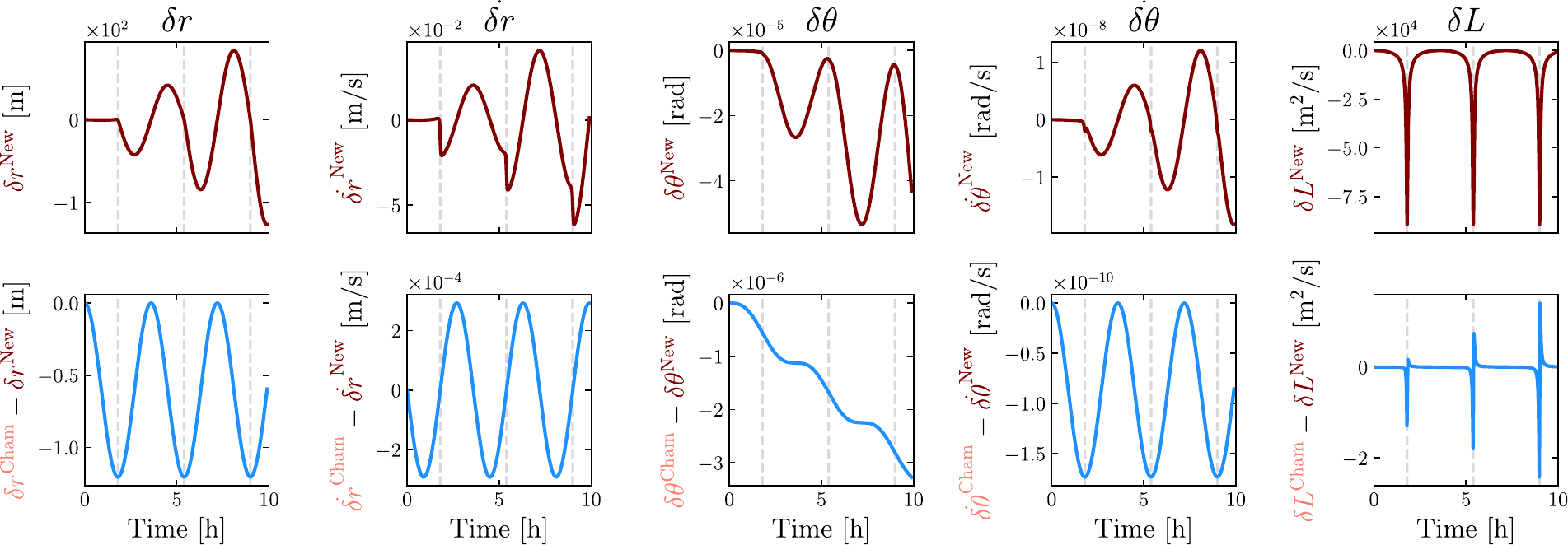}
    \caption{Orbit propagation over three Keplerian periods. The first row shows the evolution of the state vector $\mathbf{X}^{\mathrm{New}} = (\delta r^{\mathrm{New}}, \, \Dot{\delta r}^{\mathrm{New}}, \, \delta \theta^{\mathrm{New}}, \, \delta L^{\mathrm{New}})$ and $\Dot{\delta \theta}^{\mathrm{New}}$ with respect to time, where the dynamics is purely Newtonian. The second row lays emphasis on the orbital dynamics in modified gravity by showing $\mathbf{X}^{\mathrm{Cham}} - \mathbf{X}^{\mathrm{New}}$ and $\Dot{\delta \theta}^{\mathrm{Cham}} - \Dot{\delta \theta}^{\mathrm{New}}$. The vertical light-gray dashed lines correspond to the instants at which the point-like particle passes over the mountain, at $\theta =0$, in the purely Newtonian case.}
    \label{fig:Xstate}
\end{figure*}

The evolution with respect to time of the main quantities of interest are presented in Fig.~\ref{fig:Xstate}. The time spans 10 hours which encompasses roughly three full orbits. The first row of this figure shows how, in a purely Newtonian setting, the presence of the mountain breaks the Keplerian, circular motion. Some elements, such as $\delta L$ are very correlated to the passage of the point-mass above the mountain (denoted by the vertical light-gray dashed lines on each panel). The angular momentum is indeed roughly constant along the trajectory, except nearby $\theta \sim 0$ where it peaks very sharply ($L_0$ being negative, this corresponds to an increase in the absolute value of $L$). On the other hand, some other elements are \textit{irreversibly} imprinted by the mountain after the first passage above it, see e.g. $\delta r$, $\Dot{\delta r}$ or $\delta \theta$, leaving traces on the longer term. The physical intuition for this is that, although the gravity field is symmetric with respect to $\theta = 0$, the dynamics is not. Indeed, in the $(\theta > 0)$-plane, the system acquires non-zero velocity $\Dot{\delta \theta}$ which leads $\delta \theta$ to slowly drift away from zero initial state. Right after the passage of the mountain \textemdash \ that is when $\theta$ becomes negative \textemdash \ the mountain's gravity acts as a restoring force, which has the immediate effect of slowing down $\delta \theta$. But it is already too late: in the meantime, the altitude has been disturbed $(\delta r \neq 0)$ and $\theta$ continues on its run (at an increased $\omega +\Dot{\delta \theta}$ pace), so that the restoring force at $-\theta_* < 0$ is not equal to the force that disturbed the Keplerian motion at $\theta_*$. Once the symmetry is broken, the orbit can no longer be circular \textemdash \ it has a non-zero osculating eccentricity \textemdash \ which is why $(\delta r, \Dot{\delta r}, \delta \theta, \Dot{\delta \theta})$ exhibit an oscillatory behavior at approximately the Keplerian frequency. We dedicate Appendix~\ref{app:proof} to prove this point in a more rigorous way.

\begin{figure}
    \centering
    \includegraphics[width=0.85\linewidth]{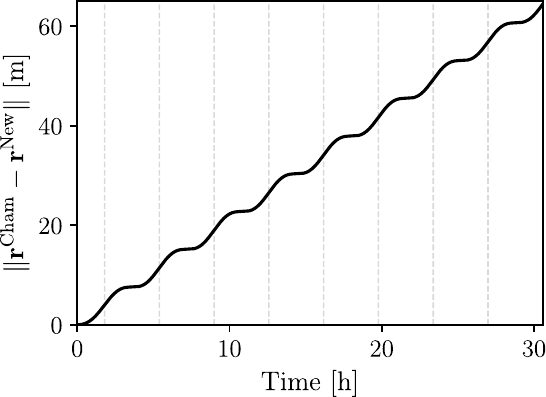}
    \caption{Distance anomaly as a function of time. The various passages above the mountain, depicted by the vertical light-gray dashed lines, correspond to the most rapid increase in this distance.}
    \label{fig:distance-anomaly}
\end{figure}

In the second row of Fig.~\ref{fig:Xstate}, we illustrate what we call the \textit{anomaly} $\mathbf{X}^{\mathrm{Cham}} - \mathbf{X}^{\mathrm{New}}$, that is simply the difference between the geodesic in modified gravity and in Newtonian gravity \textemdash \ for the same set of initial conditions. Surprisingly, apart from $\delta \theta^{\mathrm{Cham}} - \delta \theta^{\mathrm{New}}$ which undergoes a steady decline, the other elements of the anomaly seem to be periodic and remain around zero. In Fig.~\ref{fig:distance-anomaly}, we show the slow drift of the distance anomaly between the two trajectories, that is $\| \mathbf{r}^{\mathrm{Cham}} - \mathbf{r}^{\mathrm{New}} \|$. This steady increase of the distance has a mean slope of $\sim 2 \, \mathrm{m/h}$, but the rate of increase is maximized at each passage of the point-mass above the mountain where it exceed $4 \, \mathrm{m/h}$.

All these orders of magnitude relating to the anomaly should be put into perspective with the current level of precision with which we are able to determine a satellite's position and other orbital elements. This process goes under the name `\textit{Precise Orbit Determination}' (POD) and involves analyzing various observational data, often obtained from ground-based tracking stations or satellite-based instruments \textemdash \ see e.g. Refs.~\cite{stat_orbit_deter, POD-DORIS, POD-GNSS} for the implementation of these techniques and the reachable orders of magnitude in terms of precision. One of the main space geodetic techniques is \textit{Satellite Laser Ranging} (SLR) which measures the time it takes for a laser beam to travel from the ground station to a retro-reflector on the satellite and back again, providing unambiguous range measurements to millimeter precision \cite{SLR-LEO, SLR-science}. This technique is also placed at the service of fundamental physics; in this respect let us mention the recent launch of the LARES 2 satellite \cite{LARES-2} to test GR. 

Therefore, with no uncertainty on the model and initial conditions, the anomaly caused by the fifth force is of the order of a meter (see the leftmost column of Fig.~\ref{fig:Xstate}), which is around three orders of magnitude larger than the best attainable precision. At this stage of the discussion, it would seem easy to detect the fifth force.

\subsubsection{The GRACE-FO scenario}
\label{subsubsec:GFO}

The GRACE-FO\footnote{Gravity Recovery and Climate Experiment  Follow-On.} mission, currently in operation, aims at monitoring the Earth's gravitational field. It uses a pair of satellites flying on the same orbital path, approximately $220 \, \mathrm{km}$ apart. As they orbit the Earth, the spacecraft are affected by the uneven gravity field caused by the uneven distribution of mass inside the planet \textemdash \ e.g. the presence of a mountain, which produces a slightly stronger gravitational pull. As a result, the distance between the two satellites varies continuously over time. This distance variation is measured down to the micron level thanks to a microwave ranging system\footnote{GRACE-FO also employs laser-ranging interferometry for a more accurate inter-satellite ranging which can improve the separation distance measurement by a factor of more than 20 relative to the GRACE mission \cite{GFO-LRI}.} \cite{GRACE-FO}. Ultimately, the changes in the distance between the satellites are used to monitor the time variations of the Earth gravity field due to mass changes (ice melting, droughts, floods, etc.).

\begin{figure}
    \centering
    \includegraphics[width=0.9\linewidth]{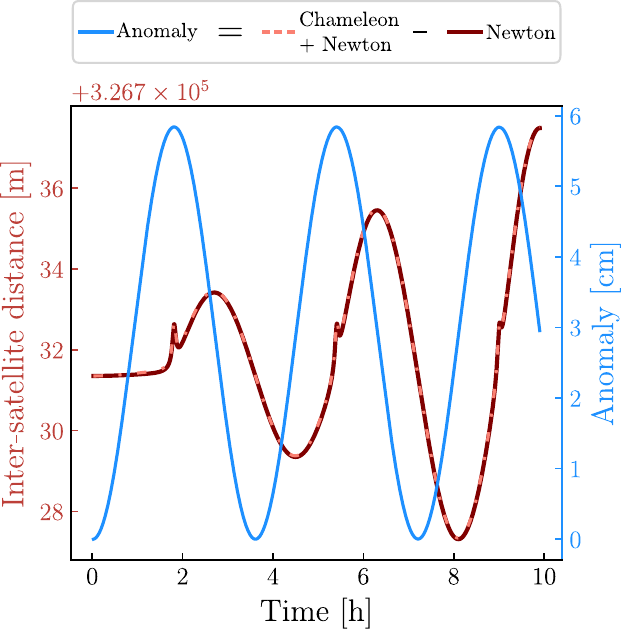}
    \caption{Left y-axis: inter-spacecraft distance with respect to time in Newtonian gravity (solid red line) and in modified gravity (dashed salmon line). Right y-axis: the blue curve corresponds to the anomaly, that is the difference between the two models. The initial time-delay between the two satellites is set to 100 seconds.}
    \label{fig:gfo-chameleon}
\end{figure}

Given the extreme level of precision GRACE-FO is able to reach in terms of ranging, we investigate whether or not fifth force effects would end up being in its sensitivity range. To do so, we simulate a pair of satellites following each other by duplicating the trajectory and shifting it in time by a few minutes, mimicking the real mission configuration. We can then reconstruct the change in inter-spacecraft distance with respect to time $d(t)$. An example of such a curve is given in Fig.~\ref{fig:gfo-chameleon} (red solid line or salmon dashed line, the two being indistinguishable by eye), where roughly three orbits have been completed. The passage above the mountain can easily be spotted on the curve by the little \textit{spikes} they spawn. They can be understood fairly intuitively: approaching the mountain's latitude, the leading satellite starts feeling a slightly stronger gravity relative to the trailing one and is pulled slightly ahead, increasing the distance between the two satellites. When the first satellite has eventually passed on the other side of the mountain (that is $\theta < 0$), it is slowed down while the trailing satellite is accelerating, resulting in a decrease in the inter-satellite distance overall. Long after the occurrence of this short-term event, this distance continues to vary in sinusoidal fashion. This is due to the fact that, as discussed above and brought to light in Fig.~\ref{fig:Xstate}, the orbit is no longer circular after the passage of the mountain and therefore the velocity varies along an orbit.

The difference between the modified gravity case $d_{\mathrm{Cham}}$ and the Newtonian case $d_{\mathrm{New}}$ can hardly be seen on those curves. It is depicted by the solid blue line on Fig.~\ref{fig:gfo-chameleon} (again called the `anomaly'). Choosing the same set of initial conditions for both models ensures that the anomaly is null at time $t = 0$. The anomaly exhibits three maxima \textemdash \ at a level of a few centimeters \textemdash \ corresponding to the passage of the pair of satellites above the mountain.

In view of these results, we may believe that chameleon-like fifth forces should be detectable with our current space technology as the anomaly is $\sim 10^4$ times larger than the sensitivity threshold of GRACE-FO. That would be true under the (unrealistic) assumptions that:
\begin{enumerate}
    \item the initial conditions are \textit{perfectly} known, that is there is no uncertainty in our initial state $\mathbf{X}_0$ prior to the propagation;
    \item the density model of the main body (the Earth) is \textit{perfectly} known.
\end{enumerate}
Neither of the two hypotheses can be fulfilled in practice. In the two forthcoming sections, we tackle these points and strongly mitigate our previous statement in regard to fifth force detectability in space.

\subsubsection{Perturbation of initial conditions}
\label{subsubsec:perturbation-IC}

Here, we investigate whether a slight modification of the initial state vector $\mathbf{X}_0 \gets \mathbf{X}_0 + \delta \mathbf{X}_0$ could account for the anomaly that we unveiled in Fig.~\ref{fig:gfo-chameleon}. For that purpose, we employ a Nelder-Mead optimizer where our objective function is
\begin{equation}
\begin{aligned}[t]
  g \colon \mathbb{R}^4 &\to \mathbb{R}_+\\
  \delta \mathbf{X}_0 &\mapsto \| d_{\mathrm{Cham}}(\mathbf{t}, \mathbf{X}_0) - d_{\mathrm{New}}(\mathbf{t}, \mathbf{X}_0 + \delta \mathbf{X}_0) \|_2 \, ,
\end{aligned}
\label{eqn:opt-func-1}
\end{equation}
where $\|\cdot\|_2$ is the two-norm in $\mathbb{R}^4$ and $\mathbf{t} = [t_0, t_1, \dots , t_N]$ is the discrete time vector with $t_N \sim 10 \, \mathrm{h}$ and $N = 10^4$. We find an optimum at
\begin{equation}
  \delta \mathbf{X}_0^{\mathrm{opt}} =
  \begin{bmatrix}
      -6.04 \times 10^{-1} & \mathrm{m} \\
      +6.16 \times 10^{-7} & \mathrm{m/s} \\
      -7.49 \times 10^{-7} & \mathrm{rad} \\
      +1.98 \times 10^{3} & \mathrm{m^2/s}
  \end{bmatrix} \, ,
\label{eqn:Xstate_ini_opt}
\end{equation}
with a residual smaller than $5 \, \mathrm{mm}$\footnote{The last entry of vector $\delta \mathbf{X}_0^{\mathrm{opt}}$ in Eq.~\ref{eqn:Xstate_ini_opt} corresponds to the perturbation in the initial angular momentum and may attract attention due to the fact it is orders of magnitude bigger than the other entries. To provide a benchmark, the unperturbed initial angular momentum is $L_0 \simeq 2.2 \times 10^{10} \, \mathrm{m^2/s}$.}. This is an extremely good fit given the characteristic length of the problem (several hundred kilometers). The conclusion to be drawn from this is clear: on this specific inter-satellite distance tracking example, an extra chameleonic acceleration cannot be distinguished from a small perturbation of the initial state vector. A brief analysis indicates that the parameter that has the biggest \textit{weight} in Eq.~(\ref{eqn:Xstate_ini_opt}) is $\delta r_0$. Now the question is how this small perturbation compares to the precision with which we have access to the initial state. As discussed previously in Sec.~\ref{subsubsec:geodesic-deviation}, it turns out that the initial radial distance $r_0$ could in principle be determined with at most centimetric precision which is smaller than the $60 \, \mathrm{cm}$ perturbation found in Eq.~(\ref{eqn:Xstate_ini_opt}). Although this does not constitute a rigorous proof, this brief study tends to indicate that the `unknown initial state' hypothesis can be ruled out.

\subsubsection{Perturbation of the mass distribution}
\label{subsubsec:perturbation-rho}

\begin{figure}
    \centering
    \includegraphics[width=0.95\linewidth]{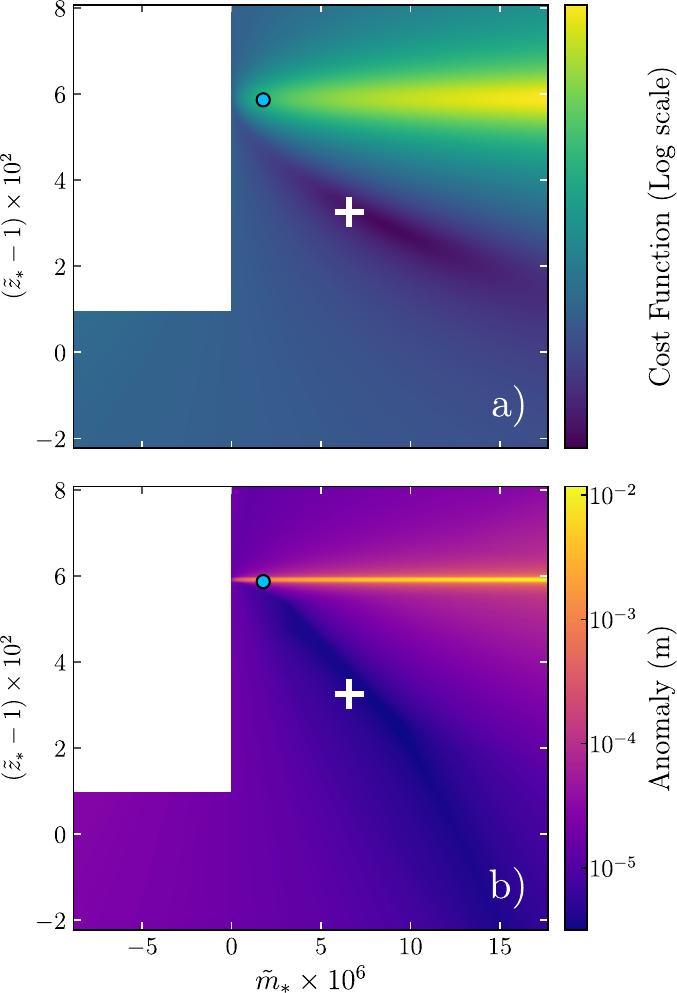}
    \caption{a) Contour plot of the objective function (in log scale) in the $(m_*,z_*)$-plane\footnote{Note that negative mass (which would correspond to an \textit{extrusion} for $\Tilde{z}_* < 1+h_m$) cannot represent the chameleon acceleration as well as positive mass.}. b) Contour plot of the anomaly in the $(m_*,z_*)$-plane. The white cross and blue circle are located at the objective function's minimum when $\Tilde{r} = 1.059$ and $\Tilde{r} = 1.111$ respectively. The area left in plain white is not physically accessible as it corresponds to a `negative extra mass' in vacuum. Warning: looking at the anomaly (bottom panel), one might expect the special case $m_* = 0$ to reduce to the case displayed on Fig.~\ref{fig:gfo-chameleon} and exhibit an anomaly of a few centimeters. The difference lies in the fact that here, the chameleon radial acceleration offset is artificially reproduced by slightly increasing the main body's mass.}
    \label{fig:mz-optim}
\end{figure}

Nonetheless, the knowledge of initial conditions is not the only potential source of degeneracy. Indeed, the mass distribution inside the main body \textemdash \ the very source of gravity \textemdash \ is perhaps the most important degree of freedom to have knowledge of. In that perspective, can the fifth force effects on a satellite be interpreted in the framework of Newtonian gravity as a slightly altered density model? In order to answer that  question, we continue in the same spirit as in Sec.~\ref{subsubsec:perturbation-IC} by constructing an optimization problem. We saw earlier on, notably in Fig.~\ref{fig:newton-ortho-profiles-techniques-comparison}, that the Newtonian potential of the mountain could very well be approximated by a point-mass. We can thus try and perturb the density model \textemdash \ and consequently the Newtonian potential \textemdash \ by adding a point-mass somewhere along the $z$-axis (see Fig.~\ref{fig:mnt-views}), as we do not wish to break the azimuthal symmetry. This simple model has only two parameters:
\begin{itemize}
    \item[--] $m_*$ the mass of the point-mass;
    \item[--] $z_*$ the $z$ coordinate of the point-mass.
\end{itemize}
The goal is then to find the pair $(m_*, z_*)$ for which Newtonian gravity best mimics the modified gravity case. Precisely, our objective function is
\begin{equation}
\begin{aligned}[t]
  f \colon \mathbb{R}^2 &\to \mathbb{R}_+\\
  (m_*, z_*) &\mapsto
  \int_0^{\pi} \left( \partial_{\theta} \Phi_* - \partial_{\theta} \Psi \right)^2 \mathrm{d}\theta\\[3pt]
  & \hphantom{\mapsto} + \int_0^{\pi} \left( \partial_r \Phi_* - \partial_r \Psi \right)^2 \mathrm{d}\theta
\end{aligned}
\label{eqn:opt-func-2}
\end{equation}
where $\Phi_*$ is the Newtonian potential created by the extra point-mass and the integral is carried out at fixed $\Tilde{r}$. We denote by $(m_*^{\Tilde{r}}, z_*^{\Tilde{r}})$ the pair that minimizes the function $f$ at radius $\Tilde{r}$. Using $\|\cdot\|_{L^2}$ to denote the $L^2$-norm over the space of square-integrable function on $[0, \pi]$, one can rewrite
\begin{equation*}
    f(m_*, z_*) = \|\partial_{\theta} (\Phi_* - \Psi)\|_{L^2}^2 + \| \partial_r (\Phi_* - \Psi)\|_{L^2}^2 \, .
\end{equation*}
Basically, we aim at approximating both the radial and orthoradial parts of the chameleon acceleration at the same time. This optimization problem being low-dimensional, we can dispense with a sophisticated optimization algorithm and do a full exploration of the parameter space instead (see Fig.~\ref{fig:mz-optim}). Note that our point-mass model \textit{cannot} reproduce the chameleon monopole (which is, in other words, a constant radial acceleration offset). Therefore, we removed it by hand before proceeding to the optimization phase. This offset is tiny: $\sim 1.4 \times 10^{-7} \, \mathrm{m/s^2}$ which corresponds to relative change of the mean density of the main body of only $\sim 9 \times 10^{-8}$. In comparison, let us mention that the Earth mass is known with a relative uncertainty of $10^{-4}$.

The results for $\Tilde{r} \in \{ 1.059, 1.111 \}$ are reported in Table~\ref{tab:ptmass-bestfit}, where $\Tilde{m}_*^{\Tilde{r}} = m_*^{\Tilde{r}} / M_{\mathrm{mountain}}$ and $\Tilde{z}_*^{\Tilde{r}} = z_*^{\Tilde{r}} / R_{\mathrm{body}}$.

\begin{table}[]
    \centering
    \begin{ruledtabular}
    \begin{tabular}{l c c}
      & $\Tilde{r} = 1.059$ & $\Tilde{r} = 1.111$ \\[7pt]
      $\Tilde{m}_*^{\Tilde{r}}$ & $6.6 \times 10^{-6}$ & $1.8 \times 10^{-6}$ \\[5pt]
      $\Tilde{z}_*^{\Tilde{r}}$ & $1.03$ & $1.06$ \\[5pt]
      $f(m_*^{\Tilde{r}}, z_*^{\Tilde{r}}) / f(0, 1)$ & $3.1 \times 10^{-2}$ & $5.6 \times 10^{-2}$ \\[5pt]
      $\|\partial_{\theta} (\Phi_* - \Psi) \|_{L^2} / \|\partial_{\theta} (\Phi_* + \Psi) \|_{L^2}$ & $0.14$ & $0.22$ \\[5pt]
      $\|\partial_{r} (\Phi_* - \Psi) \|_{L^2} / \|\partial_{r} (\Phi_* + \Psi) \|_{L^2}$ & $0.07$ & $0.08$
    \end{tabular}
    \end{ruledtabular}
    \caption{Best fit parameters of the approximation of the chameleon acceleration by a point-mass in Newtonian gravity.}
\label{tab:ptmass-bestfit}
\end{table}

To further assert the quality of the fit in quantitative terms, we compute the ratio $f(m_*, z_*)/f(0, 1)$ (where $\{m_* = 0 , z_*=0\}$ corresponds to flat profiles) as well as relative \textit{errors} in $L^2$-norm. Several comments must be made:
\begin{itemize}
  \item[--] With only a simplistic model (i.e. a single point-mass has been added to the pre-existing model, contributing to the global Newtonian potential), we manage to approximate the fifth-force profile at a given altitude with remarkable accuracy (see the various figures in Table~\ref{tab:ptmass-bestfit}).
  \item[--] This approximation is good enough to \textit{almost} reproduce the dynamics of the satellite's orbit over the mountain. In fact, we can repeat for instance the same exercise as we did in Fig.~\ref{fig:gfo-chameleon} and compute the so-called \textit{anomaly}, i.e. the difference between the ``modified gravity without extra mass" case and the ``Newtonian gravity with extra mass" case. We find it to be no greater than $15 \, \mathrm{\mu m}$. This is more than a thousand times smaller than the anomaly computed in Fig.~\ref{fig:gfo-chameleon}. This invites us to moderate the statements made earlier, since we are approaching here the precision limits of the GRACE-FO's LRI system.
  \item[--] The objection could be made that the characteristics of the point-mass associated with the best fit do not correspond to any physical reality. Indeed, taking the second column of Table~\ref{tab:ptmass-bestfit} with entries $(\Tilde{m}_* = 6.6 \times 10^{-6}, \, \Tilde{z}_* = 1.03)$ suggests that there would be a $\sim 2 \times 10^{12} \, \mathrm{kg}$ mass at an altitude of $186 \, \mathrm{km}$ (or equivalently, $123 \, \mathrm{km}$ above the mountain's top) \textemdash \ which is obviously absurd! Nevertheless, it can be seen on the top panel of Fig.~\ref{fig:mz-optim} \textemdash \ which represents the cost function (\ref{eqn:opt-func-2}) in the $(m_*,z_*)$-plane \textemdash \ that lowering a bit $z_*$ from the optimum (depicted by the white cross) while maintaining $m_*$ constant has only a slight effect on the cost function. For this reason, the dynamics is not much affected by a shift of $z_*$ towards the planet. As a matter of fact, setting $z_* = 1$ (i.e. bringing the extra mass at the planet's surface) leads to an anomaly bounded below $40 \, \mathrm{\mu m}$. The bottom panel of that same figure is intended to illustrate this phenomenon, and the strong correlation between the cost function and the anomaly is visible to the naked eye.
\end{itemize}
We can even place this extra mass at the same location as the point-mass Newtonian approximation of the mountain itself (see caption of Fig.~\ref{fig:newton-ortho-profiles-techniques-comparison}) without any major change in the dynamics. This can therefore be interpreted as slightly increasing the mountain's density, by roughly $10^{-3} \, \%$. Such a slight deviation could equivalently be attributed to the fact that the gravitational constant $G$ is only known with some certainty with four significant digits \cite{nist-G, Li2018}.

These orders of magnitude on the density must be put into perspective with our current knowledge of the Earth inner density, with all the attendant uncertainties. Despite advancements in geophysical techniques, our knowledge of mass distribution is still imperfect, for simple reasons:
\begin{enumerate}
  \item The planet's interior is out of reach. As a matter of fact, the deepest human-made hole ever dug is \textit{only} $12.3 \, \mathrm{km}$ deep (less than 0.2 \% of the Earth radius).
  \item The density is not uniform (even at fixed depth), which means extrapolation is not a valid procedure unless strong assumptions are made about the Earth's composition and structure.
  \item As it happens, we also have to rely on indirect measurements, ranging from gravitational anomalies and magnetic anomalies \cite{Chouhan2020, zhao_brief_2021, mag-anomaly-germany} to seismic analysis \cite{PREM1981, seismology1, seismology2}. All these techniques are in turn limited in both resolution and accuracy.
\end{enumerate}
On this latter point, we stress that one should be careful when trying to put constraints on a given modified gravity model, using a model of the Earth that comes from gravitational measurements in the first place. Indeed, the inversion of a gravity map into, say, a density map is model dependent (and unless contraindicated, would have been performed in a Newtonian framework). See Ref.~\cite{Bergé_2018}, where this topic is discussed at length. In this regard, let us mention a recent work \cite{kozak2023earthquakes} that proposes to use the preliminary reference Earth model (PREM) \cite{PREM1981}, which is a radial seismic model, to constrain some alternative theories to GR.

\subsubsection{Breaking the degeneracy}
\label{subsubsec:break-degeneracy}

We have seen with two simple examples that drawing a distinction between a fifth force and model uncertainties (of different natures) is no easy task. These uncertainties spearhead degeneracies, which we partially address here.

In Sec.~\ref{subsubsec:perturbation-IC}, we provided the relevant orders of magnitude of the perturbation of a satellite initial state vector necessary to alone mimic a fifth force influence. The perturbation on the initial altitude was then put in comparison against the available level of precision for LEO satellites. It turned out POD techniques are good enough to the relatively large perturbation found in Eq.~(\ref{eqn:Xstate_ini_opt}). Yet one must bear in mind that this was done on a very specific test-case and the conclusion may not generalize to others.

In Sec.~\ref{subsubsec:perturbation-rho}, we looked at how to distinguish a chameleon acceleration on top of Newtonian dynamics from a slight change of the density model in a purely Newtonian framework. We showed that it was possible to imperceptibly tweak the mass distribution in the mountain and in the planet to reproduce the chameleon acceleration profile \textit{at a given altitude} (see Table~\ref{tab:ptmass-bestfit}). This naturally raises the question of whether such a fit works at different altitudes, or rather \textit{how well}. Elements of response can be found in Table~\ref{tab:ptmass-bestfit} and Fig.~\ref{fig:mz-optim}. The first rows of Table~\ref{tab:ptmass-bestfit} bring out the fact that inferring the mountain's characteristics at two orbital radii leads to two clashing physical realities: strikingly, the extra mass inferred at $\Tilde{r} = 1.059$ is almost 4 times greater than it appears at $\Tilde{r} = 1.111$. In order to sharpen the analysis, we represent in Fig.~\ref{fig:mz-optim} by a white cross and a blue circle the cost function's minimum at $\Tilde{r} = 1.059$ and $\Tilde{r} = 1.111$ respectively, while the contour plots are performed for $\Tilde{r} = 1.059$. Following the notations introduced above, $(m_*^{1.059}, z_*^{1.059})$ and $(m_*^{1.111}, z_*^{1.111})$ are coordinates of the white cross and the blue circle respectively. Similarly, $f_{1.059}$ and $f_{1.111}$ refer to the cost functions at the two altitudes. At $(m_*^{1.111}, z_*^{1.111})$, we see that the cost function $f_{1.059}$ is much above its minimum and, in turn, corresponds to a large anomaly. Quantitatively speaking, we have
\begin{equation*}
    \frac{f_{1.059}(m_*^{1.111}, z_*^{1.111})}{f_{1.059}(m_*^{1.059}, z_*^{1.059})} \simeq 7.8 \times 10^2 \, ,
\end{equation*}
which is a big ratio and reflects that $(m_*^{1.111}, z_*^{1.111})$ does not produce a good fit of the chameleon acceleration profile at $\Tilde{r} = 1.059$. On the other hand, changing our perspective to $f_{1.111}$, we have
\begin{equation*}
    \frac{f_{1.111}(m_*^{1.059}, z_*^{1.059})}{f_{1.111}(m_*^{1.111}, z_*^{1.111})} \simeq 13 \, ,
\end{equation*}
meaning that $(m_*^{1.059}, z_*^{1.059})$ is better \textit{tolerated} by $f_{1.111}$ and $f_{1.059}$ than $(m_*^{1.111}, z_*^{1.111})$.

\begin{figure*}
    \centering
    \includegraphics[width=\linewidth]{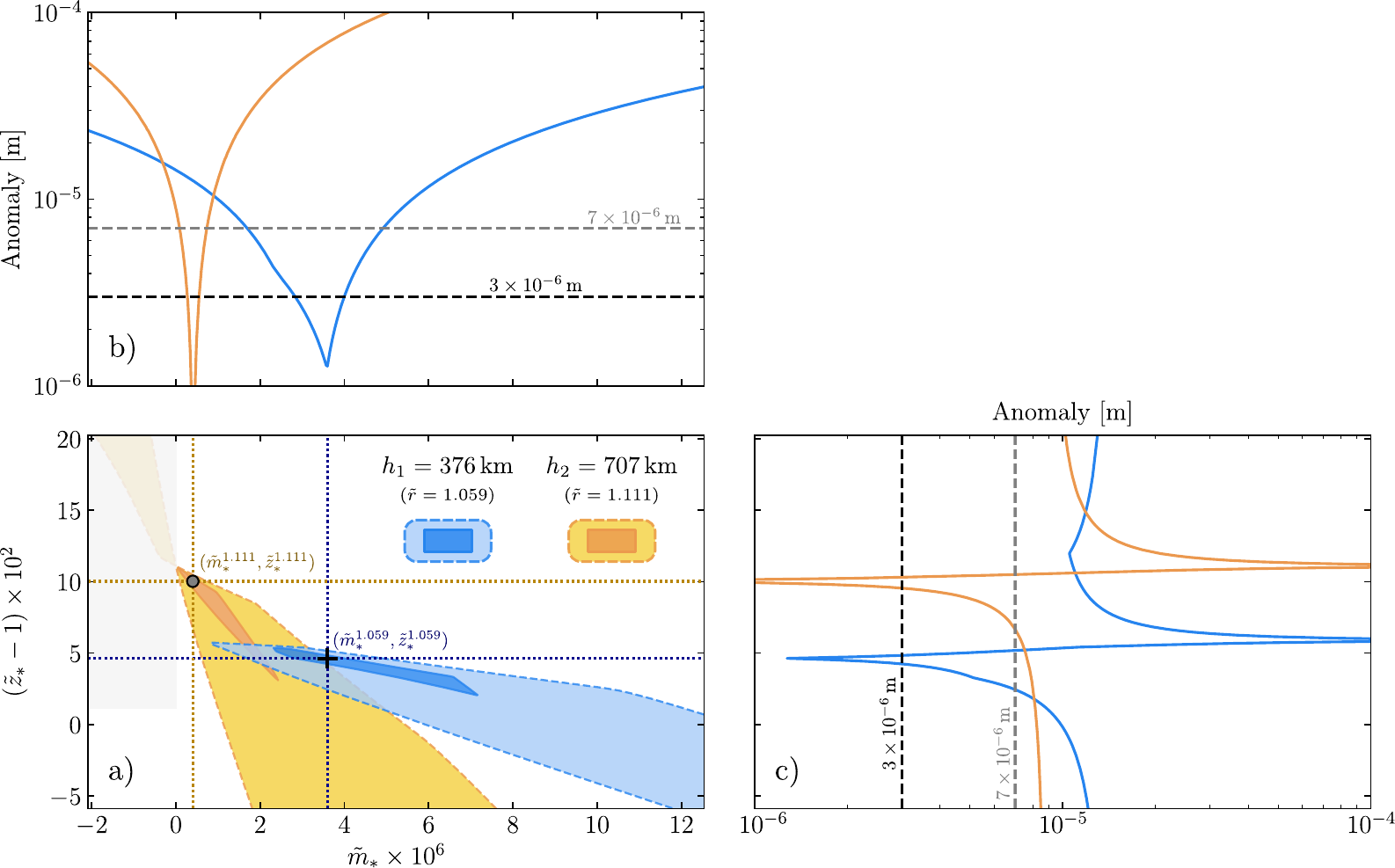}
    \caption{Tensions in the inferred mountain's characteristics. The blue elements refer to the altitude $h_1 = 376 \, \mathrm{km}$ ($\Tilde{r} = 1.059$) and the orange elements refer to the altitude $h_2 = 707 \, \mathrm{km}$ ($\Tilde{r} = 1.111$). The anomaly observed in the inter-satellite distance at $h_1$ (resp. $h_2$) is best explained in the framework of Newtonian gravity by an extra point-mass with characteristics $(m_*^{1.059}, z_*^{1.059})$ (resp. $(m_*^{1.111}, z_*^{1.111})$) depicted by the black cross (resp. gray circle) in panel a)\footnote{Note that the pairs $(m_*^{\Tilde{r}}, z_*^{\Tilde{r}})$ are different from the ones reported in Table~\ref{tab:ptmass-bestfit} for two reasons: (i) here, we minimize the anomaly in the inter-satellite distance between the ``modified gravity without extra mass" case and the ``Newtonian gravity with extra mass" case, which is different from minimizing the objective function $f$ given by Eq.~(\ref{eqn:opt-func-2}); and (ii) we allowed ourselves to modify the initial inter-satellite distance for the two considered altitudes in order to better showcase the tension.}, which minimizes the anomaly down to $1.3 \times 10^{-6} \, \mathrm{m}$ (resp. $2.6 \times 10^{-7} \, \mathrm{m}$). In panel a), the darker contours map to an anomaly below $3 \times 10^{6} \, \mathrm{m}$ while the lighter ones map to an anomaly below $7 \times 10^{-6} \, \mathrm{m}$. The gray shaded area is not physically accessible as it corresponds to a `negative extra mass' in vacuum. Panel b) (resp. c)) represent the anomaly along $\Tilde{m}_*$ at $\Tilde{z}_*^{\Tilde{r}}$ (resp. $\Tilde{z}_*$ at $\Tilde{m}_*^{\Tilde{r}}$). The anomaly peaks (maxima) visible on panel c) correspond to the horizontal feature seen previously in Fig.~\ref{fig:mz-optim}-b) around $\Tilde{z}_* = 1.06$.}
    \label{fig:tensions}
\end{figure*}

Fig.~\ref{fig:tensions} provide more visual insights into these tensions. As in panel b) of Fig.~\ref{fig:mz-optim}, we computed the anomaly as a function of the pair $(m_*, z_*)$, in a scenario where the two GRACE-FO-like satellites orbit at $\Tilde{r} = 1.059$ (associated with blue colors) and in another scenario where they orbit at $\Tilde{r} = 1.111$ (associated with orange colors). On panel a), we display two contours corresponding to anomaly thresholds of $3 \times 10^{-6} \, \mathrm{m}$ and $7 \times 10^{-6} \, \mathrm{m}$, for both altitudes. The less the blue contours overlap with the orange ones, the greater the tension. Panels b) and c) complement the figure by representing the anomaly along the dotted lines visible on panel a) which pass through the minimal anomaly for each altitude. Ideally, the performances showcased by the GRACE-FO laser-link technology would allow for an exclusion of any $(m_*, z_*)$ pair mapping to an anomaly greater than a micrometer, revealing the incompatibility between the two density models.

In conclusion of this section, the use of different altitudes in the analysis is a first step toward breaking the degeneracy. This idea was already put forward in Ref.~\cite{Bergé_2018} where the authors study the impact of a Yukawa potential on the spherical harmonic coefficients of the Earth. Precisely, the rescaled coefficients $y_{lm}$ (introduced in Sec.~\ref{subsubsec:rescaled-coeffs}) become dependent on the altitude meaning for instance that measurements of the $J_2$ zonal term at GOCE and GRACE altitudes could provide a test of the model\footnote{Note that this would also be true for the chameleon model as $y_{lm}^C$ depends on the radial coordinate $r$.}. On the whole, the difficulty lies in being able to find a set of several physical measures that would be in tension one with another when adding a fifth force to the play:
\begin{itemize}
    \item[--] The greater the tension, the tighter the potential constraints on the modified gravity model.
    \item[--] The more measurements we have, the greater the likelihood of ending-up with a significant tension. 
\end{itemize}
We tried to bring out such a tension with the GRACE-FO setup deployed at two different altitudes (see Table~\ref{tab:ptmass-bestfit} and Fig.~\ref{fig:mz-optim}). Nevertheless, this does not look practical in actual experiments. Even though we manage to create a small tension in the anomaly, one must bear in mind that our fitting model consisting of a single point-mass remains overly simplistic. More complex (and viable) density models may relieve this tension very well. In short, the ultimate goal of constraining the chameleon model with space-based geodesy is impeded by:
\begin{itemize}
    \item[--] the uncertainty on the source of gravity, that is the density model;
    \item[--] the uncertainty on the measurements themselves;
    \item[--] all the other forces acting on a satellite, ranging from atmospheric drag and solar radiation pressure, to third-body perturbation, which also come with error bars in our models. Note that these perturbing forces are also a nuisance for geodesy, hence the use of accelerometers on board satellites \cite{IEA1, IEA2}.
\end{itemize}

In Appendix~\ref{app:deltaT}, we further look at orbital periods at two different altitudes and compute their difference, in a Newtonian framework and in a modified gravity framework.

\section{Discussion and conclusion}
\label{sec:discussion}

\subsubsection*{Effect of a mountain}

This article investigated the testability of chameleon gravity by space geodesy experiments, with a focus on the influence of the local landform and the atmosphere. The motivations were twofold. First, viable regions of the chameleon parameter space all map to a screened Earth, that is only a thin-shell contributes to the fifth force. Therefore, it seemed important to study departures from spherical symmetry, hereby embodied by a mountain. Second, while published works sometimes account for the atmosphere in their study, the models implemented are simplistic (often one layer of constant density) and the determination of whether it has a thin-shell is based on rather qualitative arguments. Addressing such questions is not possible by means of analytical techniques alone due to the complexity of the physical models we wished to study, and to the nonlinear nature of the chameleon equation of motion. We thus resorted to numerical simulations \textemdash \ performed with the code \textit{femtoscope} \textemdash \ to conduct this work.

We obtained the chameleon contribution to the total gravitational potential of a mountainous planet, scanning through an extended region of the parameter space. As already pointed out in Ref.~\cite{Levy_2022}, the unscreened regime shares similarities with pure Newtonian gravity in that, in both cases, the fields are sourced by the entire mass of the main body. Consequently, the chameleon potential in the unscreened regime is roughly the same as the Newtonian potential up to an affine transform (and the same goes for the accelerations). As we enter the screened regime however, the multipole expansion of the chameleon field starts to depart from that of the Newtonian potential, revealing a distinct signature. In terms of acceleration, the chameleon acceleration vector is a bit more directed towards the mountain compared to the Newtonian acceleration. Their norm ratio remains small though, bounded from above by $\sim 10^{-6}$ at the equivalent of LEO altitudes in the atmosphere-free case\footnote{It is insightful to compare this ratio with the ratio of the Solar radiation pressure over the Newtonian acceleration, which is around $10^{-8}$ \cite{satellite_orbits}. Despite being so tiny, the Solar radiation pressure perturbing acceleration, when integrated over many orbits, is enough to cause significant drifts of orbital elements \cite{SRP-effect}. What makes the chameleon acceleration difficult to distinguish from the Newtonian acceleration is the fact that they are both sourced by the same body.}.

\subsubsection*{Effect of an atmosphere}

Based on our study of three distinct atmospheric density profiles, we found that the addition of an extra layer of air surrounding the main body can mitigate the effect of the fifth force. We showed that there exists a threshold on the value of the parameter $\alpha$. Above this threshold, the atmosphere acts as an attenuator, effectively reducing the chameleon acceleration by a certain amount compared to the case without atmosphere. Below this cutoff, the effect of the atmosphere is more drastic: any non-radial dependence of the scalar field vanishes \textemdash \ the mountain is plainly \textit{invisible}. For even smaller values of $\alpha$, the atmosphere itself becomes screened, and the chameleon field is thereupon fully determined by the atmospheric density profile. As we saw, it is even possible in this case to enhance the radial fifth force at given altitude with respect to the atmosphere-free case. This study represents a step forward with respect to previous work discussing the influence of the atmosphere. Moreover, this clearly gives the edge to bodies devoid of atmosphere when it comes to select a Solar System site for testing this screened scalar-tensor model.

\subsubsection*{Space geodesy experiments}

Our knowledge of the geopotential comes to a large extent from spaceborne geodesy. From this standpoint, we thus investigated whether constraints could be put on modified gravity models using satellites in orbit. For that purpose, we performed orbit propagations, with and without the putative fifth force, and studied the resulting anomaly\footnote{The term `anomaly' is used to refer to the difference for a given observable between the \{Newtonian gravity\} case and the \{Newtonian gravity + fifth force\} case.} on several observables (such as the variations of the distance between two satellites following each other as in the GRACE and GRACE-FO setups). While the anomalies we find are technically well within the detection range of current on-board and ground-based space-technology, we showed that uncertainties in the model for the distribution of matter are large enough to allow for degeneracies.

We laid emphasis on the fact that one way to distinguish a chameleon acceleration from a slight change of the Earth density model in a purely Newtonian framework is to rely on experiments performed at (at least) two different altitudes. Indeed, in the regime where the Earth (or any other planetary body) is screened, the chameleon acceleration does not decrease as $r^{-2}$ like the Newtonian acceleration (this is particularly stressed in Refs.~\cite{Bergé_2018, stt-ppn}). If the chameleon field \textit{actually exists}, then inferring density models under the assumption of Newtonian gravity at several altitudes should result in tensions between those models. Of course, these tensions should be accounted for in a probabilistic way, which is beyond the scope of this article. Conversely, if it were not for all the other perturbing forces that greatly complexify the model, this method could be used to put constraints on the chameleon model, and more generally on massive scalar-tensor theories.

\subsubsection*{Back-reaction of a satellite on the scalar field}

We also took into account the back-reaction of an object as small as a satellite in orbit on the scalar field. For the first time, we went beyond the various approximations found in the literature and computed the \textit{full solution} of the \{Earth + Satellite\} system using \textit{femtoscope}. This involves taking advantage of the h-adaptivity technique granted by FEM. We could then compute the overall fifth force acting on an object with a simple geometry and characteristics close to that of a real spacecraft (length-scale and density). Surprisingly, as long as the satellite is not screened and despite the background field being disturbed, the global chameleon acceleration undergone by the satellite is the same as the one a point-particle (not disturbing the background field) would experience. We provide mathematical insights into why this is the case in Appendix~\ref{app:remarkable}. In the screened regime however, the net fifth force vanishes to zero. The transition between those two regimes occurs over a narrow band in the chameleon parameter space.

\subsubsection*{Outlook}

While we focused on fifth force searches, other venues exist to test scalar-tensor theories. For instance, in any such theory involving a conformal coupling of the scalar field $\phi$ to matter fields in the Einstein frame, the gravitational redshift effect has a $\phi$-dependence (see e.g. Ref.~\cite{Khoury&WeltmanPRD} for the chameleon's contribution to this effect). We will take a deeper look at this effect in an upcoming article. Most importantly, we will tackle the question of what is actually \textit{measurable}, and with which precision. As a complement to fifth force searches where we look for dynamical effects whose amplitude depends inherently of the field's gradient, gravitational redshift (or equivalently, gravitational time-dilation) can be measured in a static configuration and is sensitive to the field's strength. Whether clocks are put into orbit (as envisaged in the ACES mission \cite{aces}) or left on Earth, they have become so precise\footnote{High-precision clocks, such as optical lattice clocks, currently achieve astonishing levels of accuracy with a fractional frequency uncertainty of approximately $10^{-19}$ to $10^{-20}$ \cite{Brewer2019, Bothwell2022}.} that their constraining power (i.e. the possibility to use this technology to rule out modified gravity models) has to be quantified. The bound given in Ref.~\cite{Khoury&WeltmanPRD} has to be revisited, given two decades elapsed since the writing of this article.

\begin{acknowledgments}
\end{acknowledgments}

HL thanks Pablo Richard for fruitful discussions about numerical computations. We acknowledge the financial support of CNES through the APR program (“MICROSCOPE” project).

\newpage
\appendix
\hphantom{one line text}
\section{Conversion of cosine and sine coefficients to bare coefficients}
\label{app:coefficient-conversion}

Let $f \colon \mathcal{S}^2 \to \mathbb{R}$ be a real-valued function on the unit sphere and $L \in \mathbb{N}^*$ a maximum spherical harmonic degree. The truncated spherical harmonic expansion, which defines an approximation $f_{\mathrm{trunc}} \simeq f$, may be written as
\begin{equation}
\begin{split}
    f_{\mathrm{trunc}}(\mathbf{n}) = \sum_{m=0}^L & \sum_{l=m}^L \Bigl[ C_{lm} \Bar{P}_{lm}(\cos \theta) \cos(m \varphi) \\
    & + S_{lm} \Bar{P}_{lm}(\cos \theta) \sin(m \varphi) \Bigr]
\end{split}
\label{eqn:truncated-sph-decomposition-stokes}
\end{equation}
with $\mathbf{n} = (\theta, \varphi)$ and $\Bar{P}_{lm}$ the normalized associated Legendre functions\footnote{In this respect, the definition of \textit{normalized associated Legendre functions} is consistent with the definition of orthonormalized spherical harmonic functions. See Table~1 from Ref.~\cite{SHTools}.} which relate to their unnormalized counterparts $P_{lm}$ via
\begin{equation}
    \Bar{P}_{lm}(x) = \sqrt{\frac{(2-\delta_{m0})(2l+1)}{4 \pi} \frac{(l-m)!}{(l+m)!}} P_{lm}(x) \, .
\end{equation}
We want to convert the cosine and sine coefficients $(C_{lm}, S_{lm})$ into the bare coefficients $f_{lm}$ that appear in the usual expansion
\begin{equation}
    f_{\mathrm{trunc}}(\mathbf{n}) = \sum_{l = 0}^L \sum_{m=-l}^{+l} f_{lm} Y_{lm}(\mathbf{n}) \, .
\label{eqn:truncated-sph-decomposition-bare}
\end{equation}
In order to express $(C_{lm}, S_{lm})$ as a function of $f_{lm}$, we start from Eq.~(\ref{eqn:truncated-sph-decomposition-bare}) and interchange the order of summations over $l$ and $m$ to obtain
\begin{widetext}
\begin{equation}
\begin{split}
f_{\mathrm{trunc}}(\mathbf{n}) & = \sum_{l=0}^{L} \sum_{m=-l}^{+l} f_{lm} Y_{lm}(\mathbf{n}) \\[3pt]
 & = \sum_{l=0}^L \left[ \sum_{m=0}^{+l} f_{lm} \Bar{P}_{lm}(\cos \theta) \cos(m \varphi) + \sum_{m=-l}^{-l} f_{lm} \Bar{P}_{l|m|}(\cos \theta) \sin (|m| \varphi) \right] \\[3pt]
  &= \sum_{l=0}^L \left[ \sum_{m=0}^{+l} f_{lm} \Bar{P}_{lm}(\cos \theta) \cos(m \varphi) + \sum_{m'=1}^{+l} f_{l,-m'} \Bar{P}_{lm'}(\cos \theta) \sin (m' \varphi) \right] \\[3pt]
  &= \sum_{l=0}^L \left[ \sum_{m=0}^{+l} C_{lm} \Bar{P}_{lm}(\cos \theta) \cos(m \varphi) + \sum_{m=0}^{+l} S_{lm} \Bar{P}_{lm}(\cos \theta) \sin (m \varphi) \right] \\[3pt]
  &= \sum_{m=0}^L \sum_{l=m}^L \left[ C_{lm} \Bar{P}_{lm}(\cos  \theta) \cos(m \varphi) + S_{lm} \Bar{P}_{lm}(\cos  \theta) \sin(m \varphi) \right] \, .
\end{split}
\label{eqn:bare2stokes-computation}
\end{equation}
\end{widetext}
The above computation is consistent if we set
\begin{equation}
\begin{split}
  C_{lm} &= f_{lm} \quad \text{if } m \geq 0 \, , \\[5pt]
  S_{lm} &= \begin{cases}
  0 & \text{if } m=0 \\
  f_{l,-m} & \text{if } m \geq 1
\end{cases} \, .
\end{split}
\label{eqn:stokes2bare-summary}
\end{equation}

\section{Verification of the scaling relation for the spherical harmonic coefficients of the Newtonian potential}
\label{app:scaling-relation}

\begin{figure}
    \centering
    \includegraphics[width=0.9\linewidth]{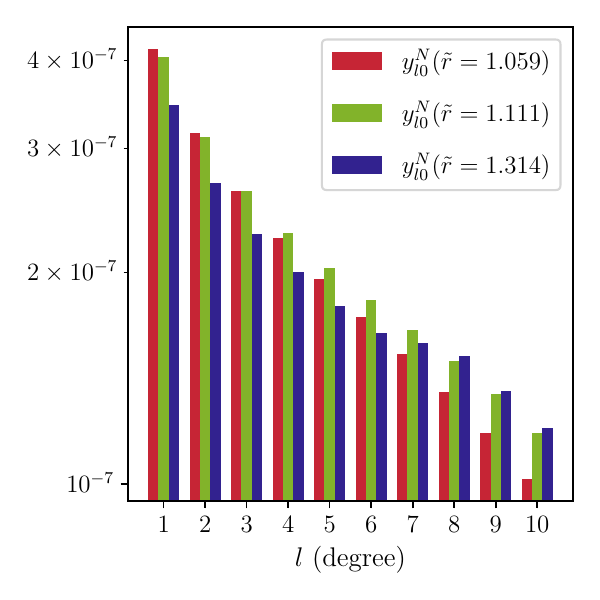}
    \caption{Verification of the scaling relation between bare spherical harmonic coefficients $\Phi_{l0}(\Tilde{r})$ and dimensionless coefficients $y_{l0}^N(\Tilde{r})$ obtained numerically. The rescaled coefficients should in principle be independent of the altitude at which they are computed.}
    \label{fig:scaling-relation}
\end{figure}

The Newtonian potential defined by Eq.~(\ref{eqn:newton-potential-integral}) is special in that its bare spherical harmonic coefficients $\Phi_{lm}(r)$ can be rescaled according to Eq.~(\ref{eqn:rescaled-newton-coeff}) which yields altitude-independent coefficients. We denote these rescaled coefficients $y_{lm}^N$. This peculiar property can be used as an additional means of test ascertaining the quality of our numerical approximations. Indeed, from our numerical $\Phi(r, \theta)$ maps of the Newtonian potential, we can compute the rescaled coefficients $y_{l0}^N$ at several altitudes and check whether or not they actually depend on the altitude. Fig.~\ref{fig:scaling-relation} shows the result of this process for $\Tilde{r} \in \{1.059, 1.111, 1.314\}$ and $l \in \{1, \cdots, 10\}$. At first sight, the scaling relation seems consistent with the numerical data at low degree. It is however more difficult to verify it at higher altitude and for higher degrees as the rescaling process involves multiplying the bare coefficients by $\Tilde{r}^{l+1}$ which quickly blows up to infinity. The bare coefficients being themselves plagued with numerical errors \textemdash \ they are derived from spherical harmonic decomposition algorithm on top of FEM computations \textemdash \ we clearly do not expect this relation to perfectly hold in this regime.

\section{Additional checks on 2D numerical computations}
\label{app:checks}

In this Appendix, we present two additional checks that were performed on all FEM computations of the chameleon field done in this article. We mainly elaborate on the ideas introduced in Sec.~\ref{subsubsec:challenges-verif}.

\subsection{Check of the radial evolution of the chameleon field}

The \{Earth + mountain\} system constitutes a small departure from spherical symmetry. Therefore, the behavior of the chameleon field in the outgoing radial direction should be close to that of the \{Earth\} system, which in turn is spherically symmetrical, and so purely radial. From a numerical viewpoint, such radial profiles are much easier to obtain than a less symmetrical case. Indeed in the former case, the Klein-Gordon equation (\ref{eqn:KG-alpha}) boils down to a simple ODE
\begin{equation}
    \alpha \frac{\mathrm{d}}{\mathrm{d}\Tilde{r}} \left( \Tilde{r}^2 \frac{\mathrm{d} \Tilde{\phi}}{\mathrm{d}\Tilde{r}} \right) = \Tilde{r}^2 \Tilde{\rho} - \Tilde{r}^2 \Tilde{\phi}^{-(n+1)} \, ,
\end{equation}
where numerical resources (density of DOFs, order of the finite elements) can be increased without blowing up the time complexity of the algorithm. As a result, we can obtain benchmark solutions at relatively low cost, for all the cases discussed in this study treated as purely radial (i.e. without mountain, all other physical parameters being equal). We denote $\Tilde{\phi}_{\mathrm{1D}}(\Tilde{r})$ such benchmarks, and $\Tilde{\phi}_{\mathrm{2D}}(\Tilde{r}, \theta)$ the 2D field profiles presented throughout the article. We then implement the following metric:
\begin{equation}
    \text{for } \Tilde{r}>1+h_m , \ \Gamma_{\Tilde{r}}  = \frac{\min\limits_{\theta \in [0, \pi]} \left| \Tilde{\phi}_{\mathrm{1D}}(\Tilde{r}) - \Tilde{\phi}_{\mathrm{2D}}(\Tilde{r}, \theta) \right| }{\left| \Tilde{\phi}_{\mathrm{1D}}(\Tilde{r}) \right|} \, .
\end{equation}
Note that this metric has the advantage of being relative, as opposed to the absolute criteria discussed in Sec.~\ref{subsubsec:challenges-verif}.

Tab.~\ref{tab:Gamma_r-all} includes $\Gamma_{\Tilde{r}}$ for radial coordinates $\Tilde{r} \in \{ 1.059, 1.111, 1.314, 4.645, 6.617 \}$ and for all ($\alpha$, atmospheric scenario) pairs considered in this work. Although there is no physical motivation for having $\Gamma_{\Tilde{r}} \equiv 0$ systematically, the fact that it remains below one part in a thousand in the vast majority of cases reflects the good agreement between the radial benchmark and $\Tilde{\phi}_{\mathrm{2D}}$. By way of comparison, applying the same metric on the Newtonian potential yields $\Gamma_{\Tilde{r}} \sim 10^{-7}$. Evaluating this metric at different altitudes is also a way to make sure that none of the 2D solutions behaves unexpectedly in the radial direction.

\begin{table*}
\renewcommand{\arraystretch}{1.2}
\begin{ruledtabular}
\begin{tabular}{ccccccc}
& $\alpha$ & $\Tilde{r}=1.059$ & $\Tilde{r}=1.111$ & $\Tilde{r}=1.314$ & $\Tilde{r}=4.645$ & $\Tilde{r}=6.617$ \\\midrule
 \parbox[t]{2mm}{\multirow{18}{*}{\rotatebox[origin=c]{90}{No-Atmosphere}}} & $10^{-4}$ & $0$ & $0$  & $0$  & $8.7 \times 10^{-7}$ & $8.0 \times 10^{-7}$ \\
& $10^{-5}$ & $0$ & $0$ & $0$ & $7.9 \times 10^{-7}$ & $7.4 \times 10^{-7}$ \\
& $10^{-6}$ & $2.6 \times 10^{-3}$ & $1.4 \times 10^{-3}$ & $5.0 \times 10^{-4}$ & $4.2 \times 10^{-5}$ & $2.7 \times 10^{-5}$ \\
& $10^{-10}$ & $0$ & $0$ & $0$ & $2.5 \times 10^{-6}$ & $1.7 \times 10^{-6}$ \\
& $10^{-11}$ & $6.9 \times 10^{-6}$ & $1.7 \times 10^{-6}$ & $1.7 \times 10^{-6}$ & $1.7 \times 10^{-6}$ & $1.7 \times 10^{-6}$ \\
& $10^{-12}$ & $6.9 \times 10^{-6}$ & $6.9 \times 10^{-6}$ & $6.9 \times 10^{-6}$ & $6.8 \times 10^{-6}$ & $6.9 \times 10^{-6}$ \\
& $10^{-14}$ & $3.9 \times 10^{-5}$ & $3.9 \times 10^{-5}$ & $3.9 \times 10^{-5}$ & $3.8 \times 10^{-5}$ & $3.8 \times 10^{-5}$ \\
& $10^{-15}$ & $1.5 \times 10^{-5}$ & $1.4 \times 10^{-5}$ & $1.4 \times 10^{-5}$ & $1.3 \times 10^{-5}$ & $1.3 \times 10^{-5}$ \\
& $10^{-16}$ & $1.3 \times 10^{-5}$ & $7.5 \times 10^{-6}$ & $3.3 \times 10^{-6}$ & $9.7 \times 10^{-7}$ & $9.0 \times 10^{-7}$ \\
& $10^{-18}$ & $1.7 \times 10^{-4}$ & $1.9 \times 10^{-4}$ & $2.1 \times 10^{-4}$ & $2.0 \times 10^{-4}$ & $2.0 \times 10^{-4}$ \\
& $10^{-20}$ & $0$ & $0$ & $1.1 \times 10^{-4}$ & $9.0  \times 10^{-6}$ & $5.1 \times 10^{-6}$ \\
& $10^{-21}$ & $0$ & $0$ & $5.1 \times 10^{-5}$ & $2.9 \times 10^{-6}$ & $1.3 \times 10^{-6}$ \\
& $10^{-23}$ & $0$ & $0$ & $1.7 \times 10^{-5}$ & $5.7 \times 10^{-9}$ & $4.5 \times 10^{-9}$ \\
& $10^{-24}$ & $0$ & $0$ & $1.1 \times 10^{-6}$ & $4.8 \times 10^{-10}$ & $4.3 \times 10^{-10}$ \\
& $10^{-25}$ & $0$ & $0$ & $0$ & $4.5 \times 10^{-11}$ & $4.9 \times 10^{-11}$ \\
& $10^{-26}$ & $0$ & $0$ & $1.1 \times 10^{-12}$ & $4.7 \times 10^{-12}$ & $4.7 \times 10^{-12}$ \\
& $10^{-27}$ & $0$ & $0$ & $0$ & $4.0 \times 10^{-13}$ & $5.8 \times 10^{-13}$ \\
& $10^{-28}$ & $5.7 \times 10^{-14}$ & $0$ & $0$ & $2.8 \times 10^{-14}$ & $4.3 \times 10^{-14}$ \\ \midrule
\parbox[t]{2mm}{\multirow{8}{*}{\rotatebox[origin=c]{90}{Tenuous}}} & $10^{-6}$ & $4.5 \times 10^{-6}$ & $4.3 \times  10^{-6}$ & $4.1 \times 10^{-6}$ & $4.0 \times 10^{-6}$ & $3.6 \times 10^{-6}$ \\
& $10^{-10}$ & $4.7 \times 10^{-6}$ & $4.7 \times 10^{-6}$ & $4.7 \times 10^{-6}$ & $4.8 \times 10^{-6}$ & $4.4 \times 10^{-6}$ \\
& $10^{-11}$ & $0$ & $0$ & $0$ & $6.5 \times 10^{-6}$ & $6.2 \times 10^{-6}$ \\
& $10^{-12}$ & $1.7 \times 10^{-3}$ & $9.4 \times 10^{-4}$ & $3.4 \times 10^{-4}$ & $4.0 \times 10^{-5}$ & $3.0\times 10^{-5}$ \\
& $10^{-14}$ & $1.0 \times 10^{-3}$ & $5.7 \times 10^{-4}$ & $2.4 \times 10^{-4}$ & $6.0 \times 10^{-5}$ & $5.4 \times 10^{-5}$ \\
& $10^{-15}$ & $2.4 \times 10^{-4}$ & $1.5 \times 10^{-4}$ & $7.4 \times 10^{-5}$ & $2.6 \times 10^{-5}$ & $2.3 \times 10^{-5}$ \\
& $10^{-17}$ & $1.9 \times 10^{-5}$ & $1.3 \times 10^{-5}$ & $8.4 \times 10^{-6}$ & $5.3 \times 10^{-6}$ & $4.7 \times 10^{-6}$ \\
& $10^{-20}$ & $1.9 \times 10^{-6}$ & $2.0 \times 10^{-6}$ &  $2.1 \times 10^{-6}$ & $2.5 \times 10^{-6}$ & $2.3 \times 10^{-6}$ \\ \midrule
\parbox[t]{2mm}{\multirow{10}{*}{\rotatebox[origin=c]{90}{Earth-like}}} & $10^{-5}$ & $7.5 \times 10^{-6}$ & $6.9 \times 10^{-6}$ & $5.7 \times 10^{-6}$ & $4.3 \times 10^{-6}$ & $3.9 \times 10^{-6}$ \\
& $10^{-6}$ & $4.5 \times 10^{-6}$ & $4.3 \times 10^{-6}$ & $4.1 \times 10^{-6}$ & $4.0 \times 10^{-6}$ & $3.6\times 10^{-6}$ \\
& $10^{-8}$ & $2.6 \times 10^{-3}$ & $1.4 \times 10^{-3}$ & $5.0 \times 10^{-4}$ & $4.7 \times 10^{-5}$ & $3.1 \times 10^{-5}$ \\
& $10^{-10}$ & $3.8 \times 10^{-7}$ & $2.8 \times 10^{-6}$ & $4.3 \times 10^{-6}$ & $4.7 \times 10^{-6}$ & $4.4 \times 10^{-6}$ \\
& $10^{-12}$ & $1.1 \times 10^{-2}$ & $6.2 \times 10^{-3}$ & $2.2 \times 10^{-3}$ & $2.2 \times 10^{-4}$ & $1.5 \times 10^{-5}$ \\
& $10^{-14}$ & $2.5 \times 10^{-3}$ & $1.4 \times 10^{-3}$ & $5.5 \times 10^{-4}$ & $8.8 \times 10^{-5}$ & $7.2 \times 10^{-5}$ \\
& $10^{-15}$ & $4.9 \times 10^{-5}$ & $3.6 \times 10^{-5}$ & $2.5 \times 10^{-5}$ & $1.9 \times 10^{-5}$ & $1.9 \times 10^{-5}$ \\
& $10^{-17}$ & $1.1 \times 10^{-5}$ & $8.3 \times 10^{-6}$ & $6.7 \times 10^{-6}$ & $5.7 \times 10^{-6}$ & $5.3 \times 10^{-6}$ \\
& $10^{-20}$ & $0$ & $9.7 \times 10^{-7}$ & $2.1 \times 10^{-6}$ & $2.4 \times 10^{-6}$ & $2.2 \times 10^{-6}$ \\
& $10^{-23}$ & $0$ & $7.7 \times 10^{-5}$ & $1.7 \times 10^{-5}$ & $7.3 \times 10^{-4}$ & $1.4 \times 10^{-3}$ \\ \midrule
\parbox[t]{2mm}{\multirow{8}{*}{\rotatebox[origin=c]{90}{Dense}}} & $10^{-6}$ & $4.3 \times 10^{-6}$ & $4.3 \times 10^{-6}$ & $4.1 \times 10^{-6}$ & $4.0 \times 10^{-6}$ & $3.6 \times 10^{-6}$ \\
& $10^{-8}$ & $1.9 \times 10^{-5}$ & $1.1 \times 10^{-5}$ & $5.9 \times 10^{-6}$ & $4.1 \times 10^{-6}$ & $3.7 \times 10^{-6}$ \\
& $10^{-10}$ & $0$ & $0$ & $4.8 \times 10^{-2}$ & $3.2 \times 10^{-3}$ & $2.1 \times 10^{-3}$ \\
& $10^{-15}$ & $0$ & $0$ & $0$ & $2.8 \times 10^{-5}$ & $ 2.7 \times 10^{-5}$ \\
& $10^{-18}$ & $0$ & $0$ & $0$ & $3.3 \times 10^{-6}$ & $2.9 \times 10^{-6}$ \\
& $10^{-20}$ & $0$ & $0$ & $0$ & $1.7 \times 10^{-6}$ & $1.6 \times 10^{-6}$ \\
& $10^{-21}$ & $0$ & $0$ & $0$ & $5.3 \times 10^{-7}$ & $6.3 \times 10^{-7}$ \\
& $10^{-25}$ & $0$ & $0$ & $0$ & $0$ & $0$ \\
\end{tabular}
\end{ruledtabular}
\caption{\label{tab:Gamma_r-all} Metric $\Gamma_{\Tilde{r}}$ for $\Tilde{r} \in \{ 1.059, 1.111, 1.314, 4.645, 6.617 \}$ and for all ($\alpha$, atmospheric scenario) pairs considered in this work.}
\end{table*}

\subsection{Check of the strong residual amplitude with respect to each term}

As mentioned in Sec.~\ref{subsubsec:challenges-verif}, the strong residual alone does not provide much insight into how good the numerical approximation is at the end of Newton's iterations. However, it is meaningful to compare locally the size of the residual against the size of each term in it, namely
\begin{equation}
    \left|\alpha \Tilde{\Delta} \Tilde{\phi}\right| \quad , \quad \left|\Tilde{\rho}\right| \quad , \quad \left|\Tilde{\phi}^{-(n+1)}\right| \, .
\label{eqn:residual-terms}
\end{equation}
A numerical approximation deemed acceptable must be such that the residual should be at least a few orders of magnitude smaller than the dominant term in (\ref{eqn:residual-terms}).

\begin{figure*}
    \centering
    \includegraphics[width=\textwidth]{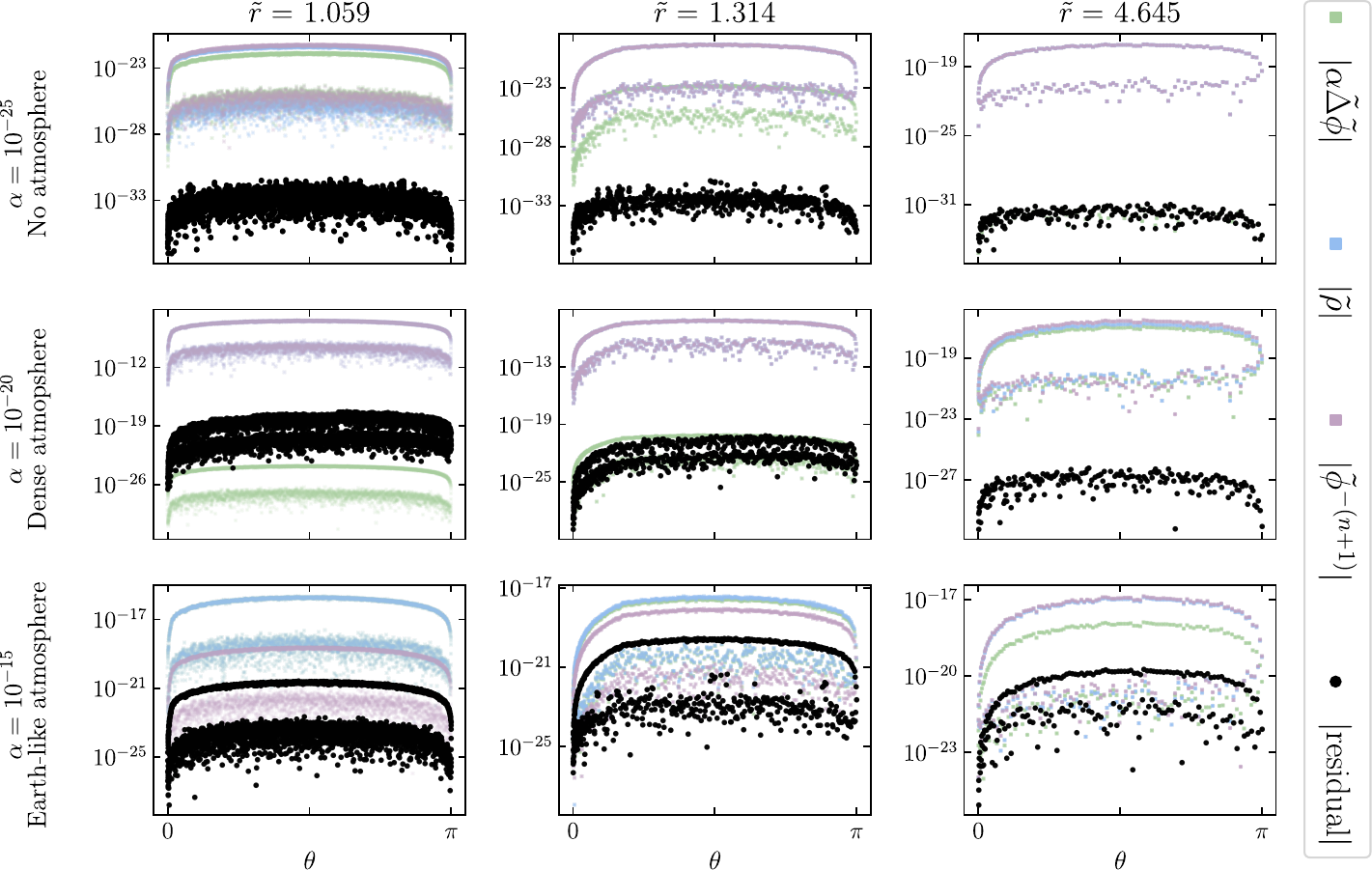}
    \caption{Representation of the strong residual (black circles) and the various terms of the dimensionless Klein-Gordon equation (\ref{eqn:KG-alpha}) (pastel-colored squares) in absolute values as a function of $\theta \in [0, \pi]$. Each column corresponds to a given radial coordinate $\Tilde{r} \in \{ 1.059, 1.314, 4.645 \}$ whereas each row corresponds to a given pair ($\alpha$, atmospheric scenario). In all cases, the absolute value of the strong residual remains at least several orders of magnitude below the dominant term of the Klein-Gordon equation, which is in line with the criterion set out in Sec.~\ref{subsubsec:challenges-verif}. The splitting of the curves associated with each term is due to the fact that we use second-order finite elements.}
    \label{fig:residual-parts}
\end{figure*}

This criterion was assessed for five specific values of $\Tilde{r} \in \{1.059, 1.111,  1.314, 4.645, 6.617\}$ on all numerical approximations discussed in this study. In Fig.~\ref{fig:residual-parts}, we show several examples of scatter plots that allowed us to do this monitoring. Each sub-panel corresponds to a given altitude and a given ($\alpha$, atmospheric scenario) pair \textemdash \ both randomly chosen \textemdash, and depicts the absolute value of the residual (black dots) \textit{vs} terms appearing in (\ref{eqn:residual-terms}) (pastel-colored squares) as functions of $\theta$. We see that the residual remains well-below the dominant term in absolute values.

\section{Spherical harmonic coefficients at different altitudes}
\label{app:sph-altitudes}

\begin{figure*}
    \centering
    \includegraphics[width=\textwidth]{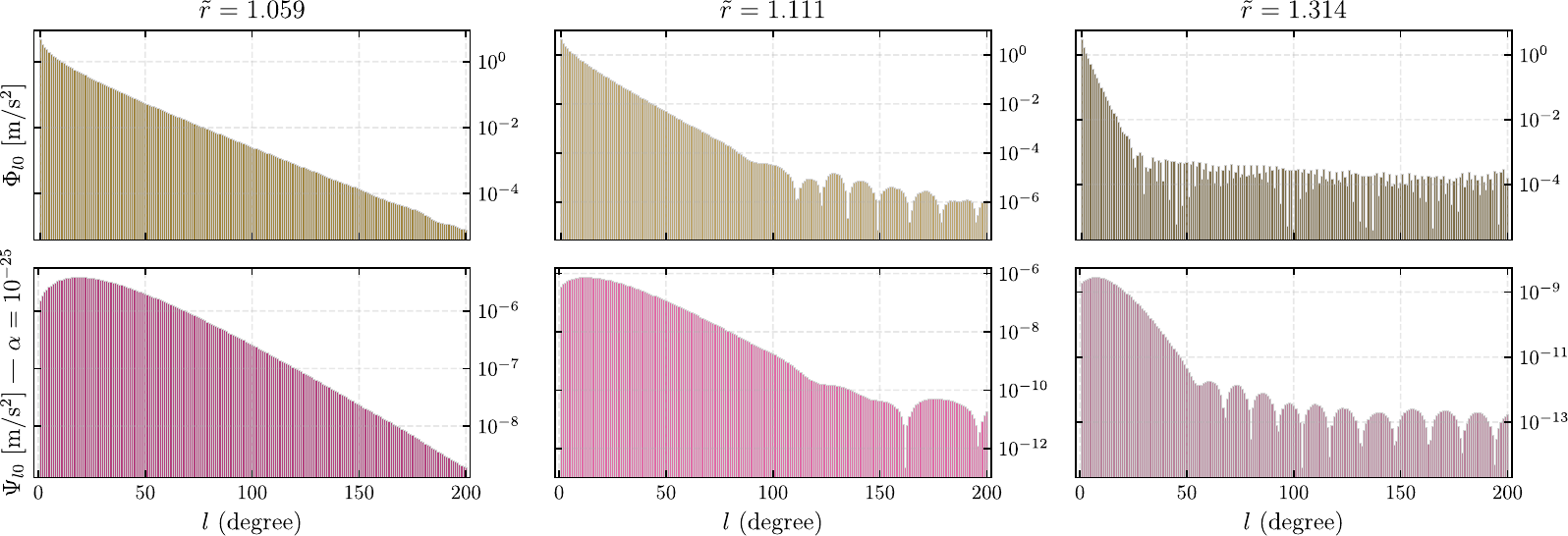}
    \caption{Spherical harmonic coefficients of the Newtonian potential (top row) and of the chameleon potential for $\alpha = 10^{-25}$ (bottom row). The spectra are computed at three different altitudes, namely $\Tilde{r} \in \{1.059, 1.111, 1.314\}$.}
    \label{fig:sph-annexe}
\end{figure*}

For the sake of comprehensiveness, we provide histograms of the spherical harmonic coefficients of both the Newtonian potential $\Phi$ and the chameleon potential $\Psi$ (up to degree $l=200$) at three altitudes in Fig.~\ref{fig:sph-annexe} (see Sec.~\ref{subsubsec:profiles-sph} in the main text). The specific shapes of both potential decompositions hold at all three altitudes, although they get squashed toward lower degrees the higher we go. The oscillations that we observe at high degrees for $\Tilde{r} \in \{1.111, 1.314\}$ are not deemed physical but can be rather attributed to numerical noise.

\section{Further insights into the fifth force experienced by an unscreened satellite}
\label{app:remarkable}

In Sec.~\ref{subsec:satellite-screening}, we computed the total chameleon acceleration undergone by a satellite in orbit. We found that, as long as the satellite was not screened, the resulting force (computed numerically by integrating the gradient of the scalar field over the whole volume occupied by the satellite) was equal to that acting on an equal mass point-like particle. In other words, the back-reaction of the satellite on the scalar field, in the unscreened regime, is such that there is no \textit{self-force} perturbing the Jordan frame geodesics. In this Appendix, we provide an explanation for this phenomenon observed through numerical simulations, based on several approximations that can be justified in the \{Earth + Satellite\} system.

\subsection{The case of Newtonian gravity}

Given two massive bodies labeled by the subscripts $i \in \{1, \, 2\}$, the total gravitational force acting on the second body is
\begin{equation*}
    \mathbf{F}_2 = - \int_{V_2} \boldsymbol{\nabla} \Phi_{N}(\mathbf{x}) \, \mathrm{d}m(\mathbf{x}) \, .
\end{equation*}
In the above expression, $V_2$ is the volume occupied by the body 2 and $\Phi_N$ is the total Newtonian potential created by the two bodies. Thanks to the linearity of the Poisson equation governing the Newtonian potential, one can apply the superposition principle $\Phi_N = \Phi_1 + \Phi_2$, where $\Phi_i$ is the potential sourced by the body $i$ alone. The $\boldsymbol{\nabla}$ operator and the integral being linear, we get
\begin{equation*}
    \begin{split}
        \mathbf{F}_2 = - \int_{V_2} \boldsymbol{\nabla} \Phi_1(\mathbf{x}) \, \mathrm{d}m(\mathbf{x}) - \int_{V_2} \boldsymbol{\nabla} \Phi_2(\mathbf{x}) \, \mathrm{d}m(\mathbf{x}) \, .
    \end{split}
\end{equation*}
Physically speaking, the first integral represents the force exerted by 1 on 2 while the second integral is the force exerted by 2 on 2, which must be zero according to Newton's third law. This can be mathematically proven fairly easily given that
\begin{equation*}
\begin{split}
    & \Phi_2(\mathbf{x}) = -G \int_{V_2} \frac{\mathrm{d}m(\mathbf{x'})}{\|\mathbf{x} - \mathbf{x'}\|} \\[3pt]
    \text{and} \ & \boldsymbol{\nabla} \left( \|\mathbf{x} - \mathbf{x'}\|^{-1} \right) = - \frac{\mathbf{x}-\mathbf{x'}}{\|\mathbf{x} - \mathbf{x'}\|^{3}} \, .
\end{split}
\end{equation*}
We thus get
\begin{equation*}
    \begin{split}
        \int_{V_2} & \boldsymbol{\nabla} \Phi_2(\mathbf{x}) \, \mathrm{d}m(\mathbf{x}) = G \int_{V_2} \left( \int_{V_2} \frac{\mathbf{x}-\mathbf{x'}}{\|\mathbf{x} - \mathbf{x'}\|^{3}} \, \mathrm{d}m(\mathbf{x'}) \right) \, \mathrm{d}m(\mathbf{x}) \\[5pt]
        &= \frac{G}{2} \int_{V_2} \int_{V_2} \frac{\mathbf{x}-\mathbf{x'}}{\|\mathbf{x} - \mathbf{x'}\|^{3}} \, \mathrm{d}m(\mathbf{x'}) \, \mathrm{d}m(\mathbf{x}) \\[5pt]
        & - \frac{G}{2} \int_{V_2} \int_{V_2} \frac{\mathbf{x'}-\mathbf{x}}{\|\mathbf{x'} - \mathbf{x}\|^{3}} \, \mathrm{d}m(\mathbf{x}) \, \mathrm{d}m(\mathbf{x'}) \\[5pt]
        &= 0
    \end{split}
\end{equation*}
In conclusion, despite disturbing the overall Newtonian potential, the body 2 experiences the force sourced by the body 1 only. Furthermore, if the body 2 is small enough that $\boldsymbol{\nabla} \Phi_1$ is approximately constant over $V_2$, we recover the point-mass approximation, i.e. $\mathbf{F}_2 \simeq - m_2 \boldsymbol{\nabla} \Phi_1 (\mathbf{x}_2)$.

\subsection{The case of the chameleon field in the unscreened regime}

The above demonstration relies mainly on the superposition principle, which is lost in the case of the chameleon field because of the nonlinear nature of the Klein-Gordon equation governing the scalar field Eq.~(\ref{eqn:KG-dim}). Nonetheless, let $\phi_{\mathrm{tot}} \coloneqq \phi_{\oplus} + \delta \phi$ be the chameleon field of the \{Earth + Satellite\} system, where $\phi_{\oplus}$ is the background field of the Earth alone. Working with the dimensionless version of the Klein-Gordon equation (\ref{eqn:KG-alpha}), we have by definition
\begin{equation*}
    \begin{cases}
        \alpha \Delta  \phi_{\mathrm{tot}} &= \left( \rho_{\oplus} + \rho_{\mathrm{Sat}} + \rho_{\mathrm{vac}} \right)(\mathbf{x}) - \phi_{\mathrm{tot}}^{-(n+1)} \\[8pt]
        \alpha \Delta \phi_{\oplus} &= \left(  \rho_{\oplus} + \rho_{\mathrm{vac}} \right)(\mathbf{x}) - \phi_{\oplus}^{-(n+1)}
    \end{cases} \, .
\end{equation*}
In the unscreened case, $\delta \phi$ can indeed represent a small perturbation with respect to the background field $\phi_{\oplus}$ \textemdash \ see e.g. the case illustrated in Fig.~\ref{fig:screening-sat}. Then, the nonlinear term can be approximated as
\begin{equation*}
    \phi_{\mathrm{tot}}^{-(n+1)} \simeq \phi_{\oplus}^{-(n+1)} - (n+1) \phi_{\oplus}^{-(n+2)} \delta \phi \, , 
\end{equation*}
so that
\begin{equation}
    \alpha \Delta \delta \phi \simeq \rho_{\mathrm{Sat}}(\mathbf{x}) + (n+1) \phi_{\oplus}^{-(n+1)} \frac{\delta \phi}{\phi_{\oplus}} \, .
\label{eqn:KG-alpha-linearized}
\end{equation}
The r.h.s. of Eq.~(\ref{eqn:KG-alpha-linearized}) can be further simplified if we assume that, at the satellite's altitude, $\phi_{\oplus}$ is close to its asymptotic value in vacuum, that is $\phi_{\oplus}(\mathbf{x}_{\mathrm{Sat}}) \sim \rho^{-1/(n+1)}$. Then we have, depending on whether $\mathbf{x} \in V_{\mathrm{Sat}}$,
\begin{itemize}
    \item[--] Inside the satellite: $\rho_{\mathrm{Sat}}(\mathbf{x}) \neq 0$ and so $$ \frac{\phi_{\oplus}^{-(n+2)} \delta \phi}{\rho_{\mathrm{Sat}}(\mathbf{x})} \sim \frac{\delta \phi}{\phi_{\oplus}} \frac{\rho_{\mathrm{vac}}}{\rho_{\mathrm{Sat}}} \ll 1 \, . $$ Consequently, Eq.~(\ref{eqn:KG-alpha-linearized}) can be legitimately approximated by a Poisson equation inside the satellite.
    \item[--] Outside the satellite: $\rho_{\mathrm{Sat}}(\mathbf{x}) = 0$. We still have $\delta \phi / \phi_{\oplus}^{n+2} \ll 1$ and $\delta \phi \to 0$ as one moves away from the satellite (while $\phi_{\oplus}^{-(n+2)}$ remains bounded) so that we essentially recover a Laplace equation.
\end{itemize}

In brief, we showed that, under some assumptions, $\delta \phi$ obeys a Poisson equation inside the satellite, and a Laplace equation outside the satellite. The Newtonian potential sourced by the satellite (denoted by $\Phi_2$ in the previous discussion) is governed by the same partial differential equation. Yet, \textit{same equations have the same solutions}, which means that $\delta \phi$ has a role similar to $\Phi_2$. Therefore, following the demonstration made in the case of the Newtonian potential above, we get
\begin{equation}
\begin{split}
    \mathbf{F}^{\mathrm{5^{th}}}_{\mathrm{Sat}} &= - \frac{\beta}{M_{\mathrm{Pl}}} \int_{V_{\mathrm{Sat}}} \boldsymbol{\nabla} \phi_{\mathrm{tot}} \, \mathrm{d}m(\mathbf{x}) \\[8pt]
    & \simeq -\frac{\beta}{M_{\mathrm{Pl}}} \int_{V_{\mathrm{Sat}}} \boldsymbol{\nabla} \phi_{\oplus} \, \mathrm{d}m(\mathbf{x}) \, ,
\end{split}
\end{equation}
QED.

\section{Mathematical proof of the absence of symmetry in the orbital dynamics}
\label{app:proof}

\paragraph*{Context \& Notations.} In Sec.~\ref{subsec:GFO}, we have seen on simulation results that, although the gravity field is exactly symmetric with respect to the line $\theta = 0$, the dynamics of a point-mass in orbit is not. Let us translate this statement into mathematical terms. Let $\{ \mathbf{X}_*(t) \, , \ t > 0 \}$ be a trajectory in phase space that is a solution of the ODE of interest, i.e. $\forall t > 0 , \, \Dot{\mathbf{X}}_*(t) = F(t, \mathbf{X}_*(t))$. At some point, the particle will pass over the mountain so that we can define $t_m$, the time at which $\theta(t_m) = 0$ for the first time. Demanding that the trajectory is symmetric with respect to $\theta = 0$ actually means
\begin{equation}
   \forall t \in [0, t_m] \, , \ \mathbf{X}_*(t) = \mathbf{X}_*(2 t_m - t) \, .
\label{eqn:symm-def}
\end{equation}
The state vector $\mathbf{X}_*(t)$ has components
\begin{equation*}
\begin{split}
    \mathbf{X}_*(t) &= \left[ \delta r(t), \, \Dot{\delta r}(t), \, \delta \theta(t), \, \delta L(t) \right] \\[5pt]
    &= \left[ x_1(t), \, x_2(t), \, x_3(t), \, x_4(t) \right] \, .
\end{split}
\end{equation*}
We furthermore recall that the vector field $F \colon (s, \mathbf{Y}) \in \mathbb{R} \times \mathbb{R}^4 \mapsto (F_1, F_2, F_3, F_4) \in \mathbb{R}^4 $ is given by
\begin{equation}
\begin{split}
    F_1 &= y_2 \\[5pt]
    F_2 &= \frac{(L_0 + y_4)^2}{(a + y_1)^3} + g(a + y_1, \theta_0 + \omega s + y_3) \\[5pt]
    F_3 &= \frac{L_0 + y_4}{(a + y_1)^2} - \omega \\[5pt]
    F_4 &= h(a + y_1, \theta_0 + \omega s + y_3) \, .
\end{split}
\label{eqn:vector-field}
\end{equation}
In the above, functions $g$ and $h$ refer to the gravitational potential partial derivatives $- \partial_r U$ and $- \partial_{\theta} U$ respectively.

\paragraph*{Theorem.} The perturbed Keplerian problem with vector field (\ref{eqn:vector-field}) is not symmetric on both sides of the mountain in the general case.

\begin{figure}
    \centering
    \includegraphics[width=\linewidth]{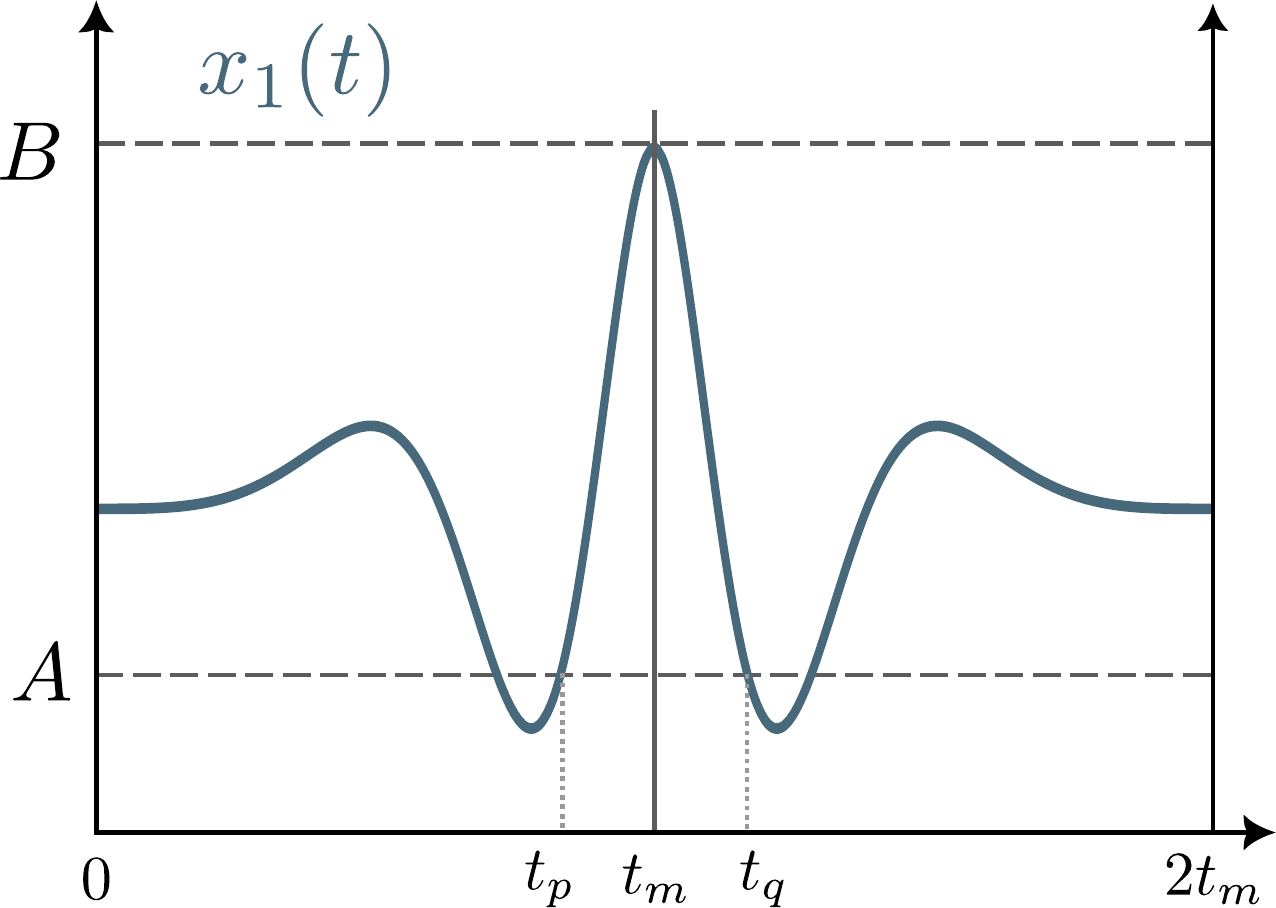}
    \caption{Visual support for the proof.}
    \label{fig:math-proof}
\end{figure}

\paragraph*{Proof by contradiction.}  Let us suppose that (\ref{eqn:symm-def}) holds and derive a (necessary) condition on function $g$ and $h$. Because $x_1$ is continuous, there exists $t_p \in [0, t_m[$ such that $x_1$ is monotonous over $[t_p, t_m]$. Letting $t_q = 2 t_m - t_p$, $x_1$ is also monotonous over $[t_m, t_q]$ due to the symmetry (\ref{eqn:symm-def}). We further set
\begin{equation*}
    \begin{cases}
        A &= x_1(t_p) = x_1(t_q) \\
        B &= x_1(t_m)
    \end{cases} \, .
\end{equation*}
Notations introduced so far are shown in Fig.~\ref{fig:math-proof}. Let us assume for now that $A \neq B$ so that $x_1$ is actually \textit{strictly} monotonous over $[t_p, t_m]$ and $[t_m, t_q]$ respectively. Then $V \coloneqq [\min(A, B), \max(A, B)]$ is not a degenerate interval and we can define
\begin{equation*}
\begin{aligned}[t]
  x_1^p \colon [t_p, t_m] &\to V\\
  t &\mapsto x_1(t) 
\end{aligned}
\hspace{1cm}
\begin{aligned}[t]
  x_1^q \colon [t_m, t_q] &\to V\\
  t &\mapsto x_1(t) 
\end{aligned}
\end{equation*}
together with their respective inverse
\begin{equation*}
\begin{aligned}[t]
  z_1^p \colon V &\to [t_p, t_m]\\
  u &\mapsto z_1^p(u) 
\end{aligned}
\hspace{1cm}
\begin{aligned}[t]
  z_1^q \colon V &\to [t_m, t_q]\\
  u &\mapsto z_1^q(u) 
\end{aligned} \, .
\end{equation*}
We will make use of the following property on the inverse functions
\begin{equation}
  \forall u \in V \, , \  z_1^p(u) = 2 t_m - z_1^q(u) \, .
\label{eqn:inverse-property}
\end{equation}
Indeed, for $u \in V$, there exist two unique times $t_{\alpha} \in [t_p, t_m]$ and $t_{\beta} \in [t_m, t_q]$ such that $u = x_1^p(t_{\alpha}) = x_1^q(t_{\beta})$. Reciprocally, $t_{\alpha} = z_1^p(u)$ and $t_{\beta} = z_1^q(u)$. Then
\begin{flalign*}
    z_1^p(u) &= z_1^p(x_1^q(t_{\beta})) = z_1^p(x_1(t_{\beta})) = z_1^q(x_1(2 t_m - t_{\beta})) \\[3pt]
    &= z_1^p(x_1^p(2 t_m - t_{\beta})) = 2 t_m - t_{\beta} \\[3pt]
    &= 2 t_m - z_1^q(u) \, .
\end{flalign*}
We then compute the integral
\begin{equation*}
    I \coloneqq \int_{t_p}^{t_q} \frac{\mathrm{d}x_1}{\mathrm{d} s}(s) \, x_2(s) \, \mathrm{d}s
\end{equation*}
by two different ways. On the one hand,
\begin{equation}
    I = \int_{t_p}^{t_q} \left| \frac{\mathrm{d} x_1}{\mathrm{d}s}(s) \right|^2 \, \mathrm{d}s
\label{eqn:I1}
\end{equation}
because $\Dot{x}_1(s) = x_2(s)$ along the trajectory. On the other hand,
\begin{equation*}
    I = \int_{t_p}^{t_m} \frac{\mathrm{d}x_1}{\mathrm{d} s}(s) \, x_2(s) \, \mathrm{d}s + \int_{t_m}^{t_q} \frac{\mathrm{d}x_1}{\mathrm{d} s}(s) \, x_2(s) \, \mathrm{d}s \, ,
\end{equation*}
from which we can make the changes of variable $u = x_1^p(s)$ in the first integral and $u = x_1^q(s)$ in the second one, yielding
\begin{equation*}
\begin{split}
    I &= \int_A^B x_2(z_1^p(u)) \, \mathrm{d}u + \int_B^A x_2(z_1^q(u)) \, \mathrm{d}u \\[5pt]
    &= \int_A^B x_2(z_1^p(u)) \, \mathrm{d}u + \int_B^A x_2(2 t_m - z_1^p(u)) \, \mathrm{d}u \\[5pt]
    &= 0 \quad \text{because of symmetry (\ref{eqn:symm-def}).}
\end{split}
\end{equation*}
From Eq.~(\ref{eqn:I1}), we immediately deduce that $\Dot{x}_1 \equiv 0$ on $[t_p, t_q]$. The fact that $x_1$ is constant contradicts our previous assumption that $A \neq B$. Therefore, $A$ has to be equal to $B$. Put in perspective with the fact that $x_1$ is monotonous on $[t_p, t_m]$ and on $[t_m, t_q]$, we have
\begin{itemize}
    \item[\sbt] $x_1$ is constant over $[t_p,  t_q]$. Let $H$ be this constant and define the radial distance $R \coloneqq a + H$.
    \item[\sbt] $\Dot{x}_2 \equiv 0$ on $[t_p, t_q]$ as well.
\end{itemize}
The final stage of this demonstration follows from the specific form of the vector field $F$. Let $s  \in [t_p, t_q]$. For convenience, we recall that $\theta = \theta_0 + \omega s + x_3(s)$ and we denote by $\partial_1$, $\partial_2$ the partial derivatives of a two-variable function with respect to the first and second variable respectively. Taking the derivative with respect to $s$ in the second equation of the ODE system $\Dot{\mathbf{X}_*(s)} = F(s, \mathbf{X}_*(s))$ yields
\begin{equation*}
\begin{split}
    & \frac{\mathrm{d}}{\mathrm{d}s} \left\{ \Dot{x}_2(s) \right\} = \frac{\mathrm{d}}{\mathrm{d}s} \left\{ \frac{[L_0 + x_4(s)]^2}{R^3} + g(R, \theta) \right\} = 0 \\
    \iff & 2 \Dot{x}_4(s) \frac{L_0 + x_4(s)}{R^3} + \left[ \omega + \Dot{x}_3(s) \right] \partial_2 g(R, \theta) = 0 \, .
\end{split}
\end{equation*}
In this last equation, we can substitute $\Dot{x}_4$ and $\Dot{x_3}$ by the r.h.s. of the ODE, yielding
\begin{equation}
    \left[ L_0 + x_4(s) \right] \left[ 2 h(R, \theta) + R \partial_2 g(R, \theta) \right] = 0 \, .
\label{eqn:pre-criterion-1}
\end{equation}
The total angular momentum $L(s) = L_0 + x_4(s) = R^2 \Dot{\theta}(s)$ cannot be zero, because otherwise the point-mass would be frozen it time. Moreover, the interval $J \coloneqq \left\{ \theta_0 + \omega s + x_3(s) \, , \ s \in [t_p, t_q] \right\}$ is not a singleton because again, the point-mass cannot remain frozen in time. Therefore, Eq.~(\ref{eqn:pre-criterion-1}) leads straightforwardly to
\begin{equation}
    \forall \theta \in J \, , \ 2h(R, \theta) + R \partial_2 g (R, \theta) = 0 \, .
\label{eqn:pre-criterion-2}
\end{equation}
Replacing functions $g$ and $h$ by their definition in relation to the gravitational potential $U$, we arrive at the final conclusion that $\forall \theta \in J$:
\begin{equation}
\begin{split}
    \partial_{\theta} \left[ 2 U(R, \theta) + R \partial_r U (R, \theta) \right] = 0 \\
    \text{i.e.} \quad 2 U(R, \theta) + R \partial_r U (R, \theta) = C^{st}
\end{split}
\label{eqn:criterion}
\end{equation}
Condition (\ref{eqn:criterion}) is very restrictive on the form of admissible gravitational potentials and one can check that the potential of the \{sphere + mountain\} system that we have been using throughout this article does not satisfy this criterion.

\paragraph*{Conclusion.} We found a necessary condition on the gravitational potential (\ref{eqn:criterion}) for the dynamics of a point-mass in orbit to be symmetric with respect to $\theta = 0$. As a remark, the potential created by a perfect sphere trivially satisfies the criterion.

\section{Orbital periods}
\label{app:deltaT}

\begin{figure*}[t]
    \centering
    \includegraphics[width=1\linewidth]{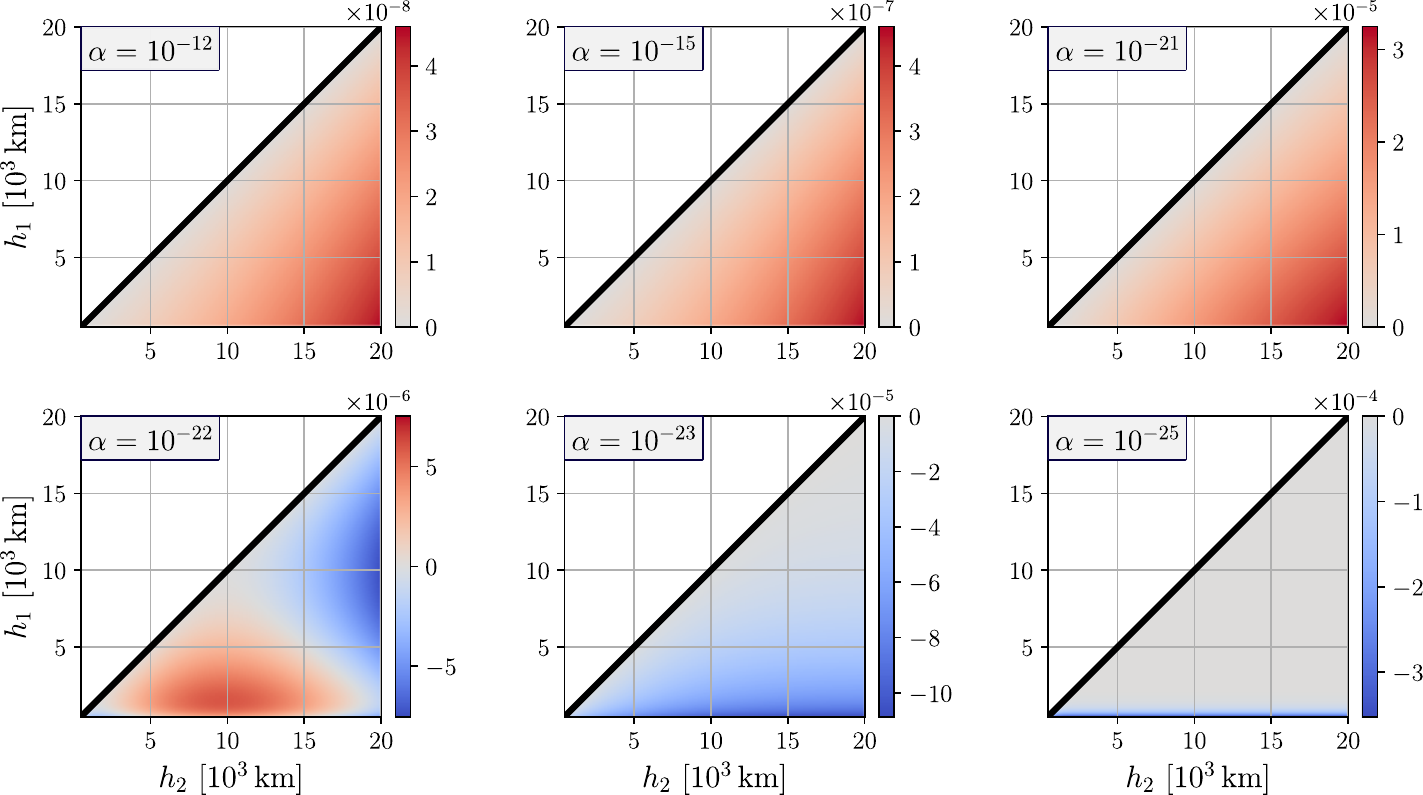}
    \caption{$\mathfrak{D} = \Delta T^{\mathrm{New}} - \Delta T^{\mathrm{cham}} \ \mathrm{[s]}$. We set $2 \times 10^4 \, \mathrm{km} \geq h_2 \geq h_1 \geq 5 \times 10^2 \, \mathrm{km}$ and the black solid line corresponds to $h_1 = h_2$. For $-\log_{10} \alpha \in \{12, 15, 21\}$, the anomaly $\mathfrak{D}$ is maximal when the two considered altitudes are far apart. The case $\alpha= 10^{-22}$ exhibits the transition between the latter regime and a new regime ($\alpha \leq 10^{-23}$) where $h_1$ has to remain low to produce a notable anomaly. This second regime arises because in the limit $\alpha \to 0$, the chameleon field quickly (i.e. at low altitude) gets close to its asymptotic value. The fifth force thus vanishes at higher altitudes, and so does the orbital period anomaly $\mathfrak{D}$.}
    \label{fig:deltaT}
\end{figure*}

In classical central force problems, the period $T$ of a satellite in circular orbit around a planet can be expressed as a function of the distance $r$ to the planet's center and the acceleration $a$ it undergoes: $T = 2 \pi \sqrt{r/a}$. A direct consequence of this formula is that the addition of the chameleon acceleration to the Newtonian one will slightly modify the orbital period. In Sec.~\ref{subsubsec:break-degeneracy}, we laid emphasis on the fact that the use of different altitudes was one possible way of circumventing the issue of model uncertainties. We can thus examine the difference in orbital period for the two gravity models, namely
\begin{flalign*}
    \Delta T^{\mathrm{New}} &= 2 \pi \left( \sqrt{\frac{R_{\oplus}+h_1}{a_{\mathrm{New}}(r_1)}} - \sqrt{\frac{R_{\oplus}+h_2}{a_{\mathrm{New}}(r_2)}} \right) \, , \\[10pt]
    \Delta T^{\mathrm{cham}} &= 2 \pi \Bigg( \sqrt{\frac{R_{\oplus}+h_1}{(a_{\mathrm{New}} + a_{\mathrm{cham}})(r_1)}} \\
    &\quad - \sqrt{\frac{R_{\oplus}+h_2}{(a_{\mathrm{New}} + a_{\mathrm{cham}})(r_2)}} \Bigg) \, ,
\end{flalign*}
with $r = R_{\oplus} + h$ and $a_{\mathrm{New}}(r_1) = \mu_{\oplus}/r^2$.

Fig.~\ref{fig:deltaT} illustrates the difference $\mathfrak{D} = \Delta T^{\mathrm{New}} - \Delta T^{\mathrm{cham}}$ (expressed in seconds) in the $(h_1, h_2)$-plane, for several values of the parameter $\alpha$. Equivalently, one can parameterize deviation from Newtonian gravity as $\Delta T^{\mathrm{New}} = \Delta T^{\mathrm{cham}} (1+\epsilon)$. Then we have $\epsilon = \Delta T^{\mathrm{New}} \mathfrak{D}$. In order to remain consistent with the rest of this article, we have fixed $\Lambda = \Lambda_{\mathrm{DE}}$ and $n=1$. The setup being spherically symmetric, the numerical computation of the chameleon fifth force can be performed with 1D finite elements. We can see that the orbital period anomaly $\mathfrak{D}$ can be greater than $10^{-5} \, \mathrm{s}$. This effect scales linearly with the number of completed orbits: after $100\, 000$ orbits, one can expect the anomaly to be of the order of a second \textemdash \ which would take approximately 20 years for a satellite orbiting $1000 \, \mathrm{km}$ above the Earth surface (ignoring all the perturbing forces otherwise present in a realistic scenario).

\bibliographystyle{apsrev4-2}
\bibliography{references}

\end{document}